\PassOptionsToPackage{pdfpagelabels=false}{hyperref} 
\documentclass[a4paper,fleqn,usenatbib]{mnras}

\usepackage{newtxtext,newtxmath}
\usepackage[T1]{fontenc}
\usepackage{ae,aecompl}
\usepackage{graphicx}
\usepackage{natbib}
\usepackage{amsmath}
\usepackage{amssymb}	

\newcommand{\kms}{km\,s$^{-1}$}
\newcommand{\iaps}{$^1$}
\newcommand{\stsi}{$^2$}
\newcommand{\liver}{$^3$}
\newcommand{\lamm}{$^4$}
\newcommand{\mpia}{$^5$}
\newcommand{\calte}{$^6$}
\newcommand{\unile}{$^7$}
\newcommand{\cnrs}{$^8$}
\newcommand{\unitou}{$^9$}
\newcommand{\nagoya}{$^{10}$}
\newcommand{\calg}{$^{11}$}
\newcommand{\colo}{$^{12}$}
\newcommand{\toron}{$^{13}$}
\newcommand{\leeds}{$^{14}$}
\newcommand{\cardiff}{$^{15}$}
\newcommand{\greno}{$^{16}$}
\newcommand{\cea}{$^{17}$}
\newcommand{\oafi}{$^{18}$}
\newcommand{\koln}{$^{19}$}
\newcommand{\eso}{$^{20}$}
\newcommand{\mpir}{$^{21}$}
\newcommand{\asdc}{$^{22}$}
\newcommand{\esac}{$^{23}$}
\newcommand{\unife}{$^{24}$}
\newcommand{\porto}{$^{25}$}
\newcommand{\unirm}{$^{26}$}
\newcommand{\diet}{$^{27}$}
\newcommand{\estec}{$^{28}$}
\newcommand{\chile}{$^{29}$}
\newcommand{\geneve}{$^{30}$}
\newcommand{\strbg}{$^{31}$}
\newcommand{\cas}{$^{32}$}
\newcommand{\laval}{$^{33}$}
\newcommand{\ira}{$^{34}$}
\newcommand{\hertf}{$^{35}$}
\newcommand{\oact}{$^{36}$}
\newcommand{\lanca}{$^{37}$}
\newcommand{\cern}{$^{38}$}
\newcommand{\oana}{$^{39}$}
\newcommand{\oats}{$^{40}$}
\newcommand{\unina}{$^{41}$}
\newcommand{\sztaki}{$^{42}$}
\newcommand{\ucl}{$^{43}$}
\newcommand{\ral}{$^{44}$}
\newcommand{\nobeyama}{$^{45}$}

\newcommand{\totfittedseds}{100922}
\newcommand{\nproto}{24584}
\newcommand{\nstarless}{76338}
\newcommand{\npre}{16667}
\newcommand{\nunb}{59671}
\newcommand{\highly}{62438}
\newcommand{\barely}{38484}

\newcommand{\recoveredblue}{912}
\newcommand{\recoveredred}{9705}
\newcommand{\nulimblue}{76215}
\newcommand{\nulimred}{9992}
\newcommand{\mirdarkfraction}{10}
\newcommand{\withdistance}{57065}
\newcommand{\withdistanceperc}{56}
\newcommand{\withdistancebadflag}{35904}
\newcommand{\percle}{0.9}
\newcommand{\perclo}{0.3}

\newcommand{\lumexponent}{{1.00}}
\newcommand{\lumcoefficient}{2.56}
\newcommand{\lumexponentmirdark}{{0.79}}
\newcommand{\lumcoefficientmirdark}{3.29}
\newcommand{\narm}{21996}
\newcommand{\niarm}{14539}
\newcommand{\ndist}{36535}
\newcommand{\sigmaslope}{0.04}
\newcommand{\sigmazero}{0.50}
\newcommand{\afitpar}{15.7 \pm 0.01}
\newcommand{\bfitpar}{0.63 \pm 0.01}
\newcommand{\fmipspacs}{0.001}
\newcommand{\fmipspacscompl}{0.004}
\newcommand{\mipsfrac}{3}
\newcommand{\mipsfraccompl}{25}
\newcommand{\tpcenn}{18.1}
\newcommand{\tpconn}{33.1}
\newcommand{\tpcezenn}{9.9}
\newcommand{\tpcozenn}{24.2}
\newcommand{\medtpe}{11.7}
\newcommand{\avetpe}{12.0}
\newcommand{\medtpo}{15.1}
\newcommand{\avetpo}{16.0}
\newcommand{\medtponmir}{14.8}
\newcommand{\avetponmir}{16.0}
\newcommand{\tskewe}{1.06}
\newcommand{\tskewo}{1.40}
\newcommand{\tkurte}{3.67}
\newcommand{\tkurto}{3.04}
\newcommand{\avetpoI}{15.8}
\newcommand{\medtpoI}{14.8}
\newcommand{\avetpoIV}{16.2}
\newcommand{\medtpoIV}{15.3}
\newcommand{\avetpeI}{11.9}
\newcommand{\medtpeI}{11.7}
\newcommand{\avetpeIV}{12.0}
\newcommand{\medtpeIV}{11.8}
\newcommand{\ginsburgfound}{22}
\newcommand{\ginsburgfoundten}{11}
\newcommand{\ginsburgmindist}{5979}
\newcommand{\ginsburgtemp}{16.4}
\newcommand{\ginsburgtempncold}{16}
\newcommand{\ginsburglonmin}{16}
\newcommand{\ginsburglonmax}{67}
\newcommand{\peaklmrpo}{2.5}
\newcommand{\peaklmrpe}{0.3}

\newcommand{\cesalim}{22.4}
\newcommand{\percsubmmo}{86}
\newcommand{\percsubmme}{99.96}
\newcommand{\lmrtentofive}{18482}
\newcommand{\molinI}{50}
\newcommand{\molinII}{20}
\newcommand{\tbmolinI}{40}

\newcommand{\tbwmolin}{55}
\newcommand{\strafi}{500}
\newcommand{\strafii}{30}
\newcommand{\strafiii}{1}
\newcommand{\strafiv}{300}
\newcommand{\nstraf}{2470}
\newcommand{\ncesapalad}{1297}
\newcommand{\ncesapaladperc}{93}
\newcommand{\percpalad}{48}
\newcommand{\npalad}{8034}
\newcommand{\ncesaDDD}{1391}
\newcommand{\ctempsigmupp}{611}
\newcommand{\ctempsigmlow}{673}
\newcommand{\ctempsigmuppperc}{48}
\newcommand{\ctempsigmlowperc}{52}
\newcommand{\numprotodens}{24397}
\newcommand{\numhiidens}{2377}
\newcommand{\nummirdarkdens}{2535}
\newcommand{\numpredens}{34021}
\newcommand{\numpredenstwenty}{30618}
\newcommand{\numpredensninety}{6804}
\newcommand{\tempfraceighty}{26}
\newcommand{\tempfracninety}{28}
\newcommand{\lmrI}{2.3}
\newcommand{\lmrIV}{2.7}
\newcommand{\minslopepre}{-1.46}
\newcommand{\maxslopepre}{-0.88}
\newcommand{\minslopeproto}{-1.23}
\newcommand{\maxslopeproto}{-0.88}
\newcommand{\numratioi}{1.63}
\newcommand{\numratioiv}{1.25}

\newcommand{\totplotted}{816}

\newcommand{\nbisector}{459}
\newcommand{\northog}{226}
\newcommand{\remainingdists}{131}
\newcommand{\remainingperc}{16}

\title[Properties of Hi-GAL clumps in the inner Galaxy]{The Hi-GAL compact source catalogue.\\I. The physical properties of the clumps in the inner Galaxy ($-71.0^{\circ}< \ell < 67.0^{\circ}$)}
   \author[D. Elia et al.]{Davide Elia,$^1$\thanks{E-mail: davide.elia@iaps.inaf.it}
S. Molinari,\iaps~
E. Schisano,\iaps~
M. Pestalozzi,\iaps~
S. Pezzuto,\iaps~ 
M. Merello,\iaps~
\newauthor
A. Noriega-Crespo,\stsi~
T.~J.~T. Moore,\liver~
D. Russeil,\lamm~
J.~C. Mottram,\mpia~ 
R. Paladini,\calte~
\newauthor
F. Strafella,\unile~
M. Benedettini,\iaps~
J.~P. Bernard,\cnrs$^,$\unitou~
A. Di Giorgio,\iaps~ 
D.~J. Eden,\liver~
\newauthor
Y. Fukui,\nagoya~
R. Plume,\calg~
J. Bally,\colo~
P.~G. Martin,\toron~
S.~E. Ragan,\leeds~
S.~E. Jaffa,\cardiff~
\newauthor
F. Motte,\greno$^,$\cea~
L. Olmi,\oafi~
N. Schneider,\koln~ 
L. Testi,\eso$^,$\oafi~
F. Wyrowski,\mpir~
A. Zavagno,\lamm~
\newauthor
L. Calzoletti,\asdc$^,$\esac~
F. Faustini,\asdc~
P. Natoli,\unife~
P. Palmerim,\lamm$^,$\porto~
F. Piacentini,\unirm~ 
\newauthor 
L. Piazzo,\diet~
G.~L. Pilbratt,\estec~
D. Polychroni,\chile~
A. Baldeschi,\iaps~
M.~T. Beltr\'an,\oafi~
\newauthor
N. Billot,\geneve~
L. Cambr\'esy,\strbg~
R. Cesaroni,\oafi~
P. Garc\'ia-Lario,\esac~
M.~G. Hoare,\leeds~
\newauthor
M. Huang,\cas~
G. Joncas,\laval~
S.~J. Liu,\iaps~
B.~M.~T. Maiolo,\unile~
K.~A. Marsh,\cardiff~
\newauthor 
Y. Maruccia,\unile~ 
P. M\`{e}ge,\lamm~
N. Peretto,\cardiff~
K.~L.~J. Rygl,\ira~
P. Schilke,\koln~
\newauthor 
M.~A. Thompson,\hertf~
A. Traficante,\iaps~
G. Umana,\oact~
M. Veneziani,\calte~
\newauthor 
D. Ward-Thompson,\lanca~
A.~P. Whitworth,\cardiff~ 
H. Arab,\strbg~ 
M. Bandieramonte,\cern~
\newauthor 
U. Becciani,\oact~
M. Brescia,\oana~
C. Buemi,\oact~
F. Bufano,\oact~
R. Butora,\oats~ 
S. Cavuoti,\unina$^,$\oana~
\newauthor 
A. Costa,\oact~
E. Fiorellino,\iaps$^,$\unirm~ 
A. Hajnal,\sztaki~
T. Hayakawa,\nagoya~ 
P. Kacsuk,\sztaki~  
P. Leto,\oact~
\newauthor 
G. Li Causi,\iaps~
N. Marchili,\iaps~
S. Martinavarro-Armengol,\ucl$^,$\ral~
A. Mercurio,\oana~
\newauthor  
M. Molinaro,\oats~    
G. Riccio,\oana~ 
H. Sano,\nagoya~
E. Sciacca,\oact~
K. Tachihara,\nagoya~
K. Torii,\nobeyama~
\newauthor 
C. Trigilio,\oact~
F. Vitello,\oact~
H. Yamamoto\nagoya
\\ \\
Author affiliations are listed at the end of the paper}

\date{Accepted XXX. Received YYY; in original form ZZZ}
\pubyear{2016}

\begin{document}

\label{firstpage}
\pagerange{\pageref{firstpage}--\pageref{lastpage}}
\maketitle

\begin{abstract}
 Hi-GAL is a large-scale survey of the Galactic plane, performed with \textit{Herschel} 
 in five infrared continuum bands between 70 and 500~$\umu$m. 
 We present a band-merged catalogue of spatially matched sources and their properties 
 derived from fits to the spectral energy distributions (SEDs) and heliocentric distances,
 based on the photometric catalogs presented in \citet{mol16a}, covering the portion of 
 Galactic plane $-71.0^{\circ}< \ell < 67.0^{\circ}$. The band-merged catalogue contains 
 \totfittedseds~sources with a regular SED, \nproto~of which show a 70~$\umu$m counterpart 
 and are thus considered proto-stellar, while the remainder are considered starless. 
 Thanks to this huge number of sources, we are able to carry out a preliminary 
 analysis of early stages of star formation, identifying the 
 conditions that characterise different evolutionary phases on a statistically 
 significant basis. We calculate surface densities to investigate the gravitational stability 
 of clumps and their potential to form massive stars. We also explore evolutionary status 
 metrics such as the dust temperature, luminosity and bolometric temperature, finding 
 that these are higher in proto-stellar sources compared to pre-stellar ones. The surface 
 density of sources follows an increasing trend as they evolve from pre-stellar to 
 proto-stellar, but then it is found to decrease again in the majority of the most evolved 
 clumps.
 Finally, we study the physical parameters of 
 sources with respect to Galactic longitude and the association with spiral arms, finding 
 only minor or no differences between the average evolutionary status of sources in the 
 fourth and first Galactic quadrants, or between ``on-arm'' and ``inter-arm'' positions. 
\end{abstract}


\section{Introduction}

The formation of stars remains one of the most important unsolved
problems in modern astrophysics. In particular, it is not clear how
massive stars ($M > 8$~M$_\odot$) form, despite their importance in the
evolution of the Galactic ecosystem \citep[e.g.,][]{fer01,bal05}. The formation 
of high-mass stars is not as well-understood as that of low-mass stars, mainly 
because of a lack of observational facts upon which models can be built. 
High-mass stars are intrinsically difficult to observe
because of their low number in the Galaxy, their large distance from
the Sun and their rapid evolution. Numerous observational surveys have
been undertaken in recent years in different wavebands to obtain a better
statistics on massive pre- and proto-stellar objects. 
\citep{jac06,sti06,law07,chu09,care09,sch09,ros10,hoa12,urq14a,moo15}. Among 
these, Hi-GAL is the only complete far-infrared (FIR) survey of the Galactic plane.

Hi-GAL \citep[\textit{Herschel} InfraRed Galactic Plane Survey,][]{mol10a} is an Open Time
Key Project that was granted about 1000~hours of observing time using the \textit{Herschel}
Space Observatory \citep{pil10}. It delivers a complete and homogeneous survey
of the Galactic plane in five continuum FIR bands between 70 and 500~$\umu$m.
This wavelength coverage allows us to trace the peak of emission of most of 
the cold ($T< 20$~K) dust in the Milky Way at high resolution for the first time, 
material that is expected to trace the early stages of the formation of stars across the 
mass spectrum. Hi-GAL data were taken using in parallel two of the three instruments 
aboard \emph{Herschel}, PACS \citep[70 and 160~$\umu$m bands,][]{pog10} and 
SPIRE \citep[250, 350 and 500~$\umu$m bands,][]{gri10}.

The present paper is meant to complete the discussions presented in \citet{mol16a}
on the construction of the photometric catalogue for the portion of Galaxy in the range 
-71.0$^{\circ} \lesssim \ell \lesssim $ 67.0$^{\circ}$, $|b|< 1.0^{\circ}$, an area 
corresponding to the first Hi-GAL proposal (subsequently extended to the whole Galactic 
plane). In particular here we explain how we went from the photometric catalogue to the 
physical one, introducing band-merging, searching for counterparts, assigning distances 
as well as constructing and fitting spectral energy distributions (SEDs). In the last 
part of the paper, in which the distribution of sources with respect to their position 
in the Galactic plane is analysed in detail, we focus on two regions, i.e. 
289.0$^{\circ}< \ell <$ 340.0$^{\circ}$ and 33.0$^{\circ}< \ell <$ 67.0$^{\circ}$. 
The innermost part of the Galactic plane, including the Galactic centre, where 
kinematic distance estimate is particularly problematic, will be discussed in
a separated paper (Bally et al, in prep.), while the two longitude ranges containing
the two tips of the Galactic bar ($340.0^{\circ}< \ell < 350.0^{\circ}$ and
$19.0^{\circ}< \ell < 33.0^{\circ}$) have been presented in \citet{ven17}.

\subsection{Brief presentation of the surveyed regions}

\subsubsection{Fourth Galactic quadrant}
In the longitude range investigated in more detail in this paper, three spiral arms are in 
view (see, in the following, Figure~\ref{galpos}), according to a four-armed spiral model 
of the Milky Way \citep[][and references therein]{urq14a}. Moving towards the Galactic
centre, for $289^{\circ} \lesssim \ell \lesssim 310^{\circ}$ the Carina-Sagittarius
arm is observed, while at $\ell \sim 310^{\circ}$ the tangent point of the Scutum-Crux
arm is encountered \citep{gar14}. The emission from this latter arm is expected to
dominate up to the tangent point of the Norma arm at $\ell \sim 330^{\circ}$. Around
this longitude, the main peak of the OB star formation distribution across the Galaxy
is found \citep{bro00}. Finally, the tangent point of the subsequent arm, the so-called
\textit{3-kpc}~arm, is located at $\ell \sim 338^{\circ}$ \citep{gar14}, i.e. very
close to the inner limit of the investigated zone.


Significant star formation activity is found in the surveyed region, as testified 
by the presence of 103 out of 481 star forming complexes in the list of \citet{rus03}, 
and of 393 star forming regions out of 1735 in the \citet{ave00} overall catalogue, 29 of 
which are H$\alpha$-emission regions of the RCW catalogue \citep{rod60}. Furthermore, 337 
out of 1449 regions with embedded OB stars of the \citet{bro96,bro00} list are
found in this region of the sky.

Hi-GAL observations of the fourth Galactic quadrant have been already used for studying 
InfraRed Dark Clouds \citep[IRDCs,][]{ega98} in the $300^{\circ} \lesssim \ell \lesssim 30^{\circ}$,
$|b| \leq 1^{\circ}$ range \citep{wil12a,wil12b}, highlighting the fundamental role of
the \textit{Herschel} FIR data for exploring the internal structure of these 
candidate sites for massive star formation. Furthermore, \citet{ven17} used the 
catalogue presented here to study the compact source population in the far tip of the long
Galactic bar. These data will be exploited, if needed, in this article as well, for instance 
for comparison between the fields studied in this paper and inner regions of the Galaxy.

Finally, the nearby Coalsack nebula \citep[$d=100-200$ pc, see references in][]{beu11} is also seen
in the foreground of our field \citep[$300^{\circ} \lesssim \ell \lesssim 307^{\circ}$,][]{wan13}.
It is one of the most prominent dark clouds in the southern Milky Way but shows no evidence
of recent star formation \citep[e.g.,][]{kat99,kai09}.

\subsubsection{First Galactic quadrant}
In the first quadrant portion investigated in more detail in this paper
($33^{\circ} \lesssim \ell \lesssim 67^{\circ}$), two spiral arms are in view, 
namely the Carina-Sagittarius and the Perseus arms. 
The Carina-Sagittarius tangent 
point is found at $\ell \simeq 51^{\circ}$ \citep{val08} near the W51 star forming 
region. From here to the endpoint of the region we are considering, only the Perseus arm 
is expected.

This area is smaller than that surveyed in the fourth quadrant and has a lower 
rate of star formation activity per unit area. \citet{rus03} 
finds 33 star-forming region in this area, and \citet{ave00} finds just 97.

Hi-GAL studies of this portion of the Galactic plane mainly focused so far on one 
of the two \textit{Herschel} Science Demonstration Phase fields, namely the one centered
around $\ell=59^{\circ}$, regarding compact source physical properties obtained
from earlier attempts of photometry of Hi-GAL maps \citep{eli10,ven13,bel13,olm13},
structure of IRDCs \citep{per10b,bat11}, source and filament large-scale disposition
\citep{bil11,mol10b,bal10}, and diffuse emission morphology \citep{mar10}. As in the 
case of the fourth quadrant, the catalogue presented here has been already used by 
\citet{ven17} for studying the clump population at the near tip of the Galactic 
bar. Finally, a recent paper of \citet{ede15} focused on two lines of sight 
centred towards $\ell=30^{\circ}$ and $40^{\circ}$, studying
arm/interarm differences in luminosity distribution of Hi-GAL sources.

\subsection{Structure of the paper}
The present paper is organised as follows: in Section~\ref{mapmaking}, the data 
reduction, source detection and photometry strategy is briefly presented, referring 
to \cite{mol10a} for further detail. In Section~\ref{phottophys} SED building, filtering, 
complementing with ancillary photometry and distances are described. In 
Section~\ref{sedfit} the use of a simple radiative model to derive the physical parameters 
of the Hi-GAL SEDs is illustrated, while the extraction of such properties from the
photometry is summarised in Section~\ref{pipeline}. The statistics of the physical 
properties is discussed in Section~\ref{results}, and the implications on the estimate of 
the evolutionary stage of sources are reported in Section~\ref{evol_par}. Finally, in
Section~\ref{sourcegal}, source properties are correlated with their Galactic positions:
a comparison between sources in the IV and I Galactic quadrants is provided and, after 
positional matching of the sources and the locations of Galactic spiral arms, the 
behaviour of arm vs inter-arm sources is briefly discussed. Further details on how the 
catalogue is organised, and on possible biases affecting the listed quantities, are 
provided in the appendices at the end of the paper.

\section{Data: map making and compact source extraction}\label{mapmaking}

The technical features of the Hi-GAL survey are presented in \cite{mol10a}, therefore 
here we limit ourselves to a summary of only the most relevant aspects. The Galactic plane was
divided into $2^{\circ} \times 2^{\circ}$ sections, called ``tiles'' that were
observed with \textit{Herschel} at a scan speed of 60\arcsec s$^{-1}$ in two orthogonal
directions. PACS and SPIRE were used in ``Parallel mode'' i.e. data were taken
simultaneously with both instruments (and therefore at all five bands). Note that
when used in Parallel mode, PACS and SPIRE observe a slightly different region of
sky. A more complete coverage is nevertheless recovered when considering
contiguous tiles; remaining areas of the sky covered only by either one of
the two are not considered for science in this paper. Single maps of the 
Hi-GAL tiles were obtained from PACS/SPIRE detector timelines using a 
pipeline specifically developed for Hi-GAL and containing the ROMAGAL map making
algorithm (\citealt{tra11}) and the WGLS post-processing \citep{pia12}
for removing artefacts in the maps.

Astrometric consistency with Spitzer MIPSGAL 24~$\umu$m data
\citep[][\url{http://mipsgal.ipac.caltech.edu/}]{care09} is obtained by
applying a rigid shift to the entire mosaic. This is obtained as the mean
shift measured on a number of bright and isolated sources common to Spitzer
24~$\umu$m and Hi-GAL PACS 70~$\umu$m data.

The further astrometric registration of SPIRE maps is then carried out by repeating
the same procedure, but comparing counterparts of the same source at 160 and
250~$\umu$m.

Compact sources were detected and extracted using the algorithm CuTEx
(\citealt{mol11a}) which is based on the study of the curvature of the
images. This is done by calculating the second derivative at any pixel
of the Hi-GAL images, efficiently damping all emission varying on
intermediate to large spatial scales, and amplifying emission concentrated
in small scales. This principle is particularly advantageous when extracting 
sources that appear within a bright and highly variable background emission. 
The final integrated fluxes are then estimated by CuTEx through a bi-dimensional 
Gaussian fit to the source profile. All details of the photometric
catalogue are presented by \citet{mol16a}, who also report estimates of
the completeness limits in flux in each band, measured by extracting a
controlled 90\% sample of sources artificially spread on a representative
sample of real images. Such limits in the regions investigated in the present
paper are discussed in Appendix~\ref{appmasslim} for their implications on the
estimate of source mass completeness limits as a function of the heliocentric
distance.


The final version of the single-band catalogues of the portion of Hi-GAL 
data presented here contain 123210, 308509, 280685, 160972, and 85460 entries in
the 70, 160, 250, 350 and 500~$\umu$m bands, respectively.

\section{From photometry to physics}\label{phottophys}
The present paper focuses on the study of compact cold objects extracted
from \textit{Herschel} data. Within this framework a final catalogue of 
objects for scientific studies has been obtained by merging the Hi-GAL
single-band photometric catalogues and filtering the resulting five-band
catalogue, applying specific constraints to the source SEDs. In the following 
sections the steps of these processes are explained in detail: band-merging, search 
for counterparts beyond the \textit{Herschel} frequency coverage, assigning distances, 
SED filtering and fitting.

\subsection{Band-merging and source selection}\label{filtering}
The first step for creating a multi-wavelength catalogue consists of 
assigning counterparts of a given source across Herschel bands. This operation 
based on iterating a positional matching \citep[cf.][]{eli10,eli13} between source 
lists obtained at two adjacent bands. In the present paper, however, instead 
of assuming a fixed matching radius as done in previous works, the matching 
region consisted of the ellipse describing the source at the longer of the 
two wavelengths\footnote{Such ellipse corresponds to the half-height section 
of the two-dimensional Gaussian fitted by CuTEx to the source profile.}.
In other words, a source has a counterpart at shorter wavelength if the
centroid of the latter falls within the ellipse fitted to the former. In this
way it is possible that more than one counterparts falls into the longer-wavelength 
ellipse. In such multiplicity cases, the association is established only with the 
short-wavelength counterpart closest to the long-wavelength ellipse centroid\footnote{For 
the 70~$\umu$m band, we also take into account of possible multiplicity for the 
estimate of the bolometric luminosity, as is explained in Section~\ref{sedfit}.}.
The remaining ones are reported as independent catalogue entries, and considered
for further possible counterpart search at shorter wavelengths. At the end of
the five band-merging, a catalogue is produced, in which each entry can contain
from one to five detections in as many bands. 

The subsequent step is to filter the obtained five-band catalogue in order to identify
SEDs that are eligible for the modified black body (hereafter grey body) fit, hence to 
derive the physical properties of the objects. This selection is based on considerations
of the regularity of the SEDs in the range 160-500~$\umu$m, since the 70~$\umu$m band
generally is expected to depart from the grey body behaviour \citep[e.g.,][]{bon10,sch12}.

First of all, as done in \citet{eli13}, only sources belonging to the common 
PACS+SPIRE area and detected at least in three consecutive \textit{Herschel} 
bands (i.e. the combinations 160-250-350~$\umu$m or 250-350-500~$\umu$m or, obviously, 
160-250-350-500~$\umu$m) were selected.

Secondly, fluxes at 350 and 500~$\umu$m were scaled according to the ratio of 
deconvolved source linear sizes, taking as a reference the size at 250~$\umu$m
\citep[cf.][]{mot10,ngu11}. This choice is supported by the fact that cold dust 
is expected to have significant emission around 250~$\umu$m; also, according to 
the adopted constraints to filter the SEDs, this is the shortest wavelength in 
common for all the selected SEDs. Finally, we searched for further irregularities 
in the SEDs such as dips in the middle, or peaks at 500~$\umu$m.

At the end of the filtering pipeline, we remain with \totfittedseds~sources. For 
each of these sources, we estimate physical parameters such as dust temperature, 
surface density and, when distance is available, linear size, mass, and luminosity, 
by fitting a single-temperature grey body to the SED. The details of this 
procedure are described in Section~\ref{sedfit}. Clearly,
the determination of source physical quantities such as temperature and mass 
is more reliable when a better coverage of the SED is available. Based on the
selection criteria listed above, sources in our catalogue can be confirmed, even 
considering the 70~$\umu$m flux, with detections at only three bands: this is the 
case for combinations 160-250-350~$\umu$m (with no detection at 70~$\umu$m) and 
250-350-500~$\umu$m (that we call ``SPIRE-only'' sources), which we consider as 
genuine SEDs, although more affected by a less reliable fit (especially the 
SPIRE-only case, in which it might be difficult to constrain the SED peak, and 
consequently the temperature). For this reason, after the SED filtering procedure, 
we further split our SEDs in two sub-catalogues: ``high reliability'' 
(\highly~sources) and ``low reliability'' (\barely~sources). 
Notice that, according to the definitions that will be provided in 
Section~\ref{protovsstarl}, all the sources in the latter list belong to
the class of ``starless'' compact sources.

In Table~\ref{numbertab} the source number statistics for the band-merged catalogue is provided,
divided in ranges of longitude \citep[identified by the two intervals studied by][]{ven17} 
and in evolutionary stages introduced in Section~\ref{protovsstarl}.

\subsection{Caveats on SED building and selection}
Building a five-band catalogue and selecting reliable sources for scientific
analysis require a set of choices and assumptions which have been described
in the previous sections. Here we collect and explicitly recall all of them
to focus the reader's attention on the limitations that must be kept
in mind when using the Hi-GAL physical catalogue:
\begin{enumerate}
\item The concept of ``compact source'' used for this catalogue refers to
unresolved or poorly resolved structures, whose size, therefore, does not
exceed a few instrumental PSFs \citep[$\lesssim 3$,][]{mol16a}. Structures 
with larger angular sizes - such as bright diffuse interstellar medium (ISM), filaments, 
or bubbles - escape from this definition and are not considered in the present catalogue.
\item The appearance of the sky varies strongly throughout the wavelength range
covered by \textit{Herschel}. The lack of a detection in a given band may be ascribed 
to a detection error, or to the physical conditions of the source as, for
instance, the case of a warm source seen by PACS but undetectable at the SPIRE
wavelengths. In this respect, the present catalogue does not aim to describe all 
star formation activity within the survey area, but rather to provide a census 
of the coldest compact structures, corresponding to early evolutionary stages 
in which internal star formation activity has not yet been able to dissipate the
dust envelope, or has not started at all. Detections at  70~$\umu$m only, or
at 70/160~$\umu$m, or at 70/160/250~$\umu$m, are expected to have counterparts
in the mid-infrared (MIR); such cases, corresponding to more evolved objects, 
surely deserve to be further studied, but this lies out of the aims of the present 
paper and is reserved for future works.

\item The band-merging procedure works fine in the ideal case of a source detected
at all wavelengths as a bright and isolated peak. Possible multiplicities, however,
can produce multiple branches in the counterpart association, so that, for instance, 
the flux at a given wavelength might result from the contributions of two or more 
counterparts detected separately at shorter wavelengths. This can also introduce
inconsistent fluxes in the SEDs and produce irregularities such that several bright 
sources present in the Hi-GAL maps might be ruled out from the final catalogue according 
to the constraints described in Section~\ref{filtering}.

\item Very bright sources might be ruled out by the filtering algorithm due to
saturation occurring at one or more bands, which produces unrecoverable gaps in the SEDs.

\item The physical properties derived from \textit{Herschel} SED analysis (see next 
sections) are global (e.g., mass, luminosity) or average (e.g. temperature) quantities 
for sources that, depending on their distance, can be characterised by a certain degree
of internal but unresolved structure (see Section~\ref{physsize}), as will be discussed
in Appendix~\ref{appdist}.
\end{enumerate}


\subsection{Counterparts at non-\textit{Herschel} wavelengths}\label{ancillary}

For every entry in the band-merged filtered catalogue we searched for counterparts at
24~$\umu$m \citep[MIPSGAL,][]{gut15} as well as at 21~$\umu$m \citep[MSX,][]{ega03} and
22~$\umu$m \citep[WISE,][]{wri10}. The fluxes of these counterparts, typically associated 
with a warm internal component of the clump, are not considered for subsequent grey-body 
fitting of the portion of the SED associated cold dust emission, but only for estimating 
the source bolometric luminosity.

In particular, we notice that the choice of sources in the catalogue of \citet{gut15} is rather 
conservative, and only a small fraction has $F_{24}<0.005$~Jy. For this reason we performed an 
additional search of sources in the MIPSGAL maps, using APEX source 
extractor\footnote{\url{http://irsa.ipac.caltech.edu/data/SPITZER/docs/dataanalysistools/tools/mopex/}}
\citep{mak05} in order to recover those sources that, from a visual inspection of the maps,
appear to be real although for some reason were not included in the original catalogue. 
Furthermore, as a cross-check, a similar procedure has been performed also with DAOFIND 
\citep{ste87}, and only sources confirmed by this have been added to the photometry list 
of \citet{gut15}. Following this procedure, approximately 2000 additional SEDs have been 
complemented with a flux at 24~$\umu$m, mostly having fluxes in the 
0.0001~Jy~$<F_{24}<$~0.001~Jy range. The risk of adding poorly
reliable sources with low signal-to-noise ratio is mitigated by the fact that 
24~$\umu$m counterparts of our \textit{Herschel} sources are considered for scientific 
analysis only if they are confirmed by a detection at 70~$\umu$m (see 
Section~\ref{protovsstarl} and Appendix~\ref{mirfakes}.

To assign counterparts to the Hi-GAL sources at 21, 22 and 24~$\umu$m, the
ellipse representing the source at 250~$\umu$m was used as the matching region,
and the flux of all counterparts at a given wavelength within this region was summed up
into one value. Indeed, possible occurrences of multiplicity can induce
a relevant contribution at MIR wavelengths in the calculation of bolometric 
luminosity of Hi-GAL sources.

On the long wavelength side of the SED we cross-matched our bandmerged
and filtered catalogue with those of \citet{cse14} from the ATLASGAL survey
\citep[870~$\umu$m,][]{sch09} and of \citet{gin13} from the BOLOCAM Galactic
Plane Survey \citep[BGPS, 1.1~mm][]{ros10,agu11}. The adopted searching
radius was 19\arcsec~for the former and 33\arcsec~for the latter,
corresponding to the full width at half maximum of the instruments at the
observed wavelengths. Out of 10861 entries in the ATLASGAL catalogue,
10517 of them lie inside the PACS+SPIRE common science area considered in this paper,
6136 of which are found to be associated with a source of our catalogue through this 
1:1 matching strategy. Similarly, 6020 out of 8594 entries of the BGPS catalogue
lie in the common science area, 4618 of which turn out to be associated with an 
entry of our catalogue. Finally, access to ATLASGAL images allowed us to extract
further counterparts, not reported in the list of \citet{cse14}, by using CuTEx.
In cases in which the deconvolved size of the ATLASGAL and/or BGPS counterpart is
larger than the one measured at 250~$\umu$m, fluxes were re-scaled according to 
the procedure described in Section~\ref{filtering}.

\subsection{Distance determination}
\label{dist_sect}
Assigning distances to sources is a crucial step in the process of giving
physical significance to the information extracted from Hi-GAL data. While reliable distance
estimates are available for a limited number of known objects, as for example \ion{H}{ii} regions
\citep[e.g.,][]{fis03} or masers \citep[e.g.,][]{gre11}, this information does not exist for the
majority of Hi-GAL sources. Therefore we adopted the scheme presented in \citet{rus11}, based 
on the Galactic rotation model of \citet{bra93}, to assign kinematic distances to a 
large proportion of sources: a $^{12}$CO (or $^{13}$CO) spectrum is extracted 
at the line of sight of every Hi-GAL source and the Velocity of the Local Standard of Rest 
$V_\mathrm{LSR}$ of the brightest spectral component is assigned to it, allowing the calculation 
of a kinematic distance. To determine the $V_\mathrm{LSR}$, the $^{13}$CO data from the 
Five College Radio Astronomy Observatory 
(FCRAO) Galactic Ring Survey \citep[GRS,][]{jac06}, and $^{12}$CO and $^{13}$CO data from the 
Exeter-FCRAO Survey (Brunt et al., in prep.; Mottram et al., in prep.) were used for the portion 
of Hi-GAL covering the first Galactic quadrant for $\ell<55^{\circ}$ and for $\ell>55^{\circ}$, 
respectively. The pixel size of those CO cubes is 22.5\arcsec, corresponding to Nyquist sampling 
of the FCRAO beam. NANTEN $^{12}$CO data \citep{oni05} were used to 
assign velocities to Hi-GAL sources in the fourth quadrant. The pixel size of these data is $4'$ 
(against an angular resolution of $2'.6$), so that more than one Hi-GAL source might fall onto the 
same CO line of sight, and the same distance is assigned to them. The spectral resolutions of the two 
data sets were 0.15~\kms{} and 1.0~\kms, respectively.

Once the $V_\mathrm{LSR}$ is determined, the near/far distance ambiguity is solved by
matching the source positions with a catalogue of sources with known distances (\ion{H}{ii} regions, 
masers and others) or, alternatively, with features in extinction maps (in this case, the near
distance is assigned). In cases for which none of the aforementioned data can be used, the ambiguity
is always arbitrarily solved in favour of the far distance, and a ``bad quality'' flag is given
to that assignment.


The additional use of extinction maps to solve for distance ambiguity \citep{rus11} 
(where applicable) can be a source of error, whose magnitude typically increases with 
increasing difference between the near and far heliocentric 
distance solutions. For the present paper, we rely on the use of extinction maps for practical 
reasons and also because, for most of the sources, no spectral line emission has yet been observed 
other than what can be extracted from the two CO surveys.

\begin{figure*}
\centering
\includegraphics[width=16cm]{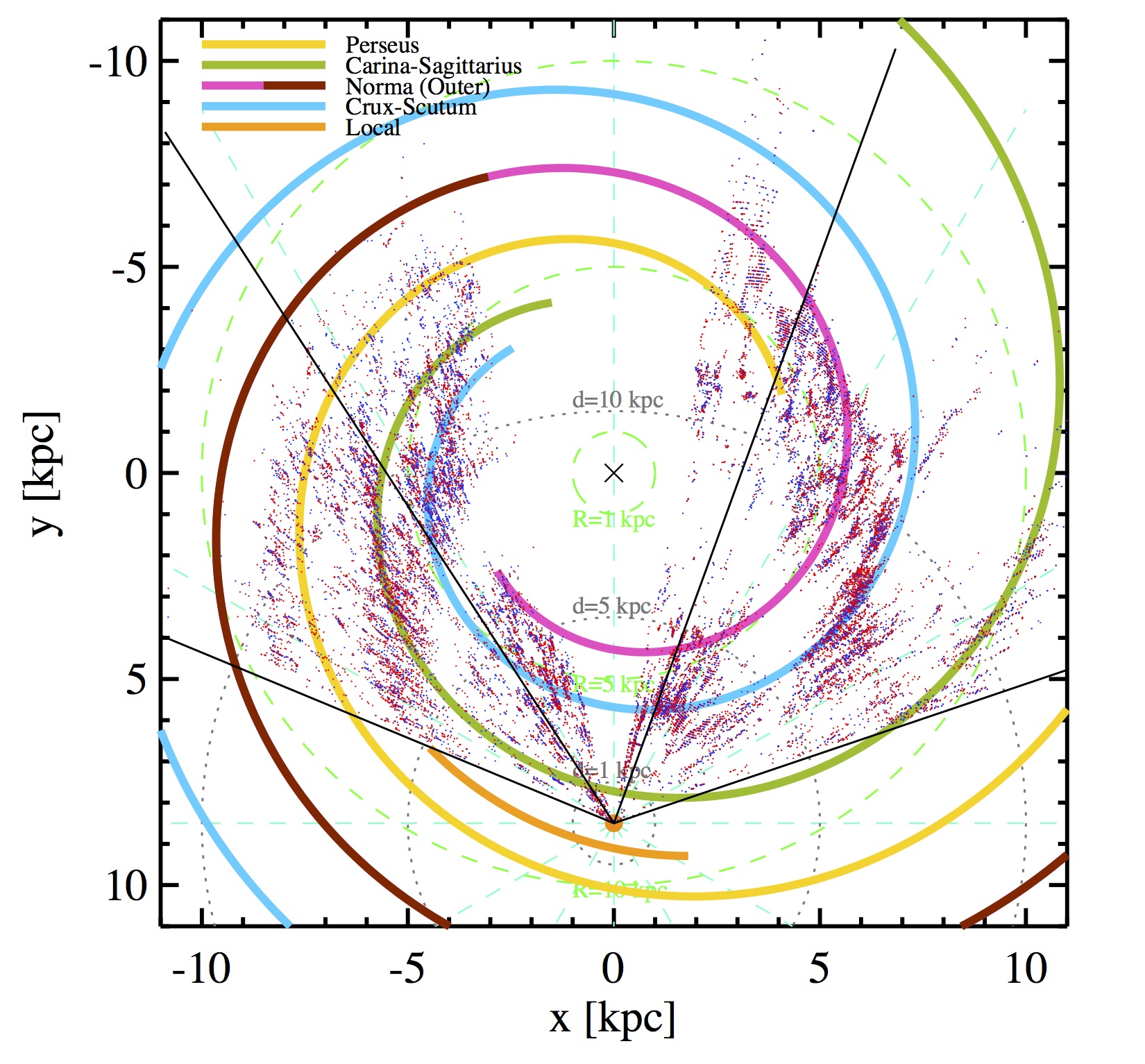}
\caption{Plot of the position in the Galactic plane of the pre- (red dots)
  and proto-stellar (blue dots) Hi-GAL objects provided with a distance estimate.
  Unbound objects are not shown to reduce the crowding of the plot. For definitions
  of pre-stellar, proto-stellar and unbound objects, see Section~\ref{protovsstarl}.
  Two pairs of black solid lines delimit the longitude ranges analyzed in this paper.
  Further sources are found closer to the Galactic centre, most of which are analyzed in 
  \citet{ven17} (see their Figure~1). In the inner zone devoid of points, distances 
  were not estimated (see text), thus only distance-independent source properties were 
  derived. The Galactic centre is indicated with a $\times$ symbol at coordinates $[x,y]=[0,0]$, 
  the Sun with an orange dot at coordinates [0,8.5], at the bottom of the plot. 
  Cyan dashed lines indicate Galactic longitude in steps of $30^{\circ}$. $R$
  are Galactocentric distances (light green dashed circles), $d$ heliocentric ones
  (grey dotted circles), respectively, with the following steps: 1, 5 and 10~kpc.
  Spiral arms, from the four-arm Milky Way prescription of \citet{hou09}, are 
  plotted with different colours, with the arc-colour correspondence reported in the 
  upper-left corner of the plot. In particular, the Norma arm is represented using 
  two colours: magenta for the inner part of the arm, and brown for the portion of it 
  generally designated as Outer arm, which starting point is established by comparison
  with \citet{mom06}. Finally the Local arm, not included in the model of \citet{hou09},
  is drawn, taken from \citet{xu16}.}
\label{galpos}
\end{figure*}

Finally, at present, no distance estimates have been obtained in the longitude range
$-10.2^{\circ}< \ell < 14.0^{\circ}$, due to the difficulty in estimating the 
kinematic distances of sources in the direction of the Galactic centre.
We were able to assign a heliocentric distance to \withdistance~sources
out of the \totfittedseds~of the band-merged filtered catalogue, i.e. \withdistanceperc\%
(see also Table~\ref{numbertab}). However, for \withdistancebadflag~of these, 
the near/far ambiguity has not been solved, since the extinction information is not
available, and the far distance is assigned by default (see above).

The distribution of sources in the Galactic 
plane is shown in Figure~\ref{galpos}. It can be seen that the available distances 
do not produce a clear segregation between high-source density regions corresponding
to spiral arm locations and less populated inter-arm regions, as will be discussed
in more detail in Section~\ref{spiralarms}. On the one hand, massive star-forming clumps are 
expected to be organized along spiral arms \citep[e.g., CH$_3$OH and H$_2$O masers observed
by][]{xu16}. On the other hand, the large number of sources present in our catalogue, 
corresponding to a large variety of physical and evolutionary conditions probed
with \textit{Herschel}, makes it likely to include also clumps located outside the
arms. Any consideration of this aspect is subject to a more correct estimate of heliocentric
distances: the work of assigning distances to Hi-GAL sources is still in progress within the 
VIALACTEA project, and a more refined set of distances (and for an increased number
of sources) will be delivered in Russeil et al. (in prep.).

\subsection{Starless and proto-stellar objects}\label{protovsstarl}
One of the most important steps in the determination of the evolutionary stages of Hi-GAL
sources is discriminating between pre- and proto-stellar sources, namely starless
but gravitationally bound objects and objects showing signatures of ongoing star formation,
respectively. Here we follow the approach already described in \citet{eli13}. If a 70~$\umu$m
counterpart is available, that object can with a high degree of confidence be labelled as
proto-stellar \citep{dun08,rag13,svo16}. This criterion works well for relatively nearby objects,
but as soon as we extend our studies to regions farther away than say 4-5~kpc, two competing
effects concur in confusing the source counts \citep[see also][]{bal17}, both affecting the 
estimates of the star formation rate (SFR). First, at large distances 70~$\umu$m counterparts of
relatively low-mass sources might be missed (and sources mislabelled) because of limits in sensitivity \footnote{For example,
applying Equation~\ref{gbthin} (presented in the following), a grey body with mass
of $50~M_\odot$, temperature of 15~K, dust emissivity with exponent $\beta=2$ with 
the same reference opacity adopted in this paper (see Section~\ref{sedfit}), and located 
at a distance of 10~kpc, would have a flux of 0.02~Jy at 70~$\umu$m, and of 0.84~Jy at 
250~$\umu$m, consequently detectable with \textit{Herschel} at the latter band but not at 
the former \citep{mol16a}.}. 
Second, since the proto-stellar label is given on the basis of the detection of a 70~$\umu$m 
counterpart, starless and proto-stellar cores close together and far away could be seen and 
labelled as a single proto-stellar clump due to lack of resolution. We address this issue in 
Appendix~\ref{confusion}.

\begin{figure*}
\centering
\includegraphics[width=16cm]{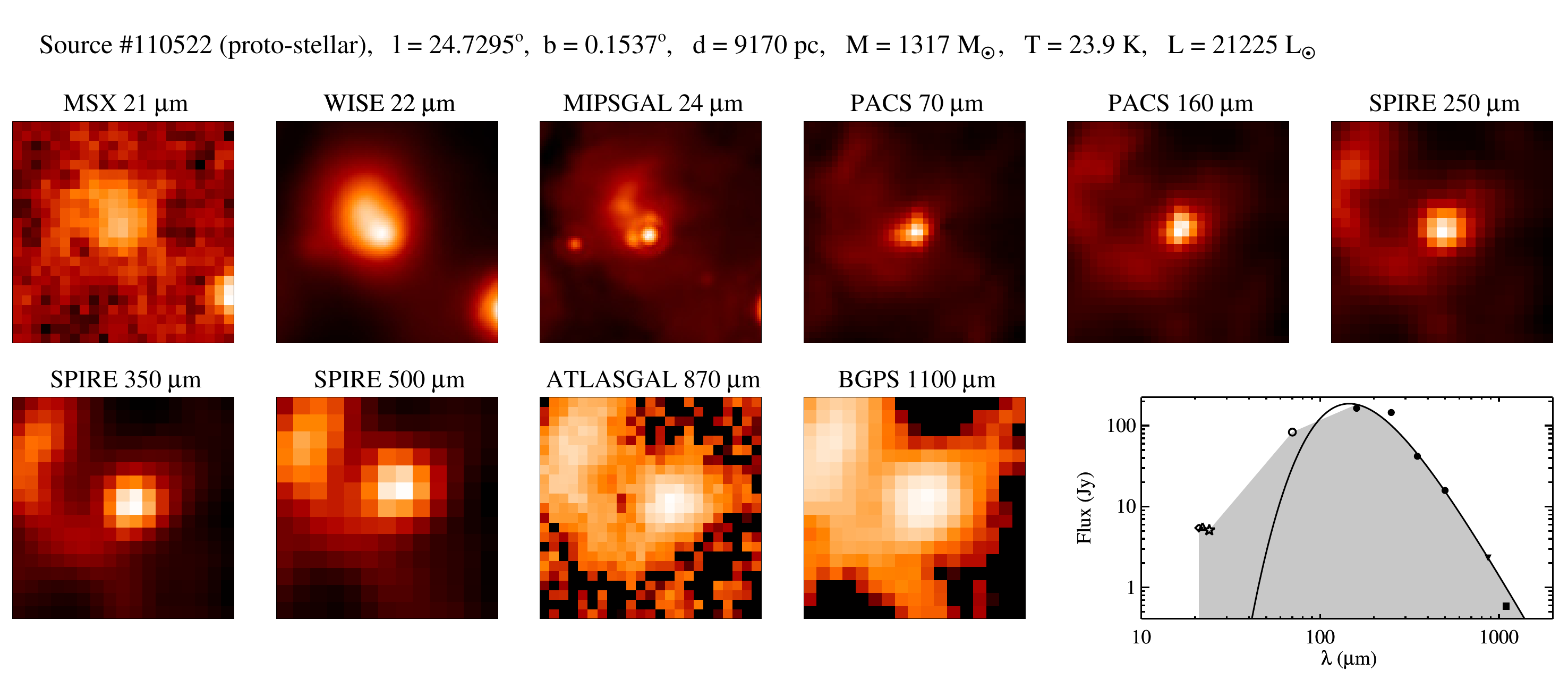}
\includegraphics[width=16cm]{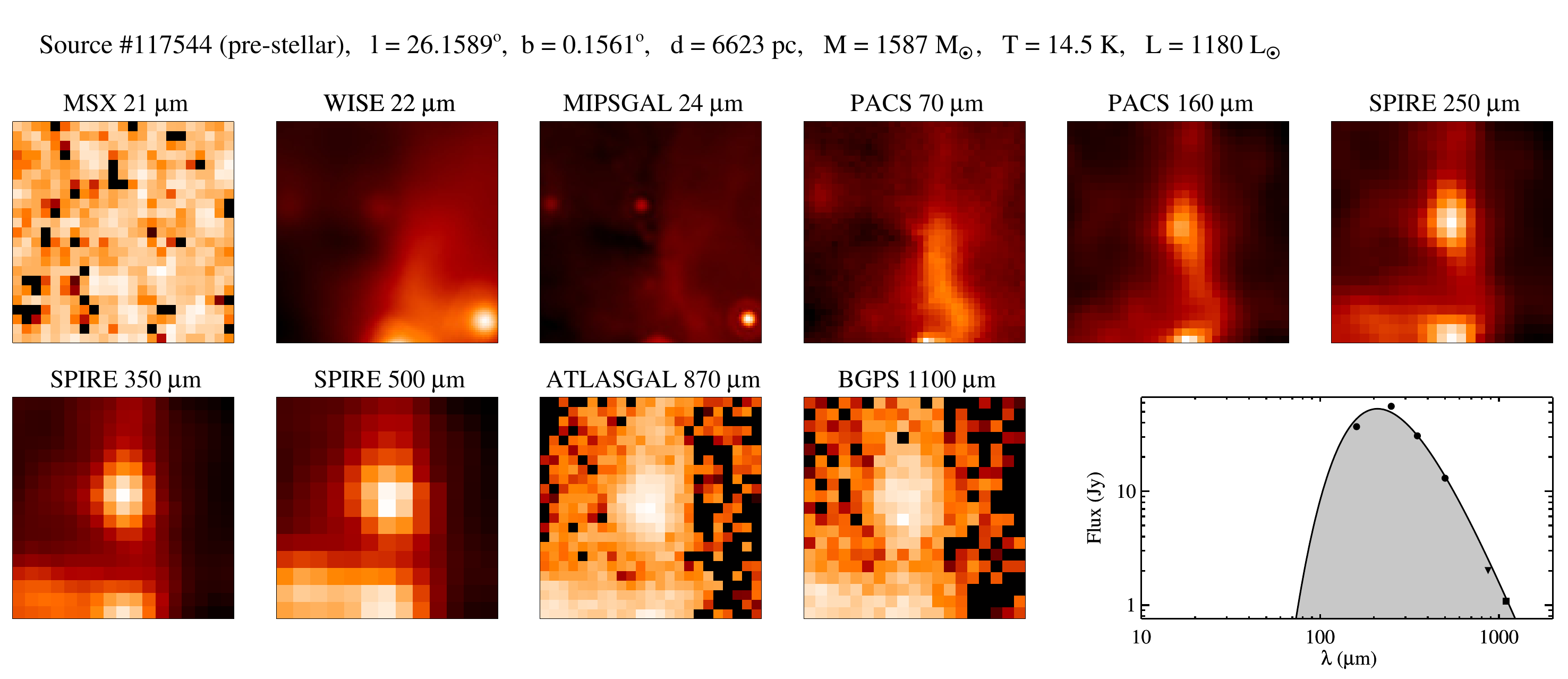}
\caption{Multi-wavelength $2\arcmin \times 2\arcmin$ images of two sources listed in our catalogue, to 
provide an example of a proto-stellar (upper 11 panels) and a pre-stellar source (lower 11 panels). 
The source coordinates and physical properties are reported above each set of panels. For each 
source position ten images at 21, 22, 24, 70, 160, 250, 350, 500, 870, and 1100~$\umu$m, taken from the
survey indicated in the title, are shown (the colour scale is logarithmic, in arbitrary units). Finally, 
the SED of each of the two sources is shown: filled and open symbols indicate fluxes which are taken 
into account or not, respectively, for the grey body fit (solid line, see Section~\ref{sedfit}). The 
grey-shaded area is a geometric representation of the integral calculated to estimate the source 
bolometric luminosity (see Sections~\ref{ancillary} and \ref{sedfit}).}
\label{sedfig}
\end{figure*}

To mitigate the first effect, we performed deeper, targeted extractions at PACS wavelengths
towards two types of sources. One type consisted of ``SPIRE-only'' sources, i.e. sources
clearly detected only at 250, 350 and 500~$\umu$m. Since SED fitting for such sources is poorly
constrained, a further extraction at 160~$\umu$m, deeper than that of \citet{mol16a},
was required in order to set at least an upper limit for the flux shortwards of what could 
be the peak of the SED. In this way, \recoveredred~further detections and \nulimred~upper 
limits at 160~$\umu$m were recovered.

The second type of sources consists of those showing a 160~$\umu$m counterpart (original
or found after deeper search) but no detection at 70~$\umu$m. To ascertain that the
starless nature of these objects is not assigned simply due to a failure of the
source detection process, we performed a deeper search for a 70~$\umu$m counterpart toward
those targets: in this way, a possible clear counterpart not originally listed in the single 
band catalogues would allow us to label the object as proto-stellar. Adopting this strategy, 
\recoveredblue~further detections and \nulimblue~upper limits at 70~$\umu$m were recovered.

Whereas the proto-stellar objects are expected to host ongoing star formation, the relation
between starless objects and star formation processes must be further examined, since 
only gravitationally bound sources fulfil the conditions for a possible future collapse. 
Here we use the so-called ``Larson's third relation'' to assess if an object can be 
considered bound: the condition we impose involves the source mass, $M$, and radius, 
$r$ (see Sections~\ref{sedfit} and \ref{results}), and is formulated as
$M(r) > 460 M_{\odot} (\mathrm{r}/\mathrm{pc})^{1.9}$ \citep{lar81}. Masses above 
this threshold identify bound objects, i.e. genuine pre-stellar aggregates.

In Table~\ref{numbertab} the statistics of the sources of the catalogue, divided into 
proto-stellar, pre-stellar and starless unbound, is reported, while in Figure~\ref{sedfig} 
the SEDs and the corresponding grey-body fits are shown for one proto-stellar and one 
pre-stellar source, for the sake of example. More details and a discussion about the  
proto- to pre-stellar source ratio are given in Section~\ref{IvsIV}.

One of the aims of this paper is to show the amount of information that can be extracted simply from 
continuum observations in the FIR/sub-mm, combining Hi-GAL data with other surveys in adjacent wavelength
bands and using spectroscopic data only to obtain kinematic distances. On the one hand, for many sources, these
data can be complemented with line observations to obtain a more detailed picture. On the other hand,
Hi-GAL produced an unprecedentedly large and unbiased catalogue containing many thousands of newly 
detected cold clumps, for which it is important to provide a first classification. The criteria we provide 
to separate different populations, although somewhat conventional in the \textit{Herschel} literature, remain 
probably too clear-cut and surely affected by biases we introduced in this section and discuss also in the
following sections of this paper. Reciprocal contamination of the samples certainly increases overlap
of the physical property distributions obtained separately for the different populations, as will
be seen in Sections~\ref{results} and~\ref{evol_par}.

\begin{table*}
\caption{Number of sources in the Hi-GAL catalog, in ranges of longitude.}
\label{numbertab}
\begin{tabular}{lccccccc}
\hline
Longitude range & \multicolumn{2}{c}{Proto-stellar} & \multicolumn{2}{c}{Highly-reliable Starless (Pre-stellar)} & \multicolumn{2}{c}{Poorly-reliable Starless (Pre-stellar)}&Total\\
                & w/ distance  & w/o distance & w/ distance  & w/o distance                          & w/ distance  & w/o distance &  \\
\hline
$-71^{\circ} \le \ell < -20^{\circ}$&         8227&        1425&       10384
 (9598)&        2978 (2265)&       10696 (7531)&        3321 (1519)&       37031
 \\
$-20^{\circ} \le \ell < -10^{\circ}$&         1752&         273&        2667
 (2506)&         611 (537)&        2548 (1855)&         591 (345)&        8442
 \\
$-10^{\circ} \le \ell < 0^{\circ}$&            0&        2154&           0 (0)&
        3357 (3139)&           0 (0)&        3417 (2544)&        8928 \\
$0^{\circ} \le \ell < 19^{\circ}$&          505&        3165&         398 (380)&
        5689 (5271)&         318 (249)&        5488 (4053)&       15563 \\
$19^{\circ} \le \ell < 33^{\circ}$&         2646&         549&        3172
 (2893)&        1260 (1068)&        3045 (2200)&        1312 (832)&       11984
 \\
$33^{\circ} \le \ell < 67^{\circ}$&         2704&        1184&        4189
 (3818)&        3149 (2548)&        3814 (2520)&        3934 (2000)&       18974
 \\
\hline
Total &        15834 &         8750 &        20810 (19195) & 
       17044 (14828) &       20421 (14355) &        18063 (11293) &      100922
 \\
\hline
\end{tabular}
\end{table*}

\section{SED fitting}\label{sedfit}

Once the SEDs of all entries in the filtered catalogue are built by assembling the photometric
information as explained above, it is possible to fit a single grey body function to its
$\lambda \geq 160$~$\umu$m portion, and therefore derive the mass $M$ and the temperature $T$ of the 
cold dust in those objects.

Many details on the use of the grey body to model FIR SEDs have been provided and discussed 
by \citet{eli16}. Here we report only concepts and analytic expressions which are appropriate 
for the present paper. The most complete expression for the grey body explicitly contains 
the optical depth:
\begin{equation}\label{gbthick}
F_{\nu}=(1-\mathrm{e}^{-\tau_{\nu}})B_{\nu}(T_d)\Omega~,
\end{equation}
recently used, e.g., in \citet{gia12}, where $F_{\nu}$ is the observed flux density at the frequency
$\nu$, $B_{\nu}(T_d)$ is the Planck function at the dust temperature $T_d$ and $\Omega$ is the 
source solid angle in the sky. The optical depth can be parametrised in turn as
\begin{equation}\label{tau}
\tau_{\nu}=(\nu/\nu_0)^\beta~,
\end{equation}
where the cut-off frequency $\nu_0=c/\lambda_0$ is such that $\tau_{\nu_0}=1$, and $\beta$ is the 
exponent of the power-law dust emissivity at large wavelengths. After constraining $\beta=2$, 
as typically adopted also in the Gould Belt \citep[e.g.,][]{kon15} and HOBYS \citep[e.g.,][]{gia12} 
consortia, and as recommended by \citet{sad13}, and $\Omega$ to be equal to the source area 
as measured by CuTEx at the reference wavelength of 250~$\umu$m \citep[cf.][]{eli13}, 
the free parameters of the fit remain $T$ and $\lambda_0$.
For these parameters we explored the ranges 5~K~$\leq T \leq$~40~K and
$5~\umu$m~$\leq \lambda_0 \leq~350~\umu$m, respectively.

The clump mass does not appear explicitly in Equation~\ref{gbthick} but can be derived from
\begin{equation}\label{mthick}
M=(d^2\Omega/\kappa_{\mathrm{ref}})\tau_{\mathrm{ref}}~,
\end{equation}
as shown by \citet{pez12}, where $\kappa_{\mathrm{ref}}$ and $\tau_{\mathrm{ref}}$ are the opacity and
the optical depth, respectively, estimated at a given reference wavelength $\lambda_{\mathrm{ref}}$.
To preserve the compatibility with previous works based on other \textit{Herschel} key-projects
\citep[e.g.][]{kon10,gia12} here we decided to adopt $\kappa_{\mathrm{ref}}=0.1$~cm$^2$~g$^{-1}$ at
$\lambda_{\mathrm{ref}}=300~\umu$m \citep[][already accounting for a gas-to-dust ratio of 100]{bec90},
while $\tau_{\mathrm{ref}}$ can be derived from Equation~\ref{tau}. The choice of $\kappa_{\mathrm{ref}}$
constitutes a critical point \citep{mar12,deh12}, thus it is interesting to show how much the 
mass would change if another estimate of $\kappa_{\mathrm{ref}}$ were adopted. The dust opacity 
at $300~\umu$m from the widely used OH5 model \citep{oss94} is $\kappa_{300}=0.13$~cm$^2$~g$^{-1}$, 
which would produce a 30\% underestimation of masses with respect to our case. \citet{pre93}
quote $\kappa_{1300}=0.005$~cm$^2$~g$^{-1}$ which, for $\beta=2$, would correspond to
$\kappa_{300}=0.094$~cm$^2$~g$^{-1}$, implying a $\sim6\%$ larger mass. Similarly, the value of
\citet{net09}, $\kappa_{250}=0.16$~cm$^2$~g$^{-1}$, would translate to $\kappa_{300}=0.11$~cm$^2$~g$^{-1}$,
practically consistent with the value adopted here. However, further literature values of $\kappa_{250}$
quoted by \citet{net09} in their Table~3, span an order of magnitude, from
$\kappa_{250}=0.024$~cm$^2$~g$^{-1}$\citep{dra07} to $\kappa_{250}=0.22-0.25$~cm$^2$~g$^{-1}$\citep{oss94},
which would lead to a factor from 6 to 0.6 on the masses calculated in this paper.

\begin{figure*}
\centering
\includegraphics[width=16.5cm]{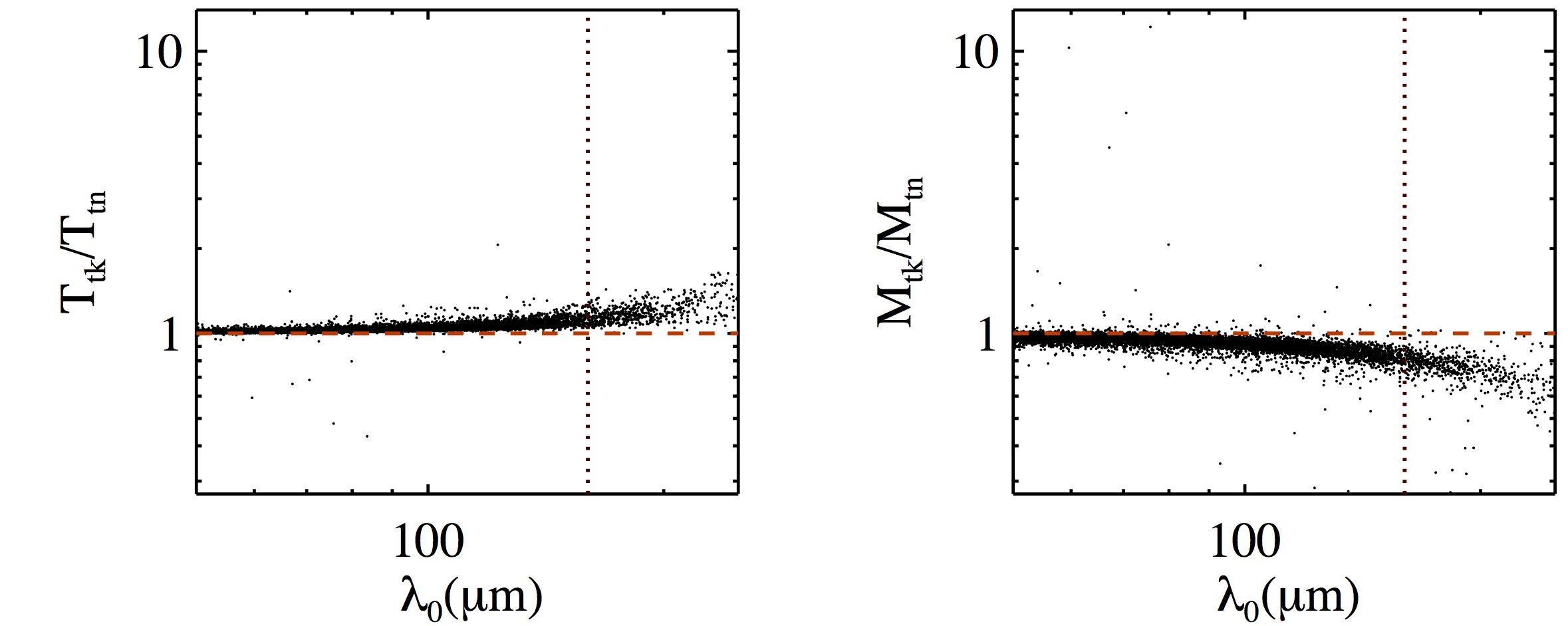}
\caption{\textit{Left}: Ratio of the dust temperatures derived by fitting Equations~\ref{gbthick}
and~\ref{gbthin} to SEDs (and denoted here with $T_\mathrm{tk}$ and $T_\mathrm{tn}$, to indicate 
an optically ``thick'' and thin regime, respectively) vs the corresponding $\lambda_0$ obtained 
through Equation~\ref{gbthick}.
The $x$-axis starts from $50.6~\umu$m, corresponding to the minimum value for which this comparison
makes sense (see text). The
$\lambda_0=160~\umu$m value is highlighted as a reference with a vertical dotted line.
The red dashed line represents the $T_\mathrm{tk} = T_\mathrm{tn}$ condition. \textit{Right}: the same as
left panel, but for the masses $M_\mathrm{tk}$ and $M_\mathrm{tn}$ derived through Equations~\ref{mthick}
and~\ref{gbthin}, respectively.}
\label{tkvsth}
\end{figure*}

For very low values of $\lambda_0$ (i.e. much shorter than the minimum of the range we consider for the
fit, namely $160~\umu$m) the grey body has a negligible optical depth at the considered wavelengths, so
that Equation~\ref{gbthick} can be simplified as follows:
\begin{equation}\label{gbthin}
F_{\nu}=\frac{M \kappa_{\mathrm{ref}}}{d^2}\left( \dfrac{\nu}{\nu_{\mathrm{ref}}}\right)^{\beta} B_{\nu}(T_d)~.
\end{equation}
\citep[cf.][]{eli10}. We note that the estimate of $\lambda_0$ does not affect our results significantly 
when it implies $\tau \leq 0.1$ at $160~\umu$m. According to Equation~\ref{tau} and for $\beta=2$, we find 
that the critical value is encountered for $\lambda_0 \sim 50.6~\umu$m. If the fit procedure 
using Equation~\ref{gbthick} provides a $\lambda_0$ value shorter than that, the fit is
repeated based on Equation~\ref{gbthin}, and the mass and temperature computed in this alternative
way are considered as the definitive estimates of these quantities for that source. In
Figure~\ref{tkvsth} we show how the temperature and mass values obtained through
Equations~\ref{gbthick} ($T_\mathrm{tk}, M_\mathrm{tk}$) and~\ref{gbthin} 
($T_\mathrm{tn}, M_\mathrm{tn}$) appear generally equivalent as long as $\lambda_0$
(obtained through the former) is much shorter than $160~\umu$m (say $\lambda < 100~\umu$m), 
since both models have extremely low (or zero) opacities at the wavelengths involved in the fit. 
Quantitatively speaking, the average and standard deviation of the $T_{tk}/T_{tn}$ ratio
for $\lambda < 160~\umu$m are 1.01 and 0.02, respectively.
However, an increasing discrepancy
is visible at increasing $\lambda_0$, highlighting the tendency of Equation~\ref{gbthin} to
underestimate the temperature and to overestimate the mass.

SED fitting is performed by $\chi^2$ optimization of a grey body model on a grid that is refined in
successive iterations to converge on the final result. The strategy of generating a SED grid to be 
compared with data also gives us the advantage of applying PACS colour corrections directly to the model 
SEDs (since its temperature is known for each of them), rather than correcting the data iteratively 
\citep[cf.][]{gia12}.

For sources with no assigned distance, a virtual value of 1~kpc was assumed, to allow
the fit anyway and distance-independent quantities (such as $T$) to be derived,
and also distance-independent combinations
of single distance-dependent quantities (as $L_{\mathrm{bol}}/M$, see Section~\ref{lum_par}).

The luminosity of the starless objects was estimated using the area under the best fitting grey body.
For proto-stellar objects, however, the luminosity was calculated by summing two contributions: the
area under the best fitting grey body starting from 160~$\umu$m and longward, plus the area of
the observed SED between 21 and 160~$\umu$m counterparts (if any) to account for MIR emission contribution
exceeding the grey body.


\section{Summary of the creation of the scientific catalogue}\label{pipeline}

The generation of the catalogue used for the scientific analysis presented in this paper can 
be summarised as follows:
\begin{enumerate}
\item Select the sources located in regions observed with both PACS and SPIRE.
\item Perform positional band-merging of single band catalogues. At the first step, the single-band
  catalogue at 500~$\umu$m is taken, and the closest counterpart in the 350~$\umu$m image,
  if available, is assigned. The same is repeatedly done for shorter wavelengths, up to 70~$\umu$m. 
  The ellipse describing the object at the longer wavelength is chosen as the
  matching region.
\item Select sources in the band-merged catalogue that have counterparts in at least three 
  contiguous \textit{Herschel} bands (except the 70~$\umu$m) and show a ``regular'' SED
  (with no cavities and not increasing toward longer wavelengths).
\item Find counterparts at MIR and mm wavelengths for all entries in the
  band-merged and filtered catalogue. Shortwards of 70~$\umu$m catalogues at 21, 22,
  and 24~$\umu$m were searched and the corresponding flux reported is the sum of all 
  objects falling in the ellipse at 250~$\umu$m. Longwards of 500~$\umu$m counterparts
  where searched by mining 870~$\umu$m ATLASGAL public data, or extracting sources
  through CuTEx, as well as 1.1\,mm BGPS data.
\item Fill the catalogue where fluxes at 160 and/or at 70~$\umu$m were missing, to improve the 
  the quality of labelling sources as starless or proto-stellar (see next step).
\item Move the selected SEDs which remain with only three fluxes in the five Hi-GAL bands
  to a list of sources with, on average, barely-reliable physical parameter estimation.
\item Classify sources as proto-stellar or starless, depending on presence or lack of a
  detection at 70~$\umu$m, respectively.
\item Assign a distance to all sources using the method described in \citet{rus11}.
\item Fit a grey body to the SED at $\lambda \geq 160~\umu$m to derive
 the envelope average temperature, and, for sources provided with a distance estimate, the mass
 and the luminosity.
\item Make a further classification, among starless sources, between gravitationally
  unbound or bound (pre-stellar) sources, based on the mass threshold suggested by the
  Larson's third law.
\end{enumerate}

The catalogue, generated as described above and constituted by two lists (``high 
reliability'' and ``low reliability'', respectively), is available for download at 
\url{http://vialactea.iaps.inaf.it/vialactea/public/HiGAL_clump_catalogue_v1.tar.gz}. 

The description of the columns is reported in Appendix~\ref{catdescription}.

\section{Results}\label{results}

\subsection{Physical size}\label{physsize}
For a source population distributed throughout the Galactic plane at an extremely wide range of
heliocentric distances, as in our case, it is fundamental to consider the effective size of these
compact objects (i.e. detected within a limited range of angular sizes) in order to assess their 
nature.

\begin{figure*}
\centering
\includegraphics[width=16.6cm]{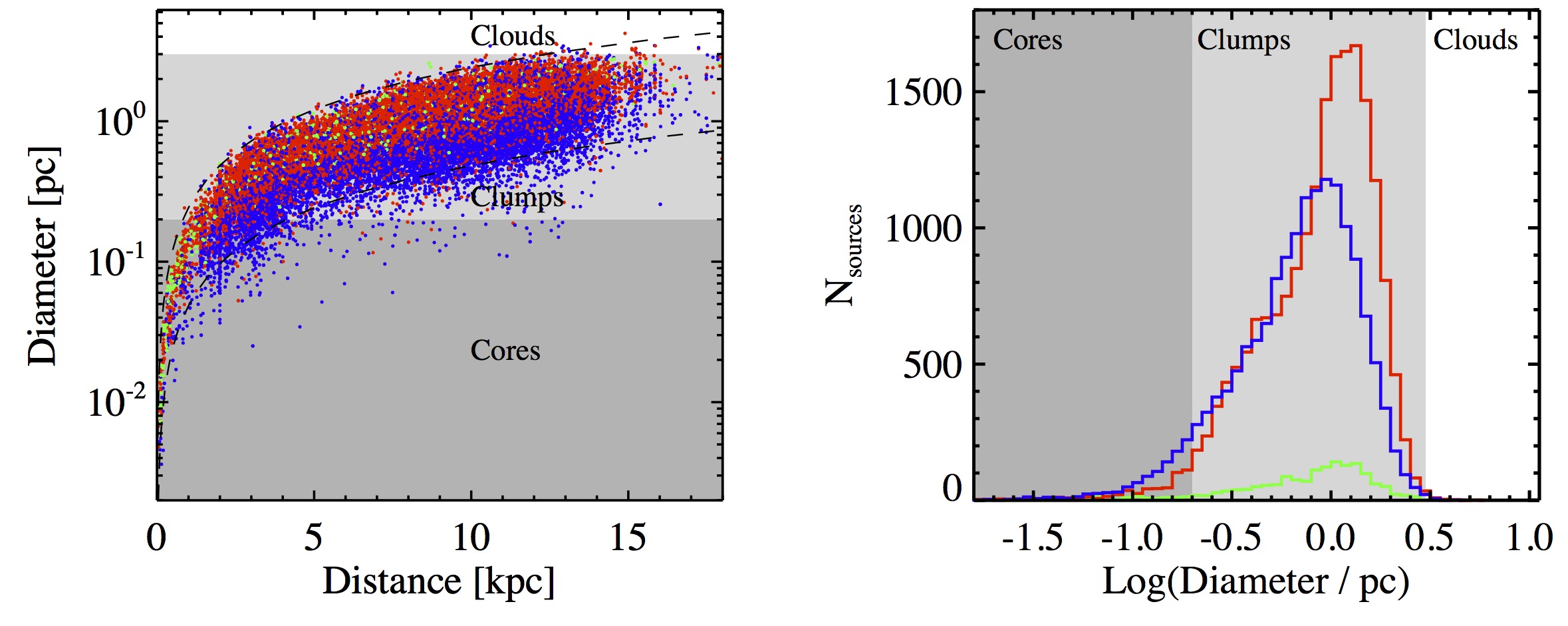}
\caption{\textit{Left}: Hi-GAL clump linear diameters, estimated at 250~$\umu$m, vs 
distances (blue: proto-stellar; red: pre-stellar;
green: starless unbound). Different background levels of grey indicate size ranges corresponding to different
object typologies (see text). The upper and lower dashed lines represent an angular size of 50$\arcsec$ and 
10$\arcsec$, respectively. \textit{Right}: Distribution of source diameters for proto-stellar,
pre-stellar, and starless unbound sources. Line and background colours are the same as in the left panel.}
\label{sizefig}
\end{figure*}

Sure enough, we can derive the linear sizes only for objects with a distance estimate, starting 
from the angular size estimated at the reference wavelength of 250~$\umu$m as the circularised 
and deconvolved size of the ellipse estimated by CuTEx. In Figure~\ref{sizefig}, left panel, 
we show the relation between the physical diameter and the distance for these sources, 
highlighting how this quantity is given by the combination of the source angular size and
its distance. Given the large spread in distance, a wide range of linear sizes is 
found, corresponding to very different classes of ISM structures.

In Figure~\ref{sizefig}, right panel, we provide the histogram of the diameter $D$ separately 
for the proto-stellar, pre-stellar and starless unbound sources, using the subdivision scheme 
proposed by \citet{ber07} (cores for $D < 0.2$~pc, clumps for $0.2 \geq D \leq 3$~pc, and clouds 
for $D>3$~pc, although the natural transition between two adjacent classes is 
far from being so sharp) to highlight how only a small portion of the Hi-GAL compact sources 
is compatible with a core classification, while most of them are actually clumps. A very 
small fraction of sources, corresponding to the most distant cases, can be considered as 
entire clouds. However, given the dominance of the clump-sized sources, is practical to refer 
to the sources of the present catalogue with the general term ``clumps''. The underlying,
generally inhomogeneous substructure of these clumps is not resolved in our observations, 
but it can reasonably supposed that they are composed by a certain number of cores and by 
inter-core diffuse medium \citep[e.g.,][]{mer15} so that, in the proto-stellar cases, we
generally should not expect to observe the formation process of a single protostar, but rather 
of a proto-cluster.

We note that the histograms in Figure~\ref{sizefig}, right panel, should not be taken as a 
coherent size distribution of our source sample, due to the underlying spread in distance.
It is not possible, therefore, to make global comparisons between the different classes
as, for example, in \citet{gia12} who considered objects from a single region, all located 
at the same heliocentric distance. The same consideration applies to the distribution of
other distance-dependent quantities.

\subsection{Dust temperature}\label{dtemp}
\begin{figure}
\centering
\includegraphics[width=9.0cm]{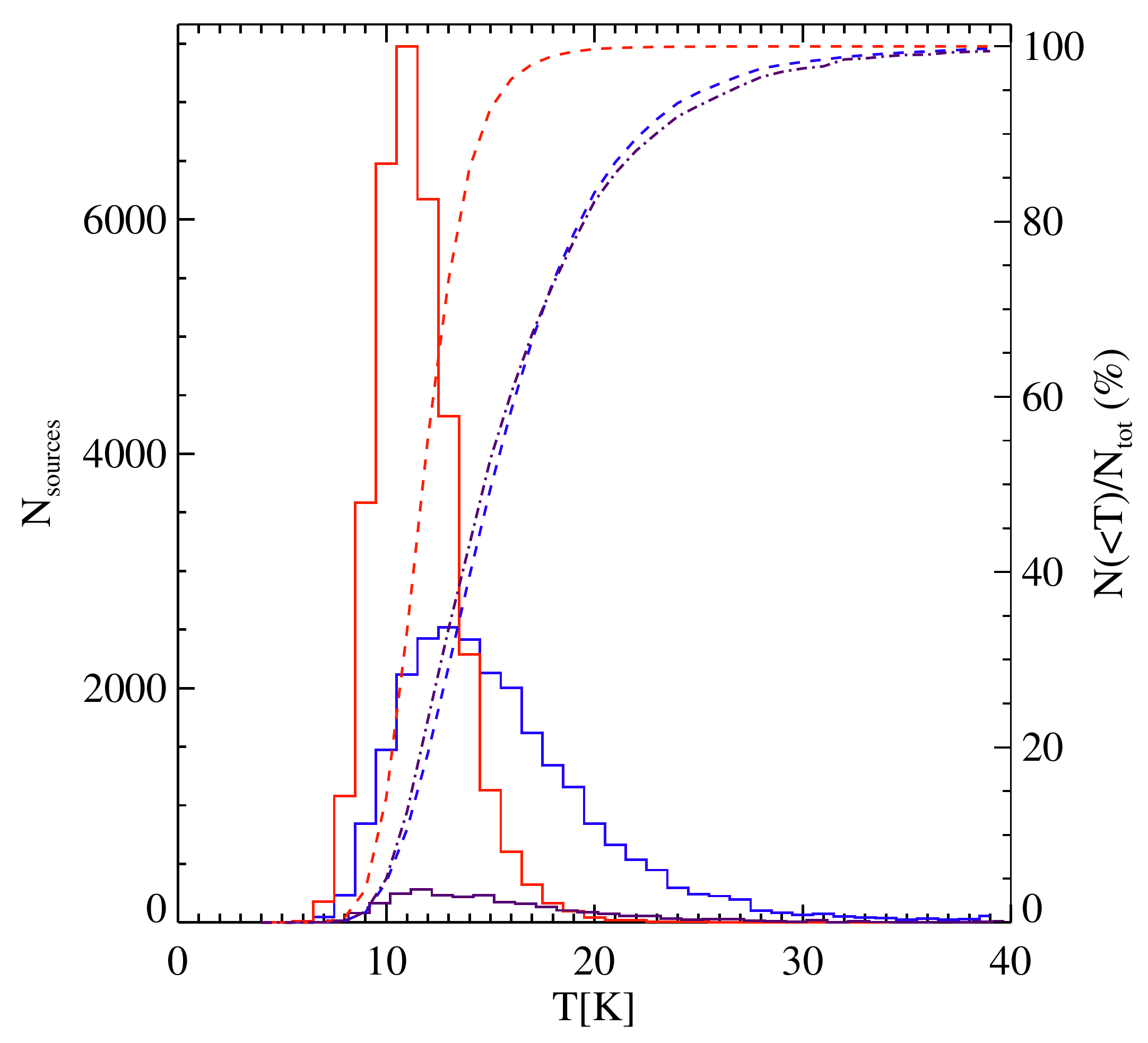}
\caption{Grey body temperature distributions for the pre- (red histogram) and proto-stellar 
(blue histogram) sources considered for science analysis in this paper. Cumulative curves 
of the same distributions are also plotted as dashed lines (and the same colours), according 
to the $y$-axis on the right side of the plot. Finally, the temperature distribution of the 
sub-sample of MIR-dark proto-stellar sources is plotted in dark purple, and the corresponding
cumulative as a dotted-dashed dark purple line.}
\label{hist_temp}
\end{figure}

The distributions of grey body temperatures of the sources are shown in Figure~\ref{hist_temp}.
As already found by \citet{gia12}, \citet{eli13}, \citet{gia13}, \citet{ven17} through Hi-GAL
observations but also by \citet{olm09} through BLAST observations, the distributions of pre- and
proto-stellar sources show some relevant differences, the latter being found towards warmer
temperatures with respect to the former. A quantitative argument is represented by the average
values $\bar{T}_{\mathrm{pre}}=\avetpe$~K and $\bar{T}_{\mathrm{prt}}=\avetpo$~K for pre- and 
proto-stellar sources, and the median values $\tilde{T}_{\mathrm{pre}}=\medtpe$~K and 
$\tilde{T}_{\mathrm{prt}}=\medtpo$~K, respectively.

Furthermore, both the temperature distributions seem quite asymmetric, with a prominent
high-temperature tail. This can be seen by means of the skewness indicator (defined as
$\gamma=\umu_3/\sigma^3$, where $\umu_3$ is the third central moment and  $\sigma$ the
standard deviation) of the two distributions: $\gamma_{\mathrm{pre}}=\tskewe$ and
$\gamma_{\mathrm{prt}}=\tskewo$, respectively. The positive skew, in this case, indicates
that the right tail is longer ($\gamma=0$ for a normal distribution), and quantifies that. 
On the other hand, the pre-stellar distribution appears more peaked than the proto-stellar 
one. The kurtosis of a distribution, defined as $\delta=\umu_4/\sigma^4$ (where $\umu_4$ is 
the fourth central moment), is useful to quantify the level of peakedness (for a normal 
distribution, $\delta=3$). In these two cases the kurtosis values are found to be quite 
different for the two distributions ($\delta_{\mathrm{pre}}=\tkurte$ and
$\delta_{\mathrm{prt}}=\tkurto$, respectively). Finally, we also plot the
cumulative distributions of the temperatures, which is another way to highlight the
behaviours examined so far. We find that $99\%$ of the pre-stellar (proto-stellar) sources 
have dust temperatures lower than $\tpcenn$~K (\tpconn~K), and
the temperature range widths required to go from~1\% to~99\% levels are \tpcezenn~K and
\tpcozenn~K, respectively.

The differences found between the two distributions are even more meaningful from the point
of view of the separation between the two classes of sources, if one keeps in mind that the
temperature is estimated from data at wavelengths longer than $160~\umu$m, hence independently 
from the existence of a measurement at $70~\umu$m, which discriminates between proto-stellar 
and starless sources in our case.

These findings can be regarded from the evolutionary point of view: while pre-stellar 
sources represent the very early stage (or ``zero'' stage) of star formation and, as
such, are characterised by very similar temperatures, proto-stellar sources are 
increasingly warmer as the star formation progresses in their interior
\citep[e.g.,][]{bat10,svo16}, so that the spanned temperature range is larger 
and skewed towards higher values. A prominent high-temperature tail should be 
regarded, in this sense, as a signature of a more evolved stage of star formation activity. 

To corroborate this view, we consider the temperatures of the sub-sample of proto-stellar 
sources of our catalogue lacking a detection in the MIR (i.e. at 21 and/or 22 
and/or 24~$\umu$m, hereafter MIR-dark sources, as opposed to MIR-bright), whose distribution 
is also shown in Figure~\ref{hist_temp}. They represent \mirdarkfraction\% of the 
total proto-stellar sources (therefore dominated by MIR-bright cases). The 
average and median temperature for this class of objects are $\bar{T}_{\mathrm{Md}}
=\avetponmir$~K and $\tilde{T}_{\mathrm{Md}}=\medtponmir$~K, respectively, i.e. halfway 
between the values found for pre-stellar sources and those for the whole sample of 
proto-stellar ones, which is dominated by MIR-bright sources.

The reader should be aware that the dust temperature discussed here is simply
derived from the grey-body fit of the SED at $\lambda \geq 160~\umu$m and represents an estimate 
of the average temperature of the cold dust in the clump. Using line tracers it is 
possible to probe the kinetic temperature of warmer environments, such as the inner part 
of proto-stellar clumps, which is typically warmer \citep[$T>20$~K, e.g.][]{mol16b,svo16} 
than the median temperature found here for this class of sources. Despite of this, as
seen in this section, the grey body temperature can help to infer the source evolutionary 
stage, and turns out to be particularly efficient in combination with other parameters,
as further discussed in Section~\ref{evol_par}.

\subsubsection{\textit{Herschel} colours and temperature}

\begin{figure}
\centering
\includegraphics[width=8.0cm]{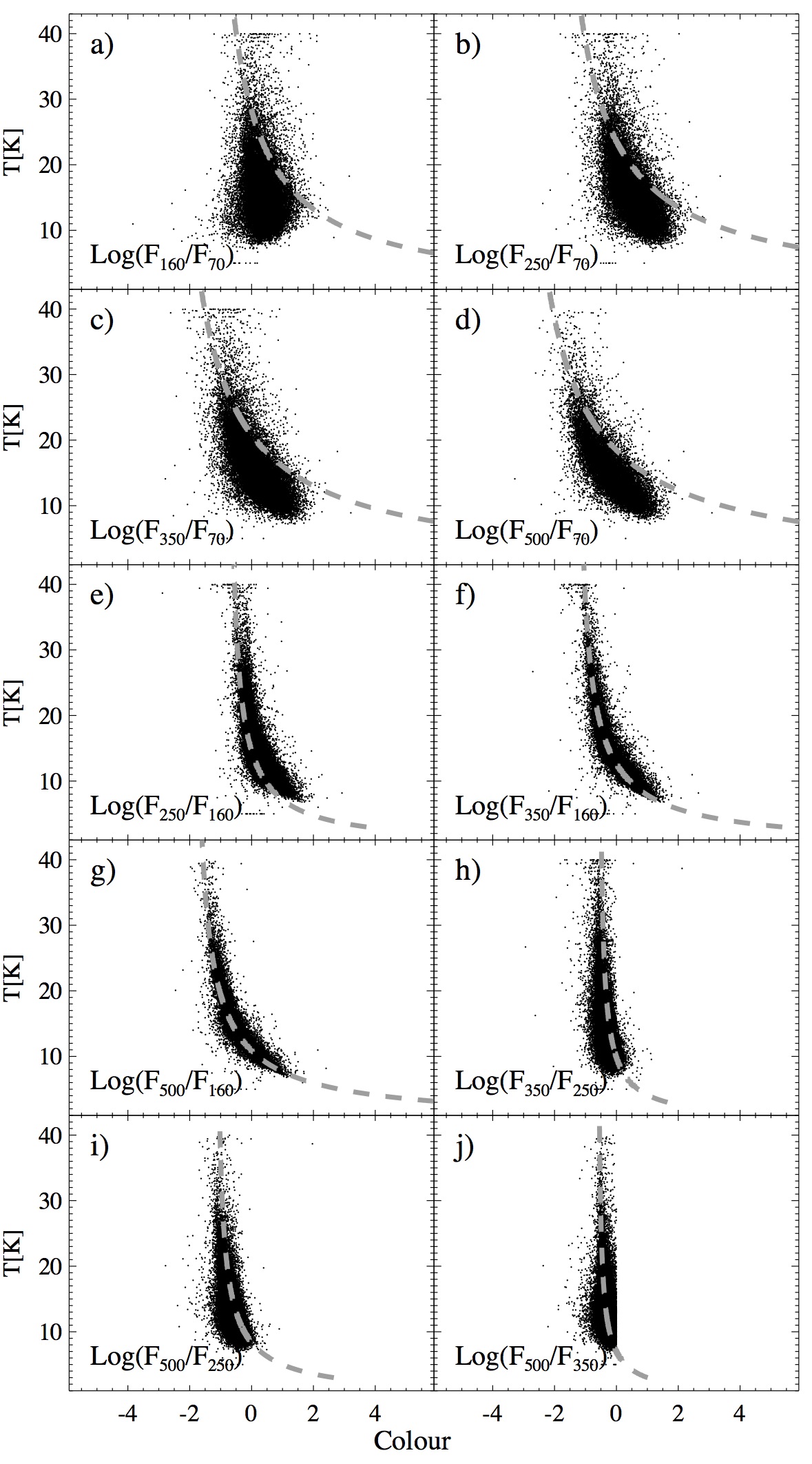}
\caption{Panels $a$-$j$: plots of source temperature (as derived through the grey body fit) vs
all colours obtainable from possible pair combinations of the five Hi-GAL wavebands. Each plot
is obtained using all sources provided with fluxes at the two involved wavelengths, specified
at the bottom of the panel. The unnatural vertical cut of the point distribution in panel $j$ 
is an artefact due to one of the filters adopted for source selection, ruling out sources with 
$F_{350} < F_{500}$ (rescaled fluxed are considered, see Section~\ref{filtering}). The grey
dashed line represents the temperature of a grey body with $\beta=2$.}
\label{tempcol}
\end{figure}

The availability of dust temperatures derived from grey body fits makes it possible 
to directly compare with 
\textit{Herschel} colours \citep[cf.][]{eli10,spe13}, to ascertain which ones are better representative 
of the temperature. In Figure~\ref{tempcol} the source temperature is plotted vs the ten possible 
colour pairs that can be obtained by combining the five \textit{Herschel} bands (for each combination, only 
sources detected at both wavelengths are displayed). We designate as colour the decimal logarithm 
of the ratio of fluxes at two different wavelengths. Since the 70~$\umu$m band 
is not involved in the 
temperature determination, the colours built from it (plots $a$ to $d$) do not show any tight 
correlation with temperature, while for the remaining six colours such correlation appears more 
evident, especially for those colours involving the 160~$\umu$m band. 
The best combination of spread of colour values (which decreases the level of temperature degeneracy,
mostly at low temperatures) 
and agreement with the analytic behaviour expected for a grey body is found for colours involving 
the flux at $160~\umu$m (panels $e$, $f$, and $g$). In particular, for the $F_{500}/F_{160}$ colour 
the grey body curve has the shallowest slope, so that we propose this colour as the most suitable 
diagnostics of the average temperature in the absence of a complete grey body fit. For the case of SPIRE-only
sources, lacking a counterpart at 160~$\umu$m, only three colours are available but their relation
with dust temperature (cf. panels from $h$ to $j$) appears to be affected by a high degree of 
degeneracy, making these colours unreliable temperature indicators.

\subsection{Mass and surface density}\label{mass_par}
Once source masses are obtained, it would be straightforward to show and analyse the resulting 
distribution (clump mass function, hereafter ClumpMF). However, since such discussion implies 
considerations about the clump mass-size relation and, somehow equivalently, the surface density,
we postpone the analysis of the ClumpMF until the end of this section, after having dealt with
those preparatory aspects.

\begin{figure*}
\centering
  \begin{tabular}{@{}cc@{}}
    \includegraphics[width=8.0cm]{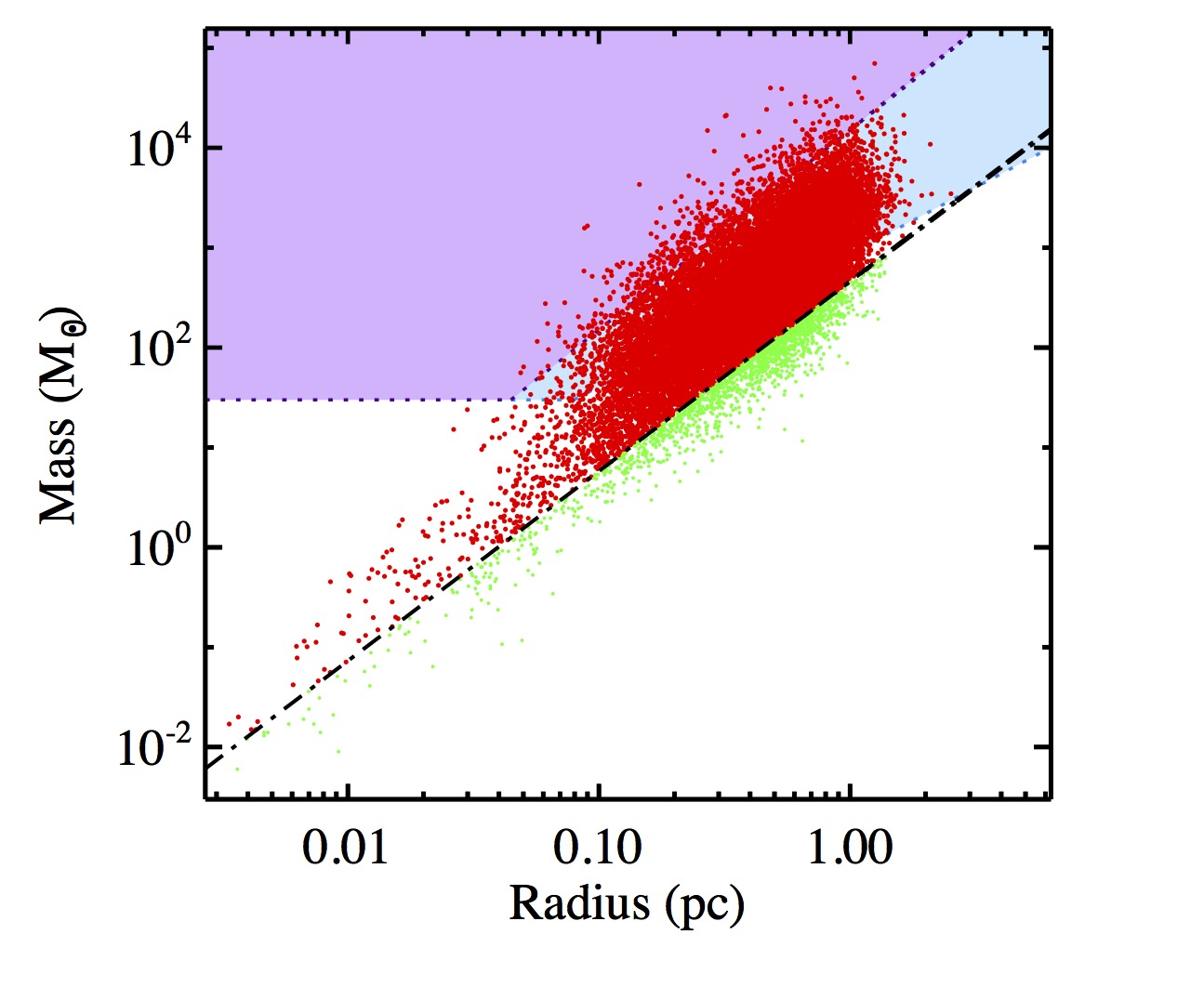} &
    \includegraphics[width=8.0cm]{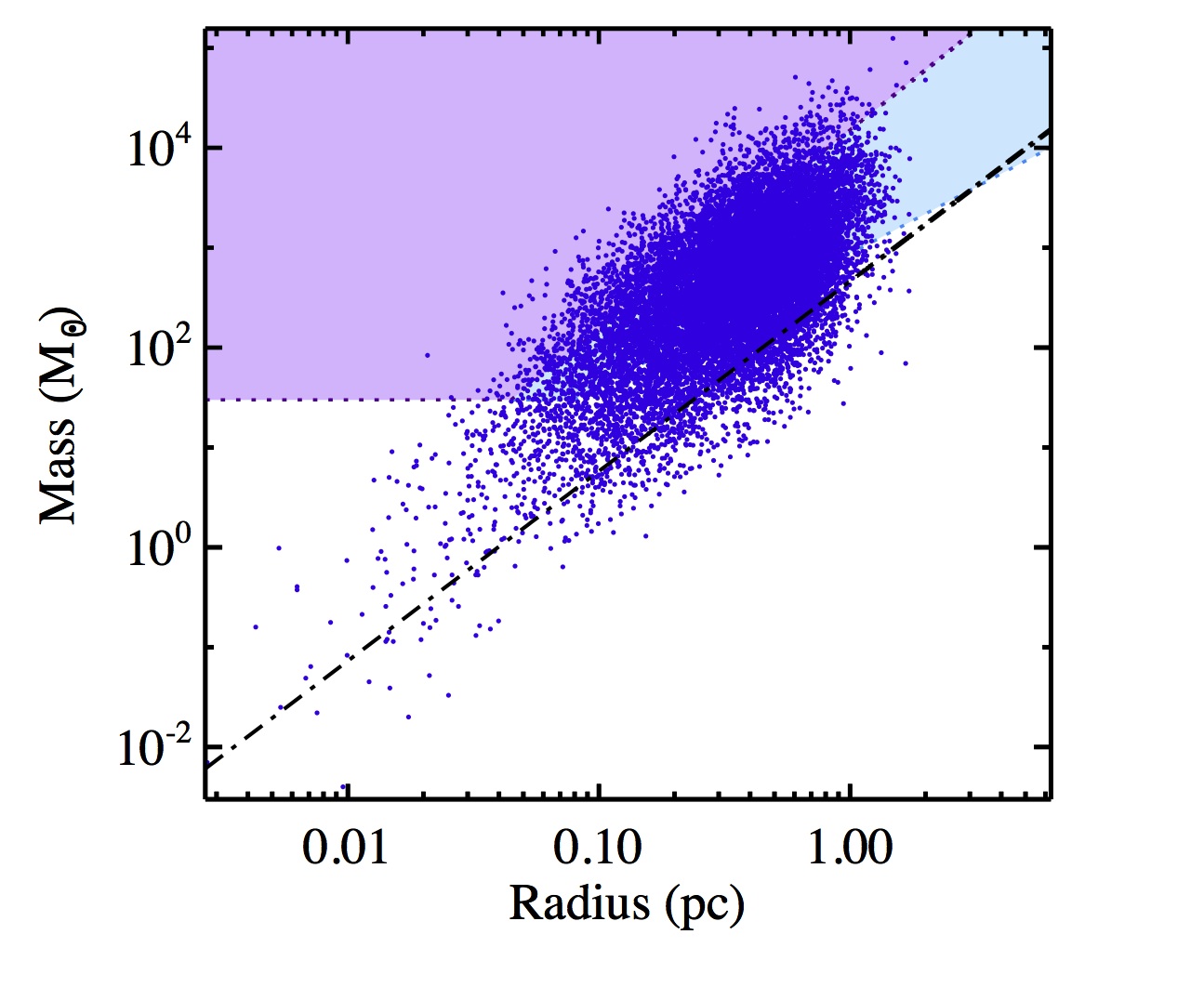} \\
    \includegraphics[width=8.0cm]{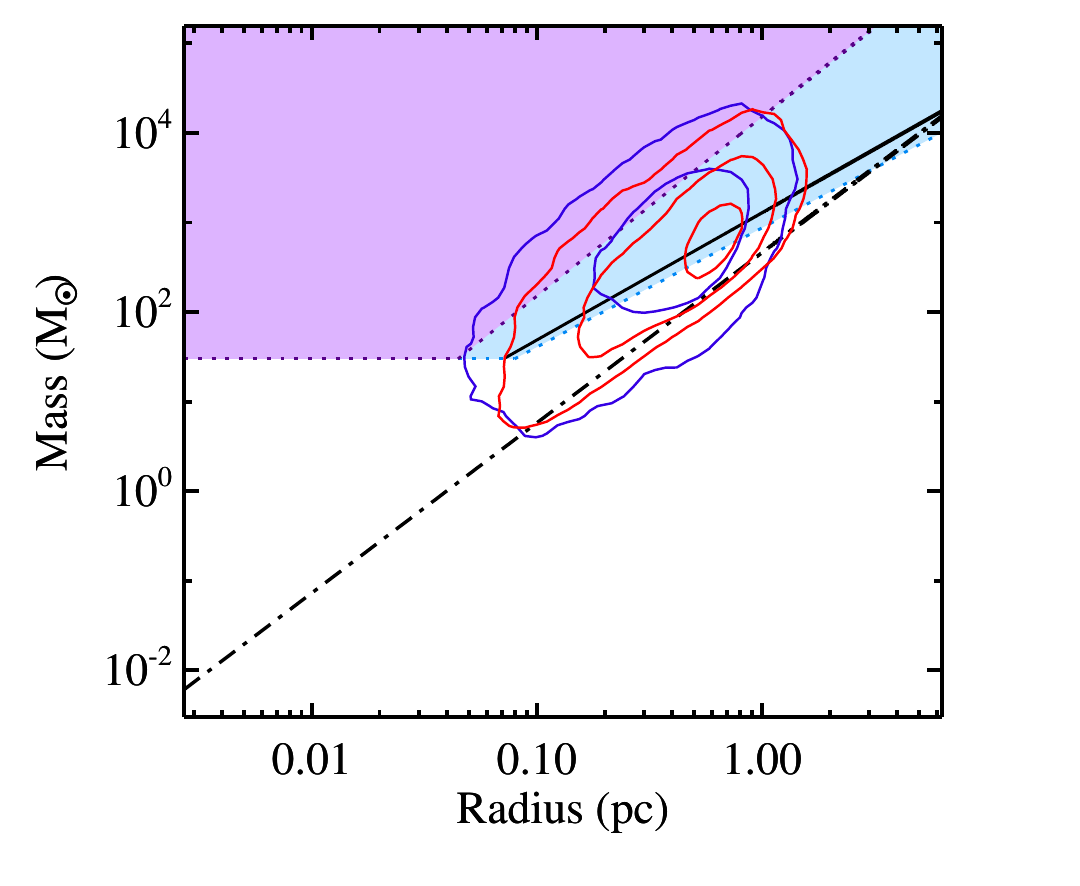} &
    \includegraphics[width=8.0cm]{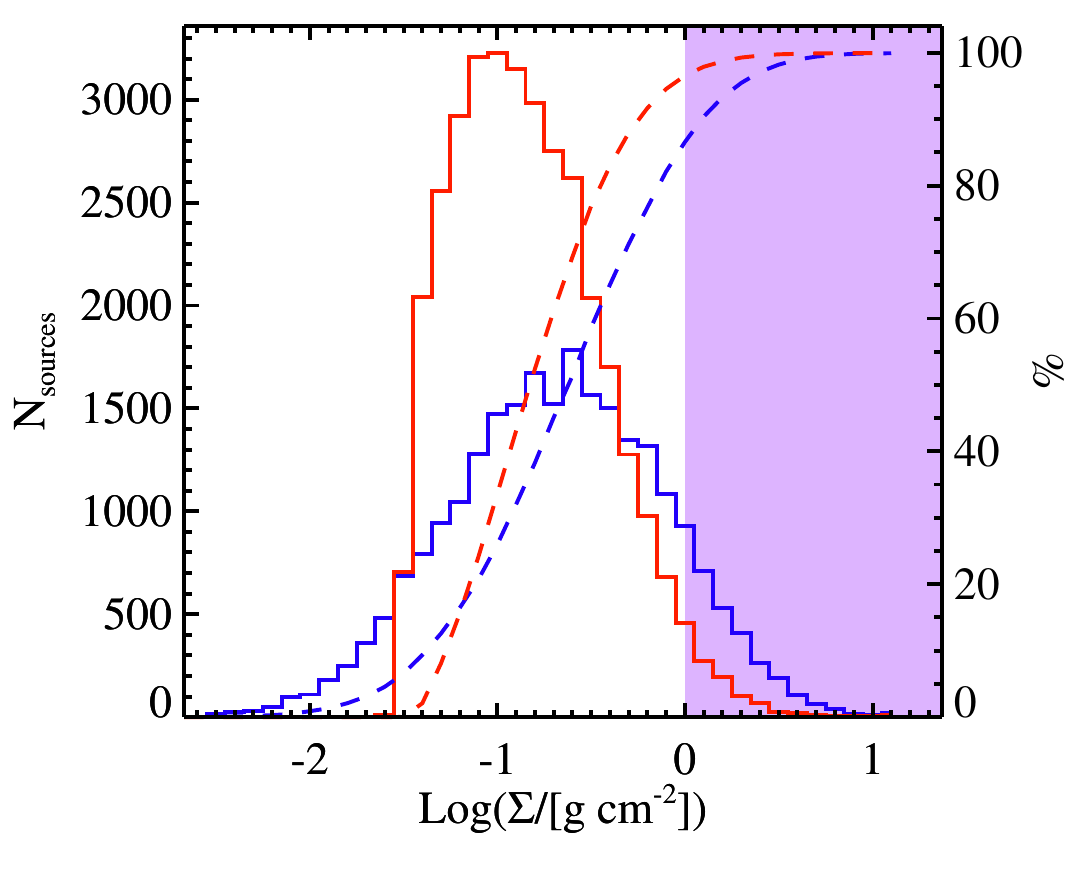}   \\
  \end{tabular}
  \caption{\textit{Top left}: Mass vs radius plot for starless sources. Pre-stellar (red) and starless
  (green) sources are separated by the line $M(r) = 460~M_{\odot}~(r/\mathrm{pc})^{1.9}$ \citep{lar81}
  (dotted dashed black line), see Section~\ref{protovsstarl}. The areas of the mass-radius plane
  corresponding to combinations fulfilling the \citet{kau10} and the \citet{kru08} thresholds for
  compatibility with high-mass star formation are filled with light blue and purple, respectively,
  the latter being contained in the former, and both delimited by a darker dotted line. Notice that,
  adopting a lower limit of $10~M_{\odot}$ for the definition of massive star, and a star formation
  efficiency factor of $1/3$ for the core-to-star mass transfer as in \citet{eli13}, these zones can
  not extend below $30~M_{\odot}$. \textit{Top right}: the same as in top left panel, but for
  proto-stellar sources. \textit{Bottom left}: source density isocontours representing the pre-
  and proto-stellar distributions displayed in the two upper panels of this figure. Densities have 
  been computed subdividing the area of the plot in a grid of $70 \times 70$ cells; once the global maxima
  for both distributions have been found, the plotted contours represent the 2\%, 20\% and 70\% of 
  the largest of those two peak values, so that the considered levels correspond to the same values
  for both distributions, to be directly comparable. The black solid line crossing the bottom part of 
  the light blue area represents the threshold of \citet{bal17}. \textit{Bottom right}: surface density
  distributions (solid histograms) for pre- (red) and proto-stellar (blue) sources. Because surface 
  density is a distance-independent quantity, all sources (with and without distance estimates) are 
  taken into account to build these distributions. The cumulative curves are plotted with
  dashed lines, normalised to the $y$-axis scale on the right. The zone corresponding
  to densities surpassing the \citet{kru08} threshold is filled with purple colour.}
  \label{masssize}
\end{figure*}

A meaningful combination of source properties is represented by the mass $M$ vs radius $r$ diagram, 
which has been shown to be a powerful tool for investigating gravitational stability of 
\textit{Herschel} compact sources, and their potential ability to form massive stars 
\citep{and10,gia12,eli13}. In both cases, in fact, requirements expressed in terms of 
surface density threshold can be translated into a simple mass-radius relation.
Figure~\ref{masssize} shows the mass vs radius distribution for the sources analyzed in the
present study. In the top left panel the starless sources are shown while, to avoid confusion, 
the proto-stellar ones are reported in the top-right panel: the Larson's relation mentioned in
Section~\ref{protovsstarl} is plotted to separate the starless bound (pre-stellar) and unbound
sources.

From this plot it is possible to determine if a given source satisfies the condition for 
massive star formation to occur, where such condition is expressed as surface
density, $\Sigma$, threshold. \citet{kru08} established a critical value of
$\Sigma_{\mathrm{crit}}=1$~g~cm$^{-2}$ based on theoretical arguments. However \citet{lop10}
and \citet{but12}, based on observational evidences, suggest the less severe values of 
$\Sigma_{\mathrm{crit}}=0.3$ and $0.2$~g~cm$^{-2}$, respectively. Also, \citet{kau10}, based 
on empirical arguments, propose the threshold $M(r)>870~M_{\odot}(\mathrm{r}/\mathrm{pc})^{1.33}$ 
as minimum condition for massive star formation. Finally, the recent analysis by
\citet{bal17} of the distance bias affecting the source classification
according to the two aforementioned thresholds, produced the further criterion 
$M(r)>1282~M_{\odot}(\mathrm{r}/\mathrm{pc})^{1.42}$. In the upper side we represent the most 
\citep{kru08} and the least \citep{kau10} demanding thresholds, respectively, to allow 
comparison with the behaviour of our catalogue sources. 

As reported in Table~\ref{masssizetab}, a remarkable fraction of sources appears to be 
compatible with massive star formation based on the three thresholds (defined as 
$\Sigma_\mathrm{KM}$, $\Sigma_\mathrm{B}$, and $\Sigma_\mathrm{KP}$, respectively), 
especially the last. 
This is further highlighted by the bottom-left panel of Figure~\ref{masssize}, which summarises 
the previous two panels reporting the source densities for both the pre- and the proto-stellar
source populations. The peak of the proto-stellar source concentration lies well inside the area 
delineated by the \citet{kau10} relation, while the pre-stellar distribution peaks at smaller 
densities. Rigorously speaking, however, such considerations on the initial conditions for star
formation should be applied only to the pre-stellar clumps, since in the proto-stellar ones
part of the initial clump mass has been already transferred onto the forming star(s) or
dissipated under the action of stellar radiation pressure or through jet ejection. In any
case, the presence of a significant number of very dense pre-stellar sources translates into an 
interestingly large sample of targets for subsequent study of the initial conditions for massive 
star formation throughout the Galactic plane (see Section~\ref{IvsIV}). For such sources,
due to contamination between the two classes described in Section~\ref{protovsstarl}, a further
and deeper analysis is requested to ascertain their real starless status, independently from
the lack of a \textit{Herschel} detection at $70~\umu$m.

Notice that the fractions corresponding to the threshold of \citet{kau10} reported in 
Table~\ref{masssizetab} appear remarkably lower than the same quantity estimated by 
\citet{wie15} for the ATLASGAL catalogue, namely 92\%. This discrepancy cannot be simply
explained by the better sensitivity of Hi-GAL: taking the sensitivity curve in the mass vs 
radius of \citet[][their Figure~23]{wie15}, we find that the majority of our sources lie 
above that curve. The main reason, instead, resides in the analytic form itself of the adopted 
threshold. As mentioned above, \citet{bal17} shown that, even in presence of dilution 
effects due to distance, sources in the mass vs radius plot are found to follow a slope 
steeper than the exponent 1.33, so that large physical radii, typically 
associated to sources observed at very far distances, correspond to masses larger than the 
\citet{kau10} power law. This is not particularly evident in our Figure~\ref{masssize}
since the largest probed radii are around 1~pc, and the large spread of temperatures 
makes the plot quite scattered. Instead, Figure~23 of \citet{wie15} contains a 
narrower distribution of points (since masses were derived in correspondence with 
only two temperatures, both higher than 20~K), extending up to $r\simeq 6$~pc: at 
$r\gtrsim 1$~pc almost the totality of sources satisfy the \citet{kau10} relation.
Clearly, another contribution to this discrepancy can be given by the scatter
produced by possible inaccurate assignment of the far kinematic distance solution
in cases of unsolved ambiguity (see Section~\ref{dist_sect}).

\begin{table}
\centering
\caption{Hi-GAL sources with mass-radius relation compatible with massive star formation according to surface density thresholds $\Sigma_\mathrm{KP}$, $\Sigma_\mathrm{B}$, and $\Sigma_\mathrm{KM}$ (see text).}
\label{masssizetab}
\begin{tabular}{lcccccc}
\hline
 & \multicolumn{2}{c}{$\Sigma>\Sigma_\mathrm{KP}$} & \multicolumn{2}{c}{$\Sigma>\Sigma_\mathrm{B}$} & \multicolumn{2}{c}{$\Sigma>\Sigma_\mathrm{KM}$} \\
 & counts & \% & counts & \% & counts & \% \\ 
\hline
Proto-stellar &        11210 & 71.3 &        10012 & 63.7 &         2062 & 13.1 \\
Pre-stellar   &        12431 & 64.8 &         9973 & 52.0 &          546 &  2.8 \\
\hline
\end{tabular}
\end{table}

The information contained in the mass-radius plot can be rearranged in a histogram
of the surface density\footnote{It is notable that if a grey body is fitted to a SED through
Equation~\ref{gbthick}, the surface density is proportional to $\tau_{\mathrm{ref}}$
(Equation~\ref{mthick}), which is in turn proportional to $\lambda_0^{-\beta}$
(Equation~\ref{tau}), with $\beta=2$ in this paper. This implies that a description based on
the surface analysis is, for such sources, equivalent to that based on the $\lambda_0$
parameter.} $\Sigma$ such as that in Figure~\ref{masssize}, bottom right. The pre-stellar source
distribution presents a sharp artificial drop at small densities due to the removal of the unbound
sources which is operated along $M \propto r^{1.9}$, i.e. at an almost constant surface density
($M \propto r^{2}$). Instead, despite the considered pre-stellar population being globally more numerous 
than the proto-stellar one, at high densities ($\Sigma \gtrsim 1$~g~cm$^{-2}$) the latter prevails 
over the former. The proto-stellar distribution, in general, appears shifted towards larger densities,
compared with the pre-stellar one, as can be seen in the different behaviour of the
cumulative curves, also shown in figure. This evidence is in agreement with the result of
\citet{he15}, based on the MALT90 survey. Further evolutionary implications will be discussed 
in Section~\ref{surfdens_par}.

We note, as discussed in Section~\ref{physsize}, that the compact sources we consider
may correspond, depending on their heliocentric distance, to large and in-homogeneous clumps with
a complex underlying morphology not resolved with \textit{Herschel}. On the one hand, this 
implies that the global properties we assign to each source do not remain necessarily constant 
throughout its internal structure, thus a source fulfilling a given threshold on the surface 
density might, in fact, contain sub-critical regions. On the other hand, in a source with a 
global sub-critical density, super-critical portions might actually be present, leading to 
mis-classifications; for this reason the numbers reported in Table~\ref{masssizetab} should 
be taken as lower limits.

\begin{figure}
\centering
\includegraphics[width=8.5cm]{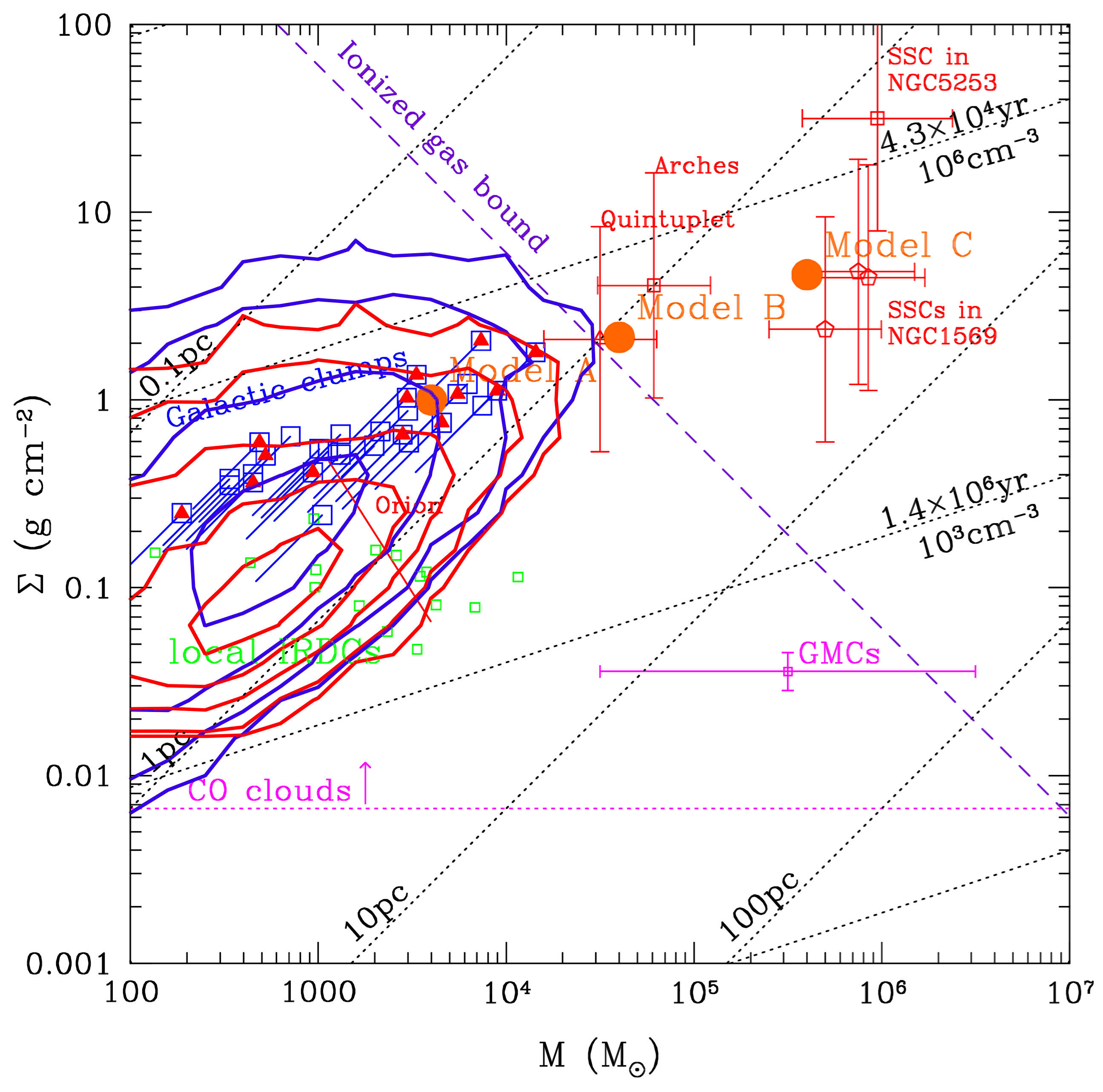}
\caption{Plot of surface density vs mass of \citet[][his Figure~1]{tan05}, with
overplotted number density contours of pre- (red) and proto-stellar (blue) 
sources of our catalogue. The source density is computed in bins of 0.2 in decimal
logarithm, and contour levels are 5, 20, 100, 250, and 500 sources per bin. The
horizontal cut at lowest densities for red contours is due to the removal of starless
unbound sources. The diagonal dotted lines represent the locii of given radii and number 
densities (and corresponding free-fall times). Typical ranges for molecular clouds are
indicated with magenta lines, while the condition for the ionised gas to remain bound 
is indicated by the blue dashed line. Finally, locations for a selection of IRDCs (green 
squares), dense star-forming clumps (blue squares), massive clusters (red symbols) and
cluster models of \citet{tan05} (filled orange circles). More details about additional 
information reported in this diagram are provided in \citet{tan05} and 
\citet[][their Figure~11]{mol14}.}
\label{sigmamass}
\end{figure}

We use the collected information about source masses and surface densities to place 
them in the $\Sigma$ vs $M$ plot of \citet{tan05}, in which different classes of
structures populate different regions. Our Figure~\ref{sigmamass} is analogous to
Figure~11 of \citet{mol14}, but the ``temporary'' data set used for that plot is replaced
here with the values from the final Hi-GAL physical catalogue. The Hi-GAL sources are 
found to lie in the regions quoted by \citet{tan05} for Galactic clumps and local IRDCs.
In the upper-right part of their distribution they graze the line representing the
condition for ionised gas to remain bound. This plot summarises the nature of the sources 
in our catalogue: clumps spanning a wide range of mass/surface density combinations, 
with many of them found to be compatible with the formation of massive Galactic clusters. 

\subsection{The ClumpMF}\label{cmfpar}
The ClumpMF is an observable that has been extensively studied for understanding the connection 
between star formation and parental cloud conditions. Although formulations are quite similar,
the ClumpMF should not be confused with the core mass function (hereafter CoreMF), typically
studied in nearby star forming regions ($d \lesssim 1$~kpc). Differences between these two 
distributions will be discussed later in this section.

 Large infra-red/sub-mm surveys generated 
numerous estimates of the ClumpMF \citep[e.g.,][]{rei05,rei06,ede12,tac12,urq14a,moo15}. Likewise, 
data from Hi-GAL have been used for building the ClumpMF in selected regions of the Galactic plane 
\citep{olm13,eli13}.

Building the mass function of a given sample of sources (Hi-GAL clumps in the present case) 
requires a sample to be defined in a consistent way. A clump mass function built from a sample 
of sources spanning a wide range of distances (as in our case) would be meaningless, 
since at large distances low-mass objects might not be detected or might be confused within 
larger, unresolved structures (see Appendix~\ref{appdist}), therefore it makes little sense 
to discuss it. Therefore, we first subdivide our source sample into bins of heliocentric 
distance, and then build the corresponding mass functions separately. In 
addition, as pointed out e.g. by \citet{eli13}, it is more appropriate to build separate ClumpMFs
for pre- and proto-stellar sources. Strictly speaking, only the mass distributions of 
pre-stellar sources is intrinsically coherent, as the mass of the proto-stellar sources 
does not represent the initial core mass, but rather a lower limit which depends on the 
current evolutionary stage of each source.

\begin{figure*}
\centering
\includegraphics[width=16.5cm]{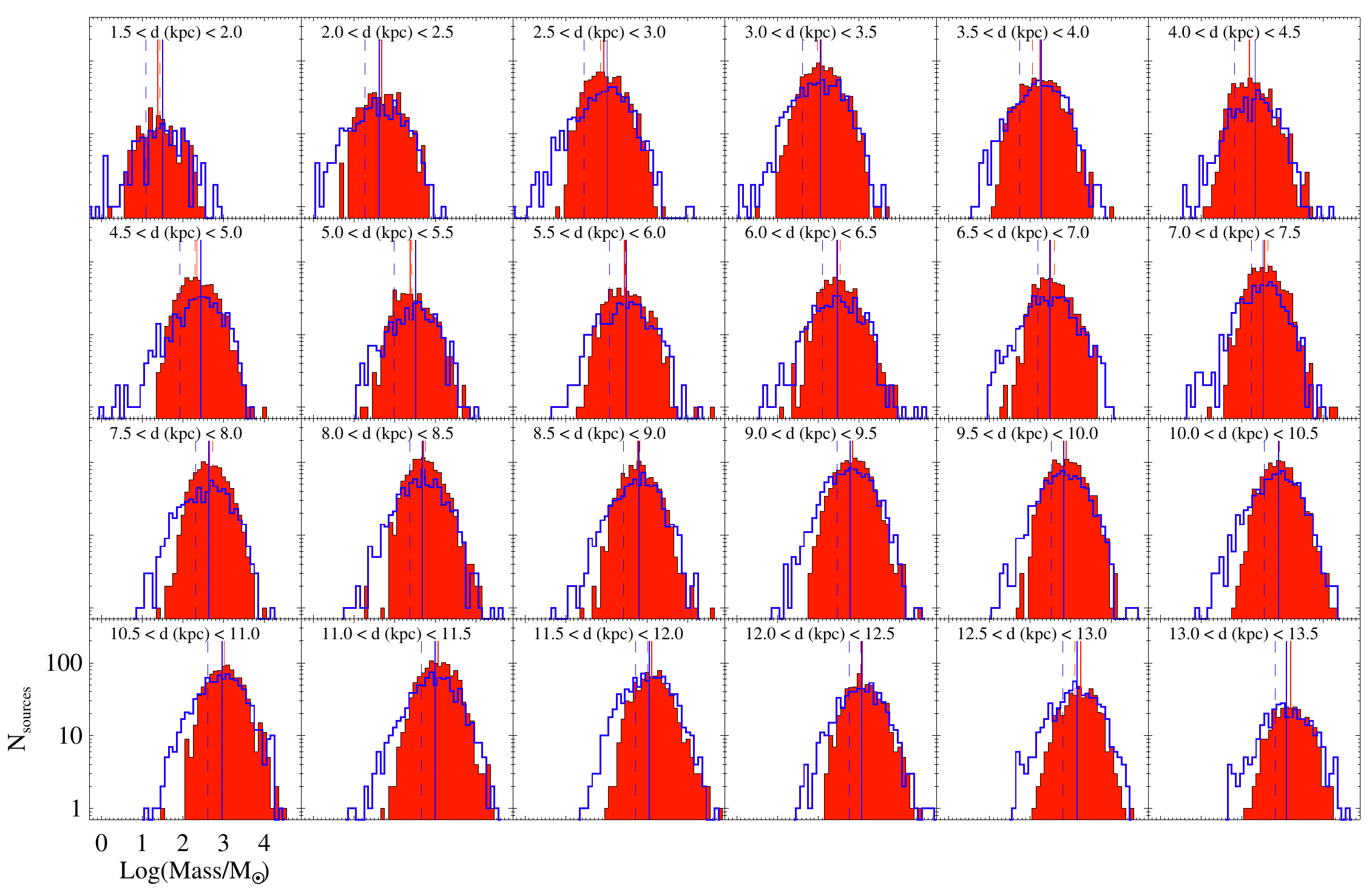}
\caption{Clump mass function for pre-stellar (red filled histogram) and proto-stellar (blue histogram)
sources, obtained in 0.5~kpc-wide heliocentric distance ranges. Each panel corresponds to the range
which is reported in the upper part. Dashed red and blue line indicate the completeness limits, and 
solid red and blue vertical line indicate the lower limit used to fit the curves in Figure~\ref{ccmf}, 
for the pre- and proto-stellar case, respectively.}
\label{hist_mass}
\end{figure*}

In Figure~\ref{hist_mass} the clump mass functions are shown. They have been calculated using sources 
provided with a distance estimate, from 1.5 to 13.5~kpc in distance bins of 0.5~kpc, in logarithmic
mass bins, and separately for pre- and proto-stellar clumps.

It can be immediately noticed that, for any distance range, the proto-stellar ClumpMF is wider 
than the pre-stellar one, so that a deficit of pre-stellar clumps with respect to proto-stellar 
ones is seen both at lowest and highest masses in each bin of distance. The former effect 
is mostly due to sensitivity: sure enough, thanks to higher temperature, a proto-stellar source 
can be detected more easily than a pre-stellar source of the same mass. For example, according 
to Equation~\ref{mcomp350} in Appendix~\ref{appmasslim}, applied to a given mass, the flux
of a source at $T=15$~K (i.e. $\sim \bar{T}_{\mathrm{prt}}$) is nearly twice that of a source 
at $T=12$~K (i.e. $\sim \bar{T}_{\mathrm{pre}}$~K).
The latter effect can be in part explained with the pre-/proto-stellar possible blending and 
misclassification at increasing distance discussed in Appendix~\ref{appdist}, which leads to
artificially overestimating the fraction of proto-stellar sources. However, from the quantitative
point of view, this effect seems not sufficient to entirely account for the complete lack of 
pre-stellar sources as massive as the most massive proto-stellar ones. A further contribution 
is surely given as well by the more rapid evolution of massive pre-stellar sources towards the 
proto-stellar status \citep[see, e.g.,][]{mot07,rag13}. To correctly examine this point, 
sub-samples which are consistent in terms of heliocentric distance must be isolated, as we 
do in fact building Figure~\ref{hist_mass}. Indeed, the structure of two clumps of, say, 
$200~M_\odot$ detected at $d=10$~kpc and $d=2$~kpc would be strongly different: the former
would be expected to be likely composed of an underlying population of low-mass cores \citep{bal17}, 
whereas the second, being better resolved by \textit{Herschel}, would be denser and less 
fragmented than the former, therefore representing a more reliable candidate for hosting 
massive star formation and having shorter evolutionary time scales. This scenario is confirmed 
in most panels of Figure~\ref{hist_mass} where, given a range of heliocentric distances, a 
lack of pre-stellar clumps with respect to the proto-stellar ones is generally found in the 
bins corresponding to the highest masses present. Importantly, this indicates that it is not 
sufficient to simply claim that massive clumps have very short lifetimes \citep{gin12,tac12,cse14}: 
indeed, a large clump mass might also result from an associated large heliocentric distance, 
which implies multiple source confusion and inclusion of diffuse emission contaminating source 
photometry \citep[see also][]{bal17}. In this respect, clump density 
has to be taken into account as well, since only high densities ensure conditions for massive 
star formation and, therefore, for a faster evolution of a clump. 

The differences observed between pre- and proto-stellar ClumpMFs are expected to be reflected
on the slope of the power-law fit of high-mass end of these distributions, i.e. 
the usual way to extract information from the ClumpMF and compare it with the stellar
initial mass function. The estimate of this slope is generally plagued by uncertainties 
due to arbitrary choice of mass bins and of the lower limit of the range to be 
involved in the fit. \citet{olm13} have presented an efficient way, based on application 
of Bayesian statistics, to overcome such issues. Here we adopt a simpler approach:
\begin{itemize}
\item The slope of the ClumpMF is derived indirectly, by estimating the slope of the 
corresponding cumulative function defined, as a function of $M$, as the 
fraction of sources having mass larger than $M$. If a ClumpMF is calculated 
in logarithmic bins and is expected to have a power-law behaviour above a certain 
value $M_\mathrm{fit}$, so that $\mathrm{d}N(M)/\mathrm{d}\log_{10}(M) \propto M^\alpha$, 
then the corresponding cumulative function (hereafter CClumpMF) has
the same exponent $\alpha$ \citep[e.g.,][]{shi03}, and the estimate of such 
slope is independent from the bin used to sample the ClumpMF. 
\item The portion of the ClumpMF used in the fit should be delimited at bottom by the turn-over 
point of the ClumpMF $M_\mathrm{peak}$, namely the peak of a log-normal best fit
\citep{cha03}. 
One has to ensure that this mass limit is larger than the completeness limit
of the distribution. 
The estimation of the mass completeness limit $M_\mathrm{compl}$ is not trivial 
in our case, since multiple bands and variable temperature and distance concur in the 
mass determination, as discussed in Appendix~\ref{appmasslim}. Equation~\ref{mcomp350} 
can be applied to compute the limit, assuming the central distance of each bin, 
the median temperatures of pre- and proto-stellar sources, and a flux 
completeness limit at $350~\umu$m $F_{\mathrm{compl,350}}$ as estimated 
by \citet{mol16a}. This completeness limit, quoted by these authors as a function 
of Galactic longitude (their Figure~9), reaches a maximum of 13.08~Jy around 
$\ell=0^{\circ}$ (a region which does not provide sources for this analysis, given the 
lack of distance information), and a minimum of 0.65~Jy in the eastmost tile of the first 
quadrant. In intermediate regions, which provide the biggest contribution in building 
our ClumpMFs, in general $F_{\mathrm{compl,350}} \lesssim 4$~Jy, which we adopt here.
In a few cases in which $M_\mathrm{peak}<M_\mathrm{compl}$, the latter
is taken as the lower limit of the fit range.
\end{itemize} 

\begin{figure*}
\centering
\includegraphics[width=16.5cm]{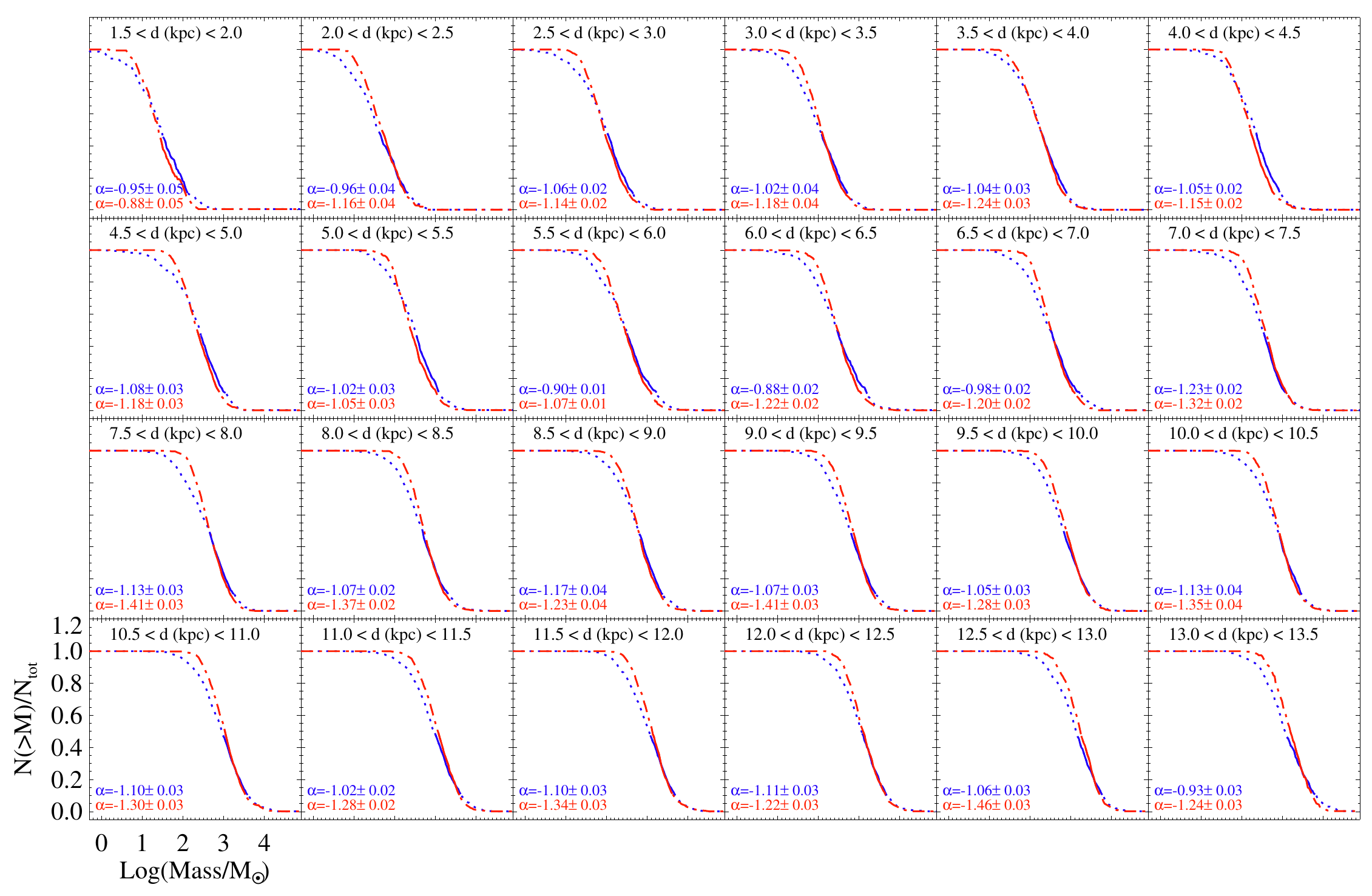}
\caption{CClumpMFs for pre-stellar (red dotted-dashed line) and proto-stellar (blue dotted 
line), obtained in the same distance ranges used for Figure~\ref{hist_mass}. The power-law (linear 
in bi-logarithmic scale) portion is highlighted with solid lines, and the corresponding slope is 
reported in the panel with the same colour coding.}
\label{ccmf}
\end{figure*}

In each panel of Figure~\ref{hist_mass} the peak of the ClumpMF and the completeness limit are 
shown for both pre- and proto-stellar distributions, while in Figure~\ref{ccmf} the corresponding 
CClumpMFs are shown. The slopes obtained through the power-law fit range from \maxslopepre~to 
\minslopepre~for the pre-stellar sources and from \maxslopeproto~to \minslopeproto~for the proto-stellar
ones. As expected from the discussion above, slopes of proto-stellar ClumpMFs are systematically 
shallower than the pre-stellar ones \citep[cf. also][]{dif10}, being the former strongly biased by 
the lack of clumps at the highest mass bins compared with proto-stellar ones. In some cases, slopes 
of pre-stellar ClumpMFs can take values even steeper than the stellar Initial Mass Function 
\citep[IMF, $\alpha_\mathrm{IMF}=-1.35$][]{sal55}, as testified also by \citet{tac12}.
On the contrary, the slopes of proto-stellar ClumpMFs remain always shallower than 
$\alpha_\mathrm{IMF}$, thus confirming the typical expectation for a generic mass 
distribution of unresolved clumps \citep{rag09,dif10,per10a,ede12,pek13}, while for 
the CoreMF a slope compatible with $\alpha_\mathrm{IMF}$ is typically 
found \citep[e.g.,][]{gia12,pol13,kon15}. On the one 
hand, this confirms, across a wide range of heliocentric distances and based on
unprecedentedly large statistics, a behaviour of the ClumpMF already known from literature.
On the other hand we caution the reader that $i$) the behaviour of the pre-stellar
ClumpMFs has to be better investigated in the future (for example by means of higher-resolution
observations of high-density pre-stellar cores, as suggested by Figure~\ref{sigmamass}), and
$ii$) the ClumpMFs discussed here are obtained regardless Galactic longitude, but simply grouping 
sources by heliocentric distance. More focused studies on selected ranges of Galactic longitude 
will enable the generation of mass distributions for even more coherent data sets, while also making it
possible to explore environmental variations when looking at, e.g., individual spiral arms, 
tangent points, and star forming complexes. This, in turn, will allow 
assessment of similarities and differences among different Galactic locations.

Finally, ClumpMF slopes at different distances allow us to test the possible effects of 
gradual lack of spatial resolution on the ClumpMF slope. This problem has been already 
investigated by \citet{rei10} by means of simulations. Despite a depletion of sources in 
low-mass bins is expected at increasing heliocentric distance, together with an increase 
in high-mass bins due to blending, \citet{rei10} did not find a progressive shallowing of 
the ClumpMF. Here we can confirm, based on observational arguments, that the slopes reported 
across various panels of Figure~\ref{ccmf} do not show any particular trend with distance. 
Therefore, whereas a clear distinction is found between the mass spectrum of cores 
\citep[typically resolved by \textit{Herschel} if located at $d \lesssim 1$~kpc,][]{gia12,bal17},
and that of clumps, with the former being steeper than the latter, 
no further systematic steepening is found for clumps observed at increasing 
distances. This might be also regarded as an indirect evidence of the self-similarity of 
molecular clouds \citep{stu98,smi08,eli14} over the investigated range of physical scales,
which, in this case, is the range of the linear sizes of compact sources located between
$d=1.5$ and 13.5~kpc.

\subsection{Bolometric luminosity and temperature}\label{lum_par}

\begin{figure*}
\centering
\includegraphics[width=16.5cm]{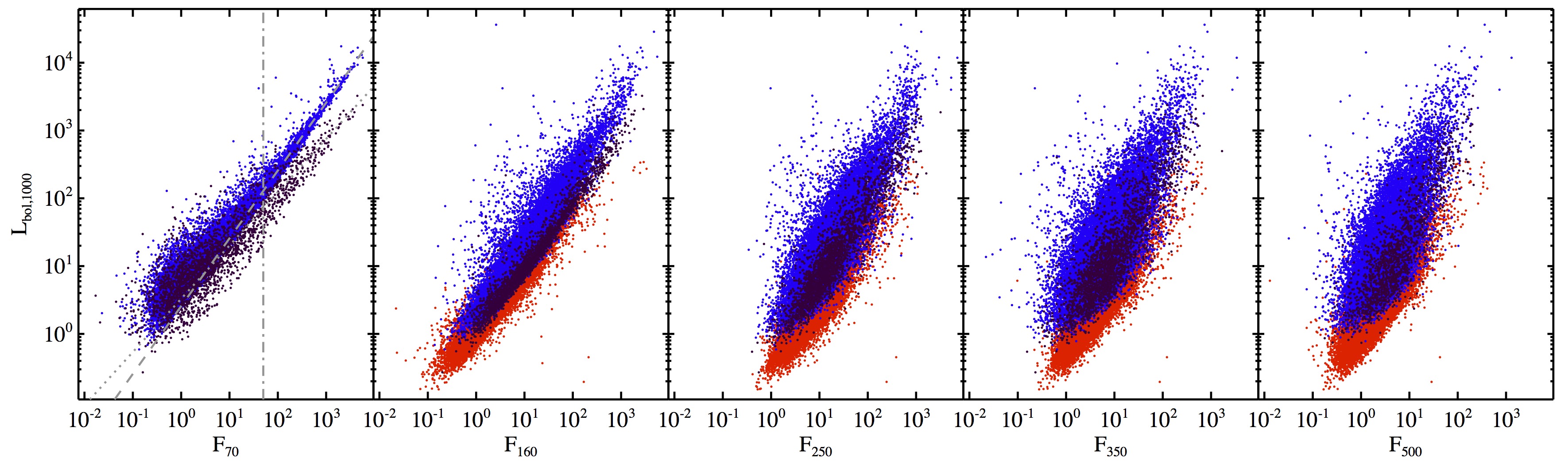}
\caption{From left to right: source bolometric luminosity re-scaled
to the virtual distance $d_v=1$~kpc vs \textit{Herschel} monochromatic
flux at 70, 160, 250, 350, and 500~$\umu$m, respectively. For clarity purposes, 
symbols of sources belonging to the same evolutionary class have been plotted all
together: proto-stellar sources (blue) are overlaid on the pre-stellar sources (red), and 
MIR-dark proto-stellar sources (dark purple) are overlaid in turn on proto-stellar sources. 
In the leftmost panel (corresponding to the 70~$\umu$m case) the power-law fit is shown for 
sources with large fluxes ($F_{70} \gtrsim 50$~Jy, grey dotted-dashed vertical line),
separately for the MIR-bright sub-sample (dashed line) and for the MIR-dark one
(dotted-line).}\label{lumflux}
\end{figure*}

Before discussing the use of bolometric luminosity, estimated as described in 
Section~\ref{sedfit}, to infer the evolutionary stage of a clump, we show 
how this quantity correlates with monochromatic \textit{Herschel} fluxes. Indeed, \citet{dun08} 
already suggested the \textit{Spitzer} flux at 70~$\umu$m as a reliable proxy of the total 
proto-stellar core luminosity, and \citet{rag12} confirmed an evident correlation
between fluxes measured at all the three PACS bands (70, 100, and 160~$\umu$m) and
bolometric luminosity of \textit{Herschel} clumps. In the five panels of 
Figure~\ref{lumflux} the bolometric luminosity, conveniently re-scaled to a common 
virtual distance $d_v=1$~kpc, is plotted vs the \textit{Herschel} flux at 
different bands. The tightest correlation is found at PACS wavelengths, especially at 
70~$\umu$m (left panel), where an overall power-law behaviour for MIR-bright sources
can be identified at large fluxes ($F_{70} \gtrsim 50$~Jy). At lower fluxes one observes a departure 
of luminosity \citep[observed also in][]{rag12} from the trend, which in this case 
represents the lower limit of the distribution. Interestingly, a secondary trend, 
similar to the main one but at a lower luminosity level and essentially due to 
MIR-dark proto-stellar sources, is observed $F_{70} \gtrsim 50$~Jy. The power-law best fit 
yields the expressions 
$L_\mathrm{bol,MIR-bright}[L_\odot]=\lumcoefficient~F_\mathrm{70}[\mathrm{Jy}]^\lumexponent$
\citep[corresponding to linear behaviour, as in][]{rag12} and
$L_\mathrm{bol,MIR-dark}[L_\odot]=\lumcoefficientmirdark~F_\mathrm{70}[\mathrm{Jy}]^\lumexponentmirdark$
for the two sub-samples in the considered range of fluxes, respectively.

The relation between bolometric luminosity and the envelope mass is particularly 
interesting as an indicator of the evolutionary status of a core/clump. The 
$L_{\mathrm{bol}}$ vs $M_{\mathrm{env}}$ diagram is a widely used tool \citep{sar96,mol08,and08,gia12,rag13,gia13,eli13} in which evolutionary tracks, essentially 
composed by an accretion phase and a clean-up phase \citep{mol08,smi14}, can 
be plotted and compared with data. In the earliest stages of
star formation, as protostar gains mass from the surrounding envelope, 
these tracks are nearly vertical, while, after the central star has reached 
the Zero Age Main Sequence (ZAMS), they assume a nearly horizontal behaviour
corresponding to dispersal of the residual clump material.

\begin{figure*}
\centering
\includegraphics[width=8.5cm]{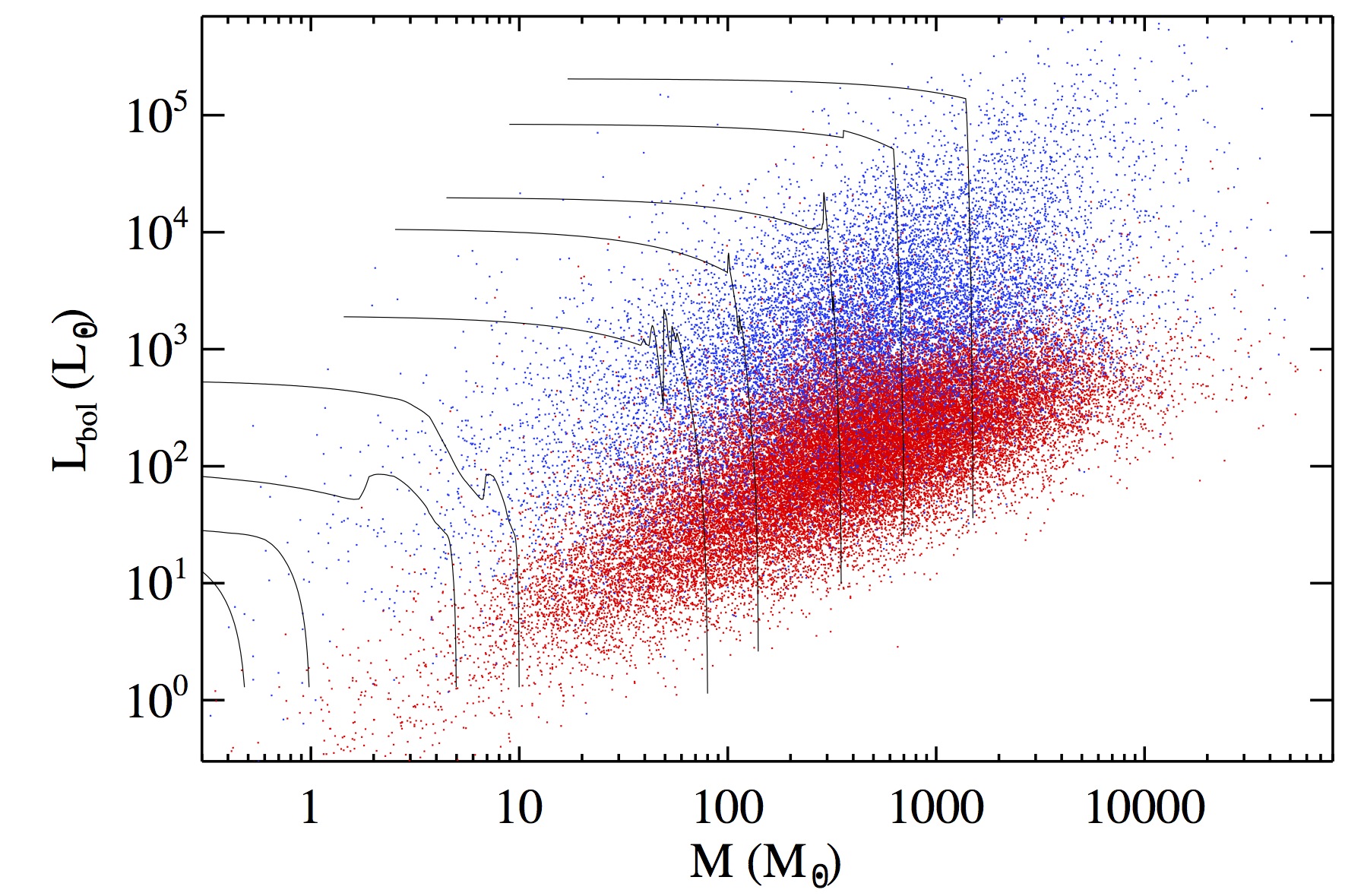}
\includegraphics[width=8.5cm]{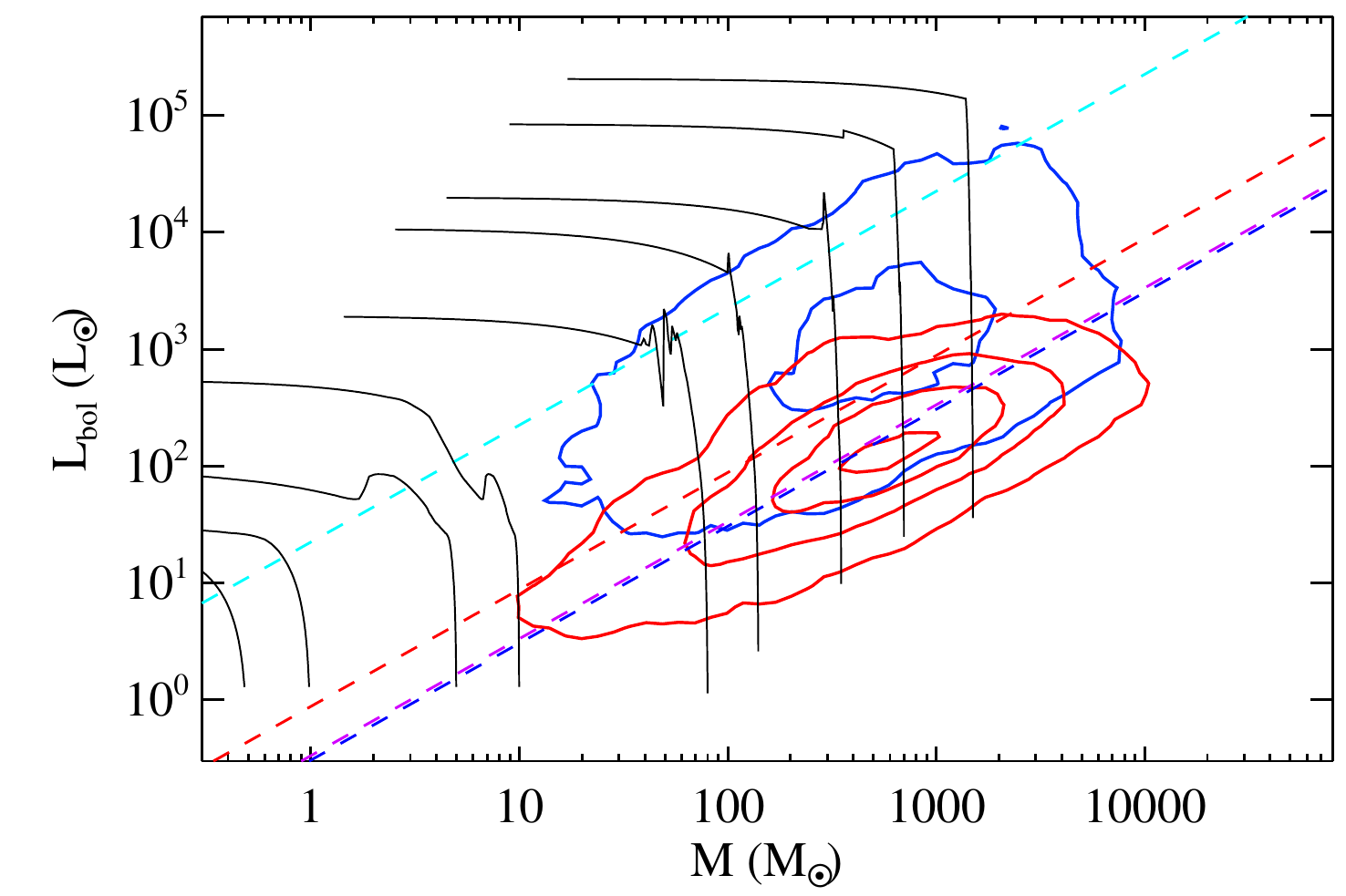}
\caption{\textit{Left}: $L_{\mathrm{bol}}$ vs $M_{\mathrm{env}}$ plot for all the sources 
considered in this paper. The black lines represent evolutionary tracks from \citet{mol08}.
\textit{Right}: The same as in the left panel, but with source density contours 
plotted instead of single sources. Contours are determined as in the bottom left
panel of Figure~\ref{masssize}, using levels of 5\%, 25\%, 50\%, and 75\% of the maximum
density found in the plot. Dashed lines correspond to some relevant percentiles of the 
$L_{\mathrm{bol}}/M_{\mathrm{env}}$ ratio for different source populations (cf. 
Figure~\ref{lmratio}): in the area of the diagram below the red line, the 90\% of 
pre-stellar sources are located, while above the blue, the dark purple and the light blue
lines the 90\% of all proto-stellar, MIR-dark proto-stellar and \ion{H}{ii}-region-compatible 
sources are located, respectively.}\label{lvsm}
\end{figure*}

In Figure~\ref{lvsm}, \textit{left}, we built the $L_{\mathrm{bol}}$ vs $M_{\mathrm{env}}$ 
plot using the bolometric luminosity obtained as described in Section~\ref{sedfit}, while 
the envelope mass is the source mass $M$ we derived through the grey body fit. For 
proto-stellar objects, $M_{\mathrm{env}}$ represents the residual mass of the parental 
clump/cloud still surrounding the embedded protostars, while for the starless sources 
it is the whole clump mass itself. Hereafter we will adopt $M_{\mathrm{env}}=M$.
The right panel of the figure clarifies, by means of density contours, how proto-stellar 
sources are spread in a large area 
corresponding to a variety of ages, encompassing even evolutionary stages closer to the 
transition between clump collapse and envelope dissolution (ZAMS). However, since the 
bolometric luminosities are computed starting from the MIR ($\sim 20~\umu$m) or 
from longer wavelengths, while the most evolved Hi-GAL sources are expected to have also 
counterparts at shorter wavelengths \citep[e.g.,][]{li12,tap14,str15,yun15}, it is likely 
that for a fraction of proto-stellar sources the evolutionary stage is underestimated, 
and the actual spread in age of this population is larger than represented here.

As better highlighted by source density contours displayed in the \textit{right} panel of Figure~\ref{lvsm},
pre-stellar sources are generally confined in a relatively narrower region corresponding to absence
of collapse, or to the earliest clump collapse phases. This result can be compared with recent Hi-GAL
works focused on smaller portions of the Galactic plane. In \citet{ven17}, the clump populations at 
the tips of the Galactic bar, extracted from the entire catalogue presented here, show a similar 
behaviour. On the contrary, a comparison with the analysis of a portion of the third Galactic quadrant
of \citet{eli13}, as well as the larger number of sources considered and the spread over heliocentric 
distances (resulting in a wider range of masses and luminosities), highlights two points. First,
the barycentre of the mass distribution is located towards higher values as a consequence of 
larger source distances involved in our sample. Second, a higher degree of overlap is seen between the 
pre- and proto-stellar source populationsthe former being found, at $M > 10~M_{\odot}$, to be
overlapped with the accretion portion of the evolutionary tracks, also populated by the proto-stellar 
sources. This does not mean necessarily that in the outer Galaxy a clearer segregation of the pre- vs 
proto-stellar clump populations is seen through this diagnostic tool \citep[see also][for another 
example of analysis of an outer Galaxy region, namely Vela-C]{gia12}, since distance effects
must also be taken into account. On average, the sources of \citet{gia12} and \citet{eli13} are 
much closer ($d=700$~pc and $d \lesssim 2200$~pc, respectively) than most sources of our 
sample, and larger distances might introduce ambiguities in the pre- vs proto-stellar 
classification (see Appendix~\ref{appdist}). Further extension of the clump property
analysis to other regions of the outer Galaxy (Merello et al., in prep.), 
and a systematic treatment of possible biases introduced by distance \citep{bal17}
will make it possible to confirm this interpretation.

\begin{figure*}
\centering
\includegraphics[width=16.0cm]{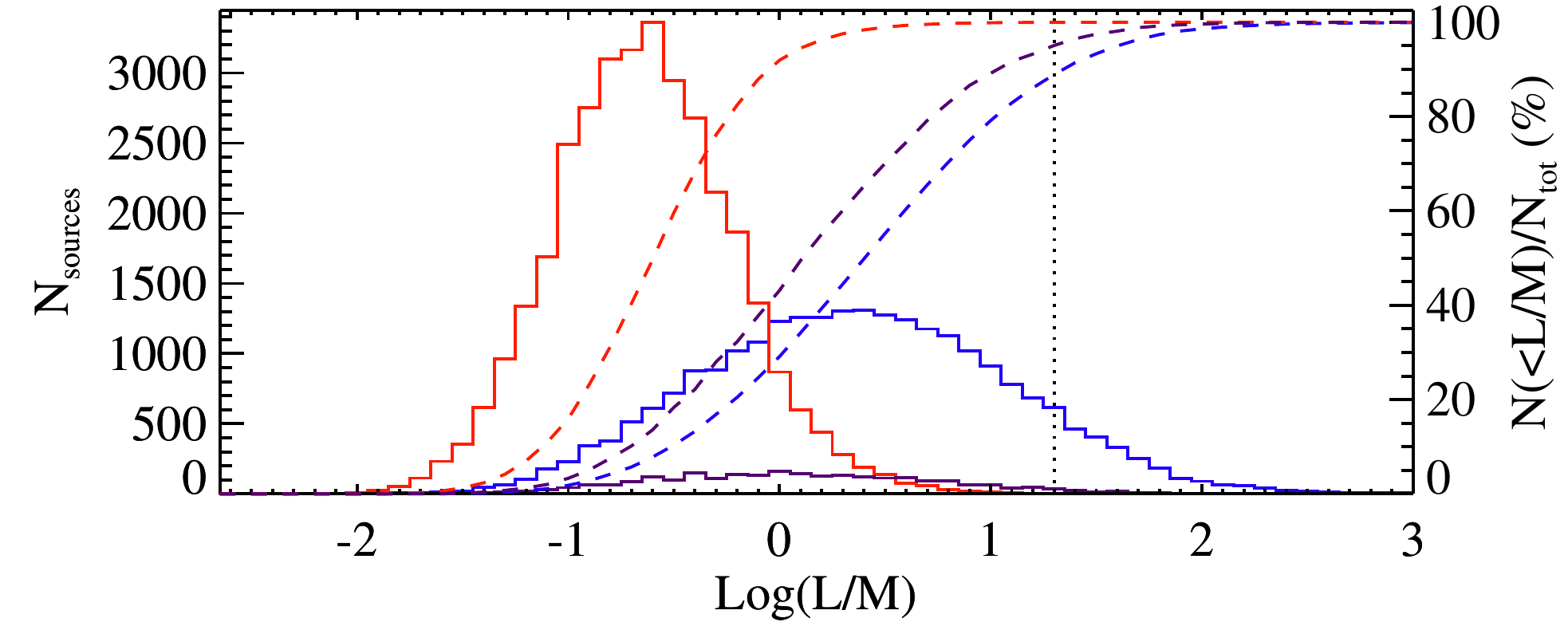}
\caption{The same as in Figure~\ref{hist_temp}, but for the $L_{\mathrm{bol}}/M$ ratio. The 
vertical dotted line represents the peak of the distribution of \citet{ces15} (see text).}
\label{lmratio}
\end{figure*}

To express the relation between $L_{\mathrm{bol}}$ and $M$ through a single indicator 
one can use their ratio $L_{\mathrm{bol}}/M$ \citep[cf., e.g.,][]{ma13,mol16b}\footnote{A 
comparison with $L_{\mathrm{bol}}/M$ ratio found by other similar surveys is provided 
in Appendix~\ref{champepos}.}, which has the advantage of being a distance-independent 
observable, allowing the use of the evolutionary analysis for catalogue 
entries devoid of a distance estimate. Figure~\ref{lmratio} shows the histograms 
of this quantity for pre- and proto-stellar Hi-GAL sources, separately. Also here 
it can be seen that the distributions of $L_{\mathrm{bol}}/M$ ratios for pre-stellar 
and proto-stellar sources appear very different: in general, pre-stellar objects show 
a more confined distribution around \peaklmrpe~$L_{\odot}/M_{\odot}$, while proto-stellar 
sources are widely distributed with a peak around \peaklmrpo~$L_{\odot}/M_{\odot}$. A 
significant overlap of the two histograms is found in any case: notice, for instance, that 
90\% of pre-stellar sources are found at $L_{\mathrm{bol}}/M<\percle~L_{\odot}/M_{\odot}$, 
while 90\% of proto-stellar sources are found at $L_{\mathrm{bol}}/M>\perclo~L_{\odot}/M_{\odot}$
(Figure~\ref{lvsm}, \textit{right}).

The larger width of the proto-stellar source distribution suggests that they are 
transition objects in an evolutionary phase between pure (yet starless) collapse 
and ``naked'' young stars without a dust envelope: the protostar(s) responsible for 
the increase in bolometric luminosity can be still embedded in a large cold dust 
envelope, responsible for keeping the mass of the entire proto-stellar 
clump high enough to reduce the $L_{\mathrm{bol}}/M$ ratio. 
Furthermore, one should keep in mind that conditions favourable to the formation 
of stars might be met only in a fraction of the entire volume of a distant 
proto-stellar clump. The observed $L_{\mathrm{bol}}/M$ ratio of a proto-stellar
object is a single observable computed from the global emission of an envelope 
containing unresolved young stellar objects (YSOs) not necessarily coeval,
but more generally at mixed stages of evolution \citep{yun15}.

From the analysis of the dust temperature distribution (Section~\ref{dtemp}) we could 
infer that MIR-dark proto-stellar sources are at an intermediate evolutionary stage 
between pre-stellar and MIR-bright proto-stellar sources. However, the 
$L_{\mathrm{bol}}/M$ metric indicates (see Figure~\ref{lmratio}) that this population 
spans a wide range of values, instead of being confined to the tail of the 
overall proto-stellar distribution. The cumulative distribution is only slightly different 
from that of the overall proto-stellar sample, and very different from that of the 
pre-stellar sample. Furthermore, the 10\% percentile of this distribution 
(Figure~\ref{lvsm}, right panel) is indistinguishable from the same percentile 
computed for the proto-stellar population. In summary, the $L_{\mathrm{bol}}/M$ ratio 
for MIR-dark proto-stellar sources is not significantly
different from that of the other proto-stellar sources.

To identify possible additional sub-classes within the proto-stellar population, the 
correlation with external evolutionary tracers can be used (as we intend to do in future 
papers based on this catalogue). For example, in the RMS survey \citep{lum13} an effort 
has been made to identify massive YSOs and ultra-compact \ion{H}{ii} 
regions by means of multi-wavelength ancillary data, both photometric and spectroscopic 
\citep[e.g.][and references therein]{urq09}. Similarly, for Hi-GAL \citet{ces15} studied, 
in the longitude range $10^{\circ} < \ell < 65^{\circ}$, the counterparts of CORNISH 
\citep{hoa12,pur13} sources, treated as bona-fide young \ion{H}{ii} regions, and distributed 
across the range $2~L_{\odot}/M_\odot < L_{\mathrm{bol}}/M < 270~L_{\odot}/M_\odot$. From the 
data of \citet{ces15} we estimate the peak of this distribution to lie around $(L/M)_\mathrm{P}\equiv
\cesalim~L_{\odot}/M_\odot$. At values larger than this peak, in principle, even more 
evolved sources are expected, so the fact that the distribution decreases beyond this peak is 
mostly due to completeness: indeed the SED filtering we apply, mostly based on availability 
of detections at 250 and 350~$\umu$m, causes the removal of a large number of evolved 
sources from our physical catalogue. 


The locus corresponding to $(L/M)_\mathrm{P}$ is reported also in Figure~\ref{lvsm}, right panel, 
as a light blue dashed line: if we compare it with a similar diagram in \citet{mol08}, we notice 
that their ``IR sources'', namely the SEDs fitted with an embedded ZAMS envelope (many 
of which compatible with the presence of an ultra-compact \ion{H}{ii} region), lie in 
the region of the diagram above this threshold.

Since our catalogue covers an area of the sky larger than that observed by the CORNISH survey 
to date, it would be useful to develop a \textit{Herschel}-based method - verified by means of 
independent tracers - able to identify possible \ion{H}{ii}-region candidates in other parts
of the Galactic plane, such as the the fourth quadrant. For this reason the sources with 
$L_{\mathrm{bol}}/M \geq (L/M)_\mathrm{P}$ will henceforth be treated as 
``\ion{H}{ii}-region candidates''.

Additional information can be obtained from the study of the source luminosity. For example, 
the $L_{\mathrm{smm}}/L_{\mathrm{bol}}$ ratio is a distance-independent evolutionary indicator 
frequently discussed in the study of the low-mass star formation. \citet{and00} characterised
Class~0 objects \citep{and93} through the ratio of their sub-millimeter luminosity, 
calculated for $\lambda \geq 350 \umu$m, and their bolometric luminosity, establishing 
a minimum threshold of 0.005 for identifying an object of this class. Subsequently, 
\citet{mau11} refined this value to 0.01, which, if one prefers to use the inverse of 
this ratio \citep[as, e.g.,][and in this paper]{beu10}, translates into the condition 
$L_{\mathrm{bol}}/L_{\mathrm{smm}}<100$. Obviously, the Class~0/I/II/III is meaningful 
only for single low-mass protostars, while in our study we address sources which do not 
correspond to single YSOs and, furthermore, in a relevant fraction of cases these might 
host massive star formation. Therefore in our case the $L_{\mathrm{bol}}/L_{\mathrm{smm}}$ 
ratio cannot be used to identify Class~0 sources \citep[see also][]{fal13}, rather we can 
only safely state that sources whose SEDs have a significant contribution of emission at 
sub-millimetric wavelengths are at an early stage of star formation. 

\begin{figure*}
\centering
\includegraphics[width=16.0cm]{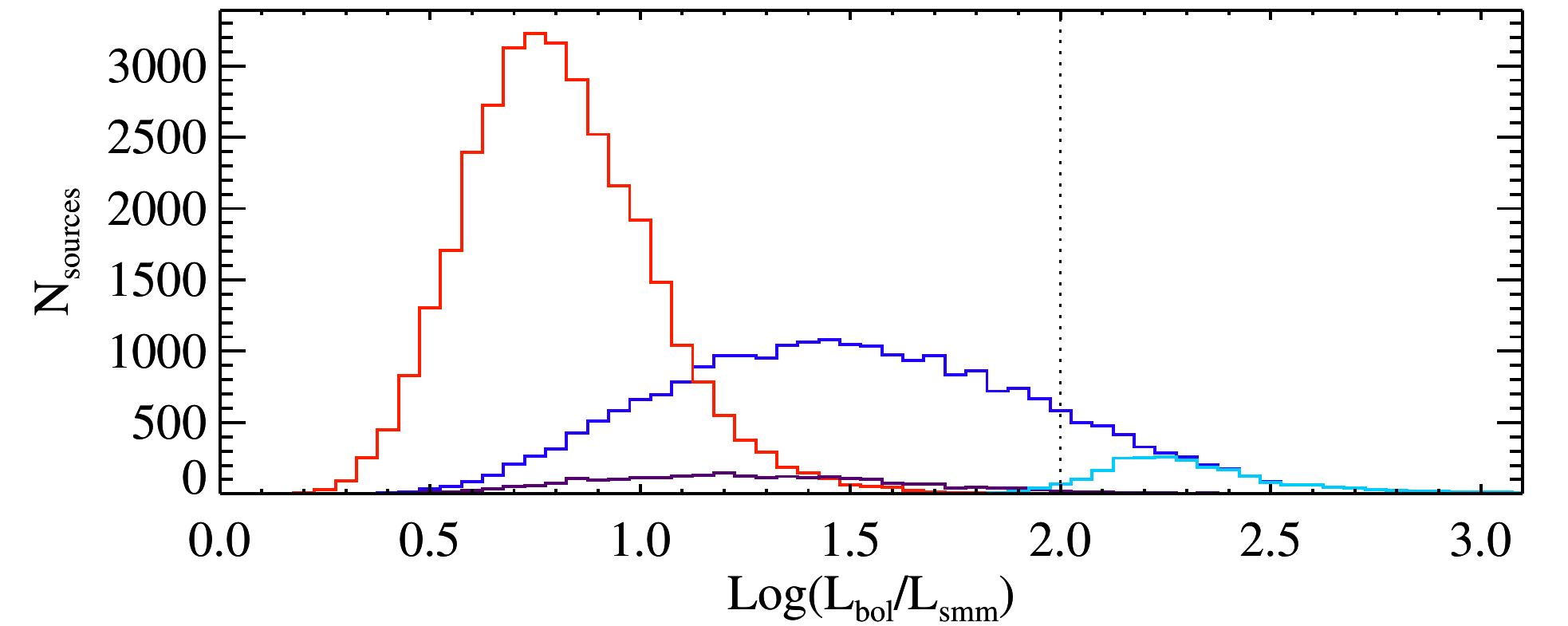}
\caption{Distributions of $L_{\mathrm{bol}}/L_{\mathrm{smm}}$ ratio for science analysis in this 
paper: pre- and proto-stellar ones are represented with a red and blue histogram, while the 
MIR-dark and \ion{H}{ii}-region candidate sub-samples of the proto-stellar ones are
represented with a dark purple and a light blue histogram, respectively. The dotted vertical 
line represents the threshold for identifying Class~0 YSO in the low-mass regime of star 
formation \citep{mau11}.}
\label{lsubmm}
\end{figure*}

In Figure~\ref{lsubmm} the distributions of $L_{\mathrm{bol}}/L_{\mathrm{smm}}$ for 
pre- and proto-stellar sources
are shown, again exhibiting a different behaviour of the two: the former (having about \percsubmme\%
of sources below the critical value of 100) peaks at a value smaller than the latter, which, in turn, 
has only \percsubmmo\% below the threshold. Since many of the sources of our catalogue belong to
well-known star forming regions where the presence of sources more evolved than Class~0 has been
assessed, these results clearly indicate that an evolutionary classification based only 
on \textit{Herschel} photometry (even extended to shorter wavelengths down to $\sim20~\umu$m),
generally leads to a biased classification of most sources as Class~0 (or, more precisely, as 
their high-mass equivalents), as already pointed out by \citet{gia12} and \citet{eli13}.
Looking in more detail at the two sub-classes of proto-stellar sources introduced above,
the MIR-dark sources how a spread similar to that encountered in Figure~\ref{lmratio}, while the 
\ion{H}{ii}-region candidates constitute the right tail of the proto-stellar distribution,
with most bins located above the threshold value of 100. 

\begin{figure}
\centering
\includegraphics[width=9.0cm]{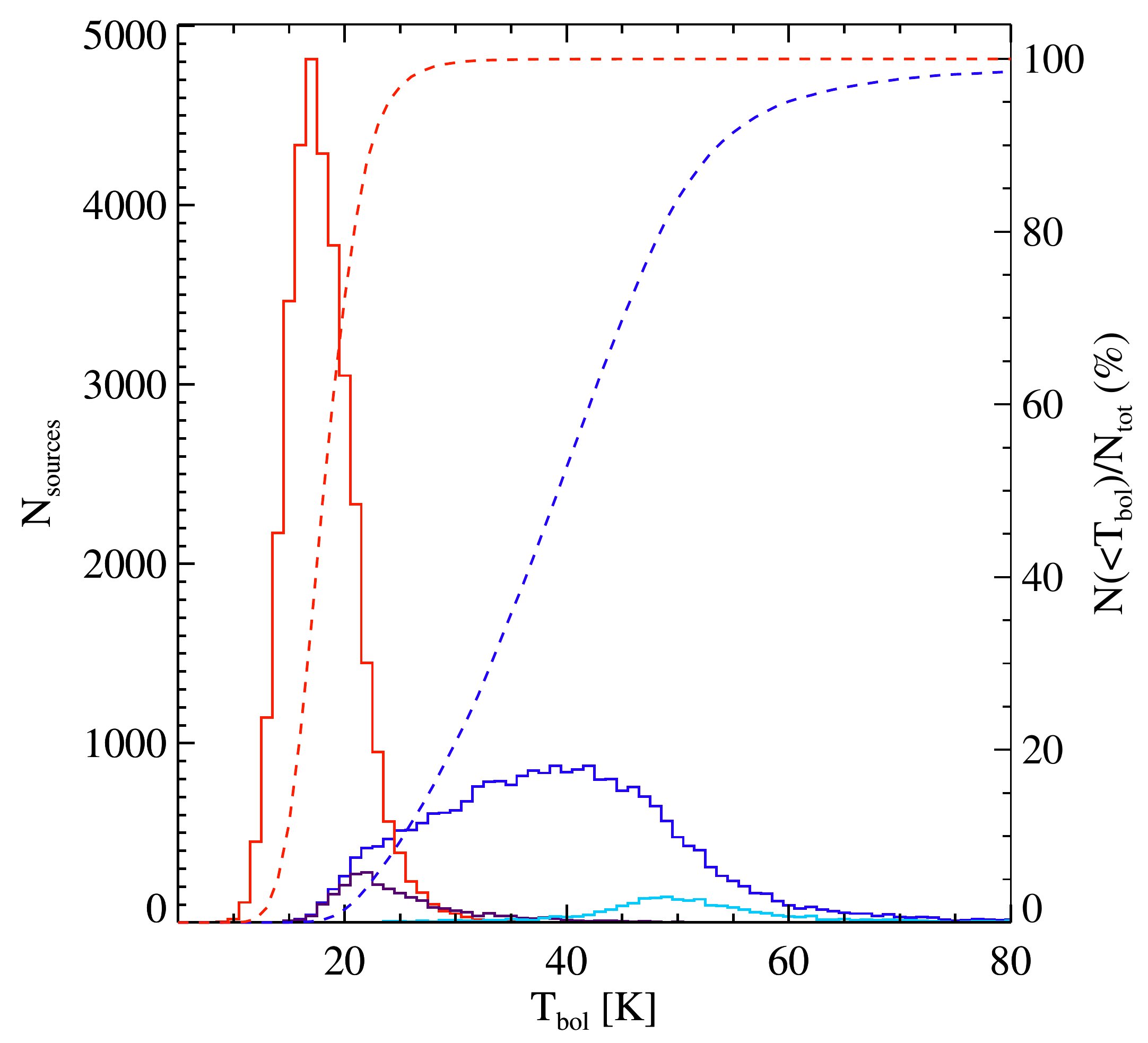}
\caption{The same as in Figure~\ref{hist_temp}, but for the bolometric temperature.}
\label{tbol}
\end{figure}

An even higher degree of segregation between pre- and proto-stellar sources can be found 
in the bolometric temperature distribution shown in Figure~\ref{tbol}, being defined by 
\citet{mye93} as:
\begin{equation}\label{tboleq}
T_{\mathrm{bol}}=1.25 \times 10^{-11}~\mathrm{K}\times \frac{\int_0^\infty \nu~F_{\nu}~d\nu}{\int_0^\infty~F_{\nu}~d\nu}\;.
\end{equation}
This diagnostic has been used recently by \citet{str15} by using Hi-GAL data, but building SEDs 
with a wider range of photometric data points, going from \textit{Spitzer}-IRAC bands,
which can lead to a shift of bolometric temperatures towards values as $T_{\mathrm{bol}} \sim 
100-1000$~K, much larger than those found here ($T_{\mathrm{bol}} \lesssim 100$~K). In fact, 
similarly to the $L_{\mathrm{smm}}/L_{\mathrm{bol}}$ ratio, 
bolometric temperatures cannot be used to infer the Class~0/I/II/III source classification 
\citep{lad84,lad87,and93}, according to which almost all sources of our catalogue would fall in 
the Class~0 range \citep[$T_{\mathrm{bol}} < 70$~K,][]{che95}. At the same time, by revealing 
a net separation between pre- and proto-stellar clumps and a pronounced spread inside the 
distribution of the latter, bolometric temperatures can still be used to highlight the variety 
in evolutionary stage of the Hi-GAL sources. 

Looking at the histogram of the MIR-dark proto-stellar sources in Figure~\ref{tbol}, it 
is evident that these sources represent the low-$T_{\mathrm{bol}}$ tail of the overall 
proto-stellar distribution. This is expected, given the $T_{\mathrm{bol}}$ definition 
(see Equation~\ref{tboleq}), since small MIR fluxes correspond to low $T_{\mathrm{bol}}$ 
values. On the contrary, the \ion{H}{ii}-region candidates are located towards
the highest probed bolometric temperatures, although they do not constitute the totality
of proto-stellar sources at these temperatures. This suggests that bolometric temperature 
and $L_{\mathrm{bol}}/M$ are not perfectly coupled, as it will be discussed in 
Section~\ref{evolbol}.

In general, typical bolometric temperatures are found to range from $\sim10$~K (pre-stellar sources)
to $\sim80$~K, a range incompatible with average values for the clump populations of \citet{mul02}, 
78~K, and \citet{ma13}, 113~K. A direct comparison with these two works, however, is 
not appropriate due to the different spectral ranges covered, with the MIR range dominated, in those 
cases, by IRAS fluxes \citep[see Appendix~\ref{champepos} for a more detailed comparison between the 
clump properties derived by][and those in the present work]{ma13}.

\section{An evolutionary scenario}\label{evol_par}
In this section we synthesize the physical quantities described above with the aim to formulate an 
evolutionary classification scheme for proto-stellar sources.

\subsection{Herschel colours}\label{hcols}

\begin{figure}
\centering
\includegraphics[width=9.0cm]{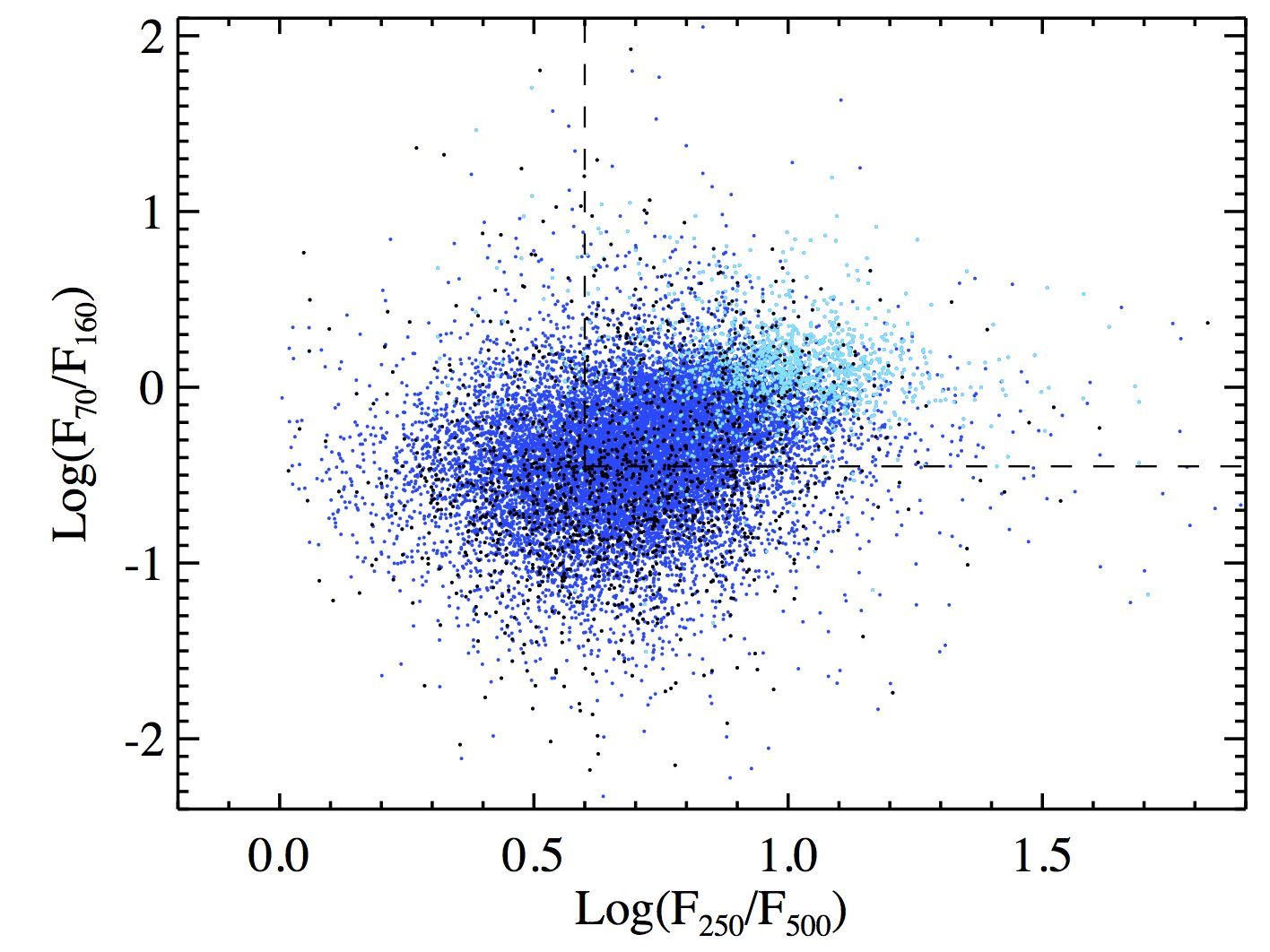}
\caption{Plot of the $F_{70}/F_{160}$ vs $F_{250}/F_{500}$ colours for the sub-sample of 
proto-stellar sources (blue dots) provided with fluxes at these four bands. The positions 
of sources with no detection in the MIR are overplotted as black dots, while 
sources identified as \ion{H}{ii} regions (see the text), are overplotted as light
blue dots. The dashed lines delimit the area identified by \citet{pal12} to contain 
\ion{H}{ii} regions.}
\label{paladinicol500}
\end{figure}

Examples of \textit{Herschel}-based colour-colour diagrams are found in literature as tools 
to perform source evolutionary classification. While here we can not apply the prescriptions
such as those given in \citet{ali10} or \citet{spe13}, intended for classifying single low-mass 
YSOs in nearby star forming regions, it is however interesting to consider and extend the analysis 
of \citet{pal12} of a sample of 16~\ion{H}{ii} regions observed in Hi-GAL around $\ell=30^{\circ}$. 
In Figure~\ref{paladinicol500}, following these authors, we build the $F_{70}/F_{160}$ vs 
$F_{250}/F_{500}$ diagram for the proto-stellar sources of our sample with fluxes at these 
four wavelengths. \citet{pal12} identify a region at the top-right of the diagram, in which 
\textit{Herschel} fluxes of analysed \ion{H}{ii} regions are found to lie (despite some contamination 
by non \ion{H}{ii} regions). We find that \npalad~of the considered sources (i.e. \percpalad\% of 
the total) are located in this area of the diagram. We highlight with a different colour the positions of sources fulfilling the condition 
$L_{\mathrm{bol}}/M \geq (L/M)_\mathrm{P}$, that we assume to be \ion{H}{ii} region candidates.
We find that the great majority of such sources (\ncesapalad~out of \ncesaDDD, i.e.~the 
\ncesapaladperc\%) lie well inside the region defined by \citet{pal12}.

It is interesting to explore also the behaviour of the sub-sample of proto-stellar MIR-dark 
sources, which are expected to correspond to an earlier evolutionary phase. These appear not to 
be confined to the \ion{H}{ii}-region locus and are generally more spread out. This indicates 
that this choice of \textit{Herschel} colours does not appear to be correlated with the presence 
of one or more MSX/WISE/\textit{Spitzer} counterparts.

\begin{figure}
\centering
\includegraphics[width=9.0cm]{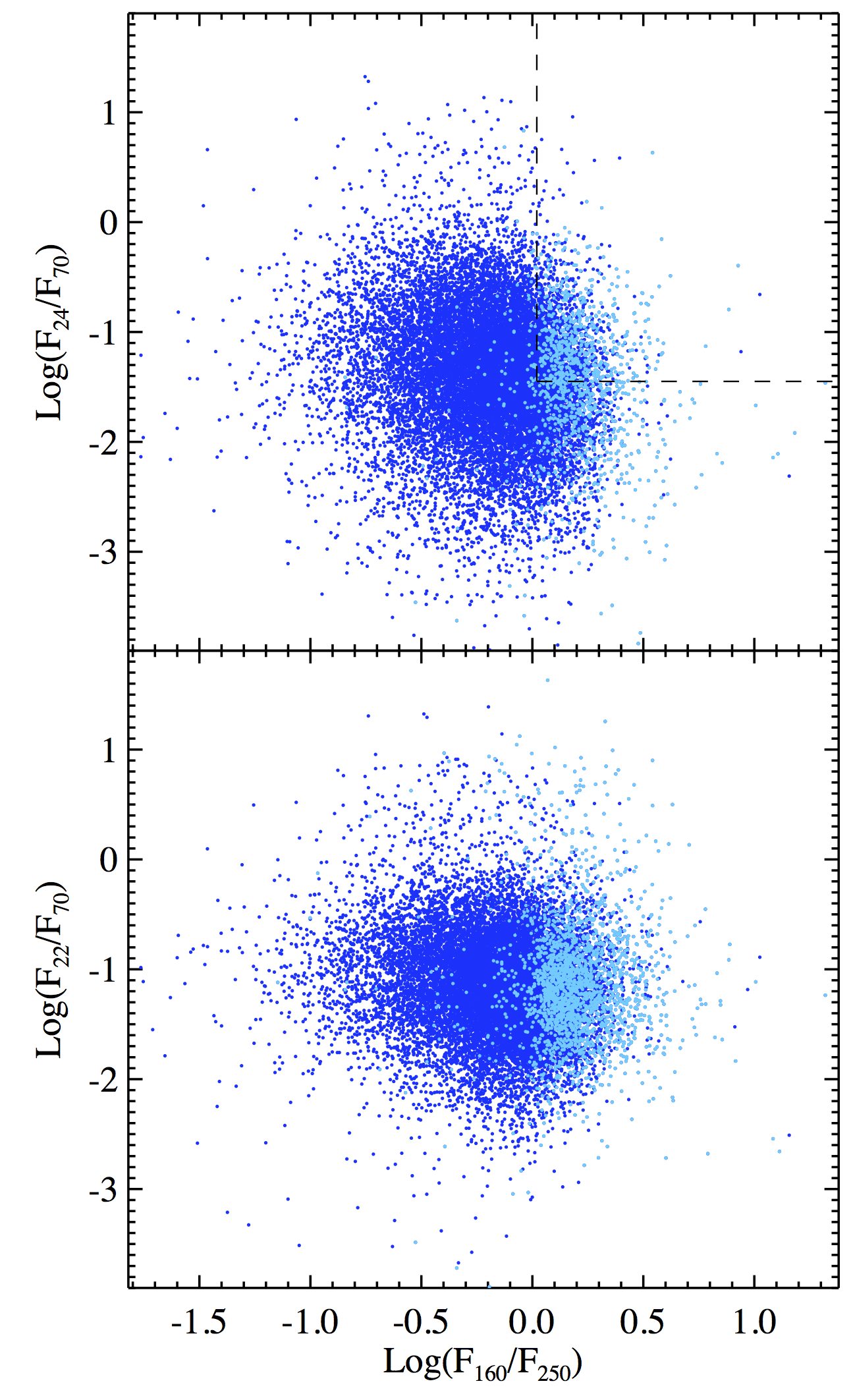}
\caption{The same as Figure~\ref{paladinicol500}, but for the colours
$F_{24}/F_{70}$ (\textit{top}) and $F_{22}/F_{70}$ (\textit{bottom})
vs $F_{70}/F_{160}$. Each of the two plots is based on the
sub-sample of proto-stellar sources provided with the four 
fluxes requested to build it. In the top panel, the dashed 
lines delimit the area occupied by \ion{H}{ii} regions in the
analogous colour-colour diagram of \citet{pal12}.}
\label{paladinicol24}
\end{figure}

Another diagnostic used by \citet{pal12} is $F_{24}/F_{70}$ vs $F_{160}/F_{250}$ 
diagram, thus involving the \textit{Spitzer}-MIPS flux at $24~\umu$m, which is shown 
in Figure~\ref{paladinicol24}, top panel. As sources with $F_{24}\gtrsim 2$~Jy 
tend to be saturated in MIPS \citep{car08}, we also construct a $F_{22}/F_{70}$ 
vs $F_{160}/F_{250}$ diagram using WISE $22~\umu$m data (Figure~\ref{paladinicol24}, 
bottom panel). 
In both panels, data appear highly scattered, though the sub-sample of 
sources identified as \ion{H}{ii}-region candidates still occupy a smaller
area of this space, where Log$(F_{160}/F_{250})>0$, similarly to \citet{pal12}. 
The large scatter in the $y$ direction, however, demonstrates that a MIR counterpart 
does not provide leverage to separate out early-stage sources from a more evolved 
population in these diagrams.

\subsection{Bolometric quantities}\label{evolbol}
Here we explore the relations between the parameters and attempt to use these to 
better define the evolutionary sequence of proto-stellar sources.


\begin{figure*}
\centering
\includegraphics[width=18.0cm]{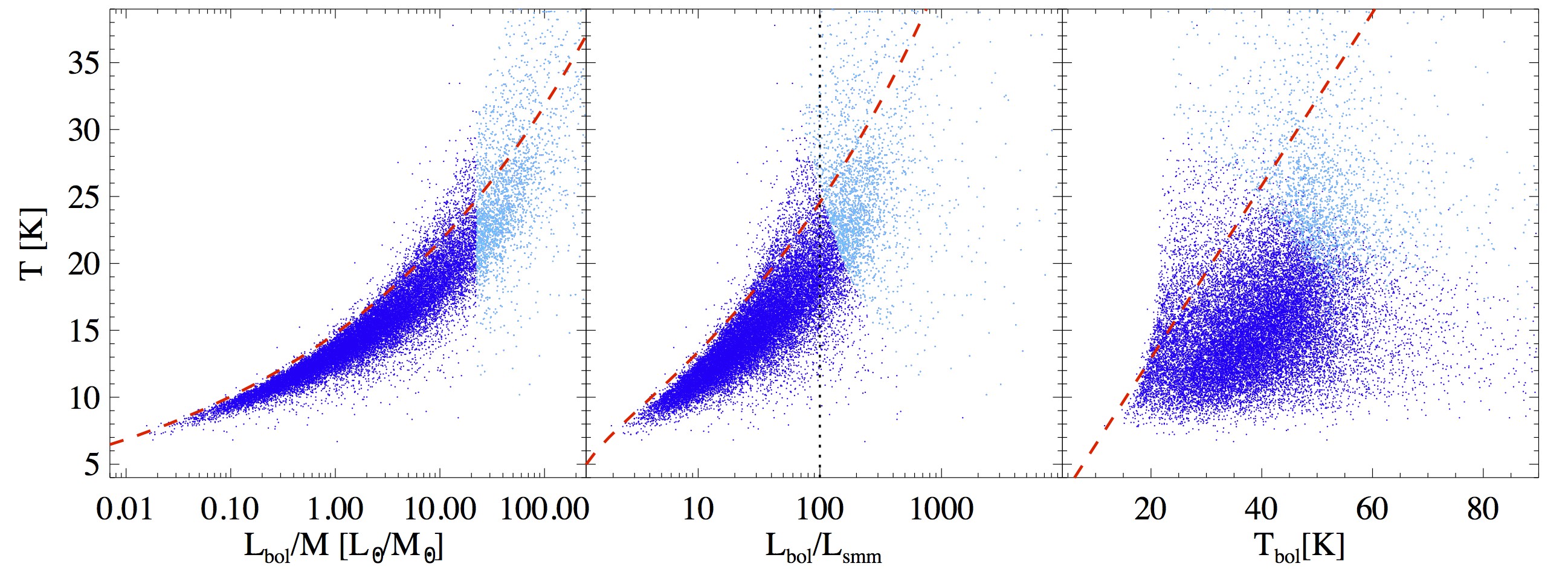}
\caption{Plot of temperature $T$ of proto-stellar sources (blue dots) vs the 
$L_{\mathrm{bol}}/M$ ratio (\textit{left}), the $L_{\mathrm{smm}}/L_{\mathrm{bol}}$
ratio (\textit{centre}), and the bolometric temperature $T_{\mathrm{bol}}$ (\textit{right}), 
respectively. The sub-sample of sources compatible with a \ion{H}{ii} region is plotted 
in cyan. Red dashed lines represent, in each panel, the expected behaviour of an optically 
thin grey body (Equation~\ref{gbthin}) with $\beta=2$, as derived by \citet{eli16}. In 
the centre panel, the vertical dotted line is the same as in Figure~\ref{lsubmm}.}
\label{evol_temp}
\end{figure*}

In Section~\ref{dtemp} a mild segregation between pre- and proto-stellar clump
grey-body temperatures was found. In Figure~\ref{evol_temp}, we directly
compare the dust temperature with evolutionary indicators such as $L_{\mathrm{bol}}/M$,
$L_{\mathrm{bol}}/L_{\mathrm{smm}}$, and $T_{\mathrm{bol}}$, introduced in
Section~\ref{lum_par}. As expected, all three quantities increase as a 
function of $T$. Also, at a given temperature most of proto-stellar clumps 
show an excess of these quantities with respect to the theoretical behaviour of 
a grey body with $\beta=2$. 

\begin{figure}
\centering
\includegraphics[width=8cm]{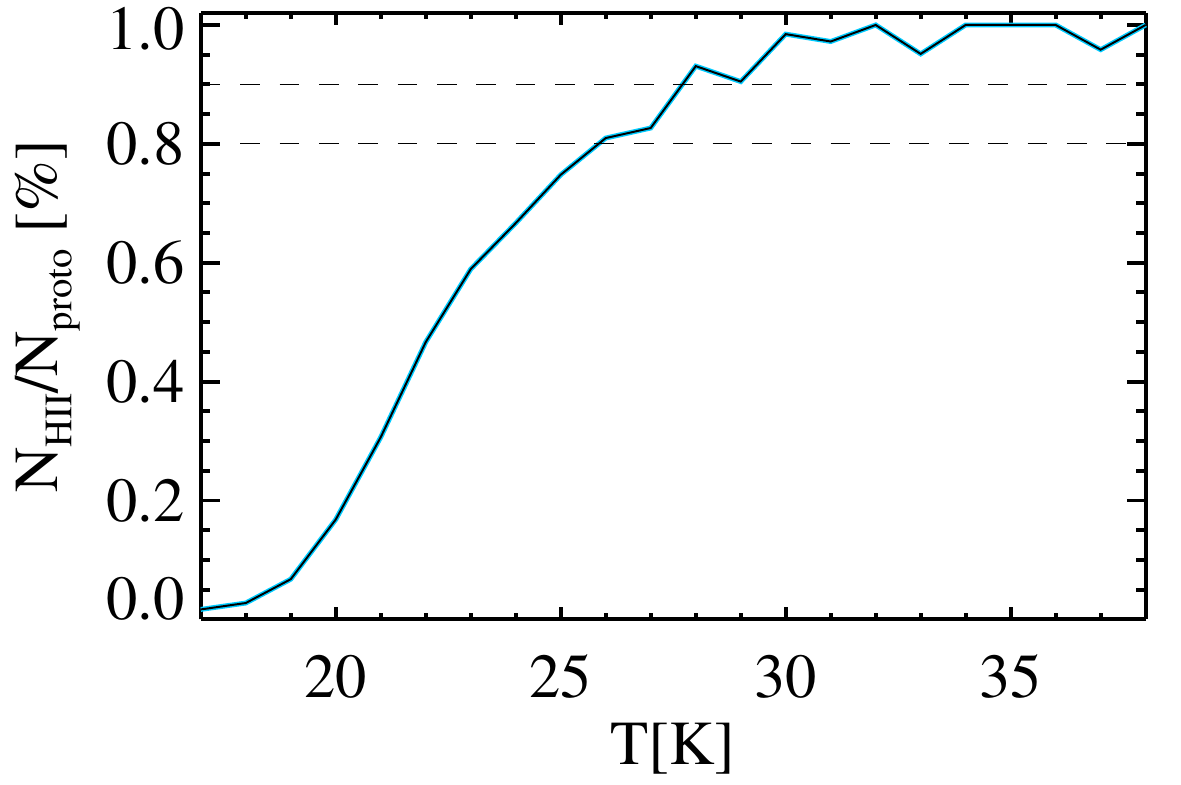}
\caption{Fraction of proto-stellar sources classified as \ion{H}{ii}-region 
candidates ($L_{\mathrm{bol}}/M \geq
(L/M)_\mathrm{P}$) over the total number of the proto-stellar ones vs temperature.
The two horizontal dashed lines indicate the 80\% and 90\% levels.}
\label{tempfraction}
\end{figure}

Two general considerations emerge from the
three panels of this figure: since the average temperature estimated for
proto-stellar sources is dominated by the cold envelope modelled as a grey body, it
subtends a certain degree of degeneracy of evolutionary stages, which can be resolved 
thanks to indicators (reported on the $x$-axes of the three panels), which are derived 
by including emission at wavelengths shorter than $70~\umu$m. On the other hand, at the 
behaviour of some of the more evolved sources in our sample, such as those compatible with 
a \ion{H}{ii} region based on their $L_{\mathrm{bol}}/M$ (see Section~\ref{lum_par}), 
their temperature is definitely high. In Figure~\ref{tempfraction}, the fraction of
proto-stellar sources classified as \ion{H}{ii}-region candidates out of the total 
number of proto-stellar sources per 
bins of temperature is shown: at $T \gtrsim \tempfraceighty$~K, 80\% of the proto-stellar sources
have a $L_{\mathrm{bol}}/M$ greater than the threshold required for classifying it as a
\ion{H}{ii}-region candidate, while 90\% level is achieved at $T \gtrsim \tempfracninety$~K. 
This is in good agreement with the estimate of $T=25$~K by \citet{hof00} for clumps 
hosting ultra-compact \ion{H}{ii} regions.

\begin{figure}
\centering
\includegraphics[width=8cm]{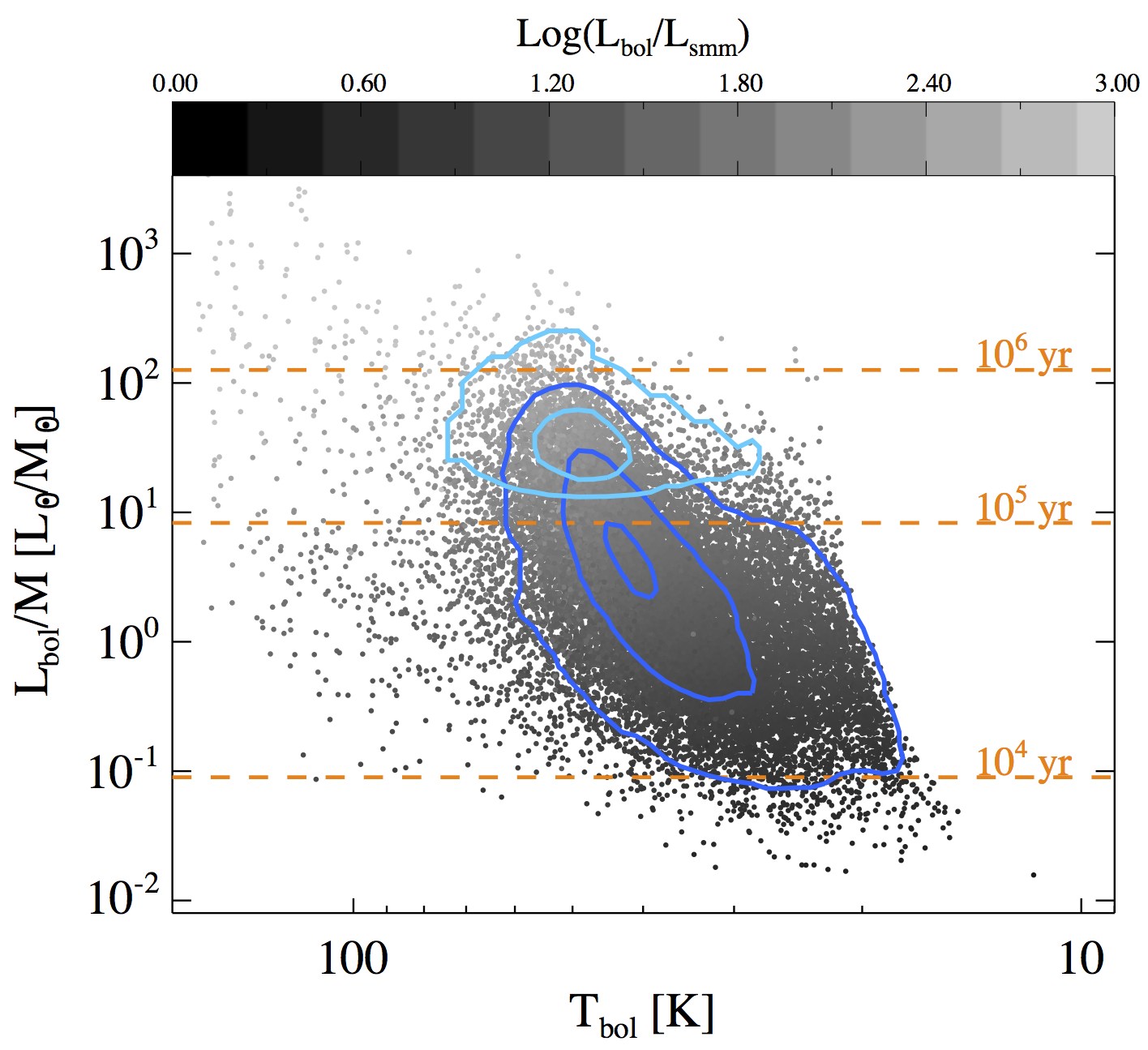}
\caption{Plot of $L_{\mathrm{bol}}/M$ vs $T_{\mathrm{bol}}$ for proto-stellar sources. The 
dot colour depends on the $L_{\mathrm{smm}}/L_{\mathrm{bol}}$ ratio of the source, ranging 
logarithmically from 1 to 1000 (from the darkest to the lightest level of grey present in 
the plot, respectively; for values outside this range the colour scale saturates). Blue 
and light blue contours represent the areas enclosing, from the innermost to the outermost
contour, the 10\%, 50\%, and 90\% of all the proto-stellar sources, and of the 
\ion{H}{ii}-region candidates, respectively. Dashed orange lines correspond to 
$L_{\mathrm{bol}}/M$ achieved after $10^4$, $10^5$, and $10^6$~yr along the rightmost 
evolutionary track of Figure~\ref{lvsm}.}
\label{tbollmr}
\end{figure}

To show more directly the mutual connections among $L_{\mathrm{bol}}/M$,
$L_{\mathrm{bol}}/L_{\mathrm{smm}}$, and $T_{\mathrm{bol}}$, in Figure~\ref{tbollmr} we
plot $L_{\mathrm{bol}}/M$ vs $T_{\mathrm{bol}}$ for proto-stellar sources, 
while also showing $L_{\mathrm{bol}}/L_{\mathrm{smm}}$ by 
means of a colour scale. The source density contours, both for the entire sample of
proto-stellar sources and for the \ion{H}{ii}-region candidates, guide the reader
through the most crowded areas of the plot.

The evolutionary tracks of \citet{mol08}, reported in Figure~\ref{lvsm}, can be used for
providing a rough estimate of the time elapsed since the beginning of the collapse 
given a value of the $L_{\mathrm{bol}}/M$ ratio\footnote{Of course, such correspondence 
depends on the considered track, in turn characterised by the clump initial mass $M_0$ 
at the time $t=0$. Furthermore, tracks are obtained for a clump forming a single star, 
a quite unrealistic case, especially at high clump masses.}. In Figure~\ref{tbollmr}
$L_{\mathrm{bol}}/M$ values of $10^4$, $10^5$, and $10^6$~yr corresponding 
to the highest mass track, i.e. the one starting at an initial mass $M_0=1500~M_\odot$, 
are displayed as horizontal lines. According to this rough timescale, the figure suggests 
that a large fraction of proto-stellar sources ($\lmrtentofive$ objects) would correspond
to an age of $t<10^5$~yr, while \ion{H}{ii}-region candidates are found to be
well above that value. Comparing with proto-stellar (Class~0~+~I) phase 
lifetimes of a few $10^5$~yr quoted by \citet{dun14} for cores in the Gould Belt,
we find that the majority of proto-stellar sources present in Figure~\ref{tbollmr}
correspond to a shorter evolutionary time: this indicates a faster evolution for 
clumps hosting formation of more massive stars, as a significant part of Hi-GAL 
clumps are generally supposed to be (cf. Section~\ref{mass_par}).

Recently \citep{mol16b} calibrated the $L_{\mathrm{bol}}/M$ through 
CH$_3$C$_2$H(12-11) line observations, finding that $i$) this line is detected 
at $L_{\mathrm{bol}}/M > 1 L_{\odot}/M_\odot$ (which should correspond to a temperature
$T<30$~K in the inner part of the clump), and $ii$) the temperature indicated by
this tracer starts to increase at $L_{\mathrm{bol}}/M > 10 L_{\odot}/M_\odot $, where 
this threshold is interpreted as the first appearance of ZAMS star(s) in the clump.
Comparing this scenario with our statistics, we find that \molinI\% of our proto-stellar
clumps have $L_{\mathrm{bol}}/M$ between 1 and 10 $L_{\odot}/M_\odot$, with an average 
bolometric temperature of \tbmolinI~K. Only \molinII\% of proto-stellar clumps, 
however, have $L_{\mathrm{bol}}/M > 10 L_{\odot}/M_\odot$, while - as we expect 
(since $(L/M)_\mathrm{P} > 10L_{\odot}/M_\odot$) - all of the \ion{H}{ii}-region candidates,
are above this threshold, having an average bolometric temperature of \tbwmolin~K.

The $L_{\mathrm{bol}}$ vs $T_{\mathrm{bol}}$ is a diagnostic used to characterise the YSOs 
from the evolutionary point of view \citep{mye98}. Recently, \citet{str15} applied it to a 
population of Hi-GAL sources all belonging to the same region (Vela-D), assumed to 
be located at the same heliocentric distance. 
On the one hand, it is possible to remove the dependence on distance
by taking into account $L_{\mathrm{bol}}/M$ instead of $L_{\mathrm{bol}}$. The 
$L_{\mathrm{bol}}/M$ vs $T_{\mathrm{bol}}$ source behaviour in Figure~\ref{tbollmr} is 
qualitatively similar to Figure~9 of \citet{str15}, given the strongly different amount of 
sources and the different quantity reported on the $y$-axis, as explained above. A similar 
range of $T_{\mathrm{bol}}$ is found, as well as some spread in $L_{\mathrm{bol}}/M$. This 
spread becomes remarkably large at high $T_{\mathrm{bol}}$, highlighting that these two 
indicators can, in a few cases mostly concentrated at high bolometric temperatures, 
give conflicting information about the evolutionary stage of a source. 
\begin{figure}
\centering
\includegraphics[width=8cm]{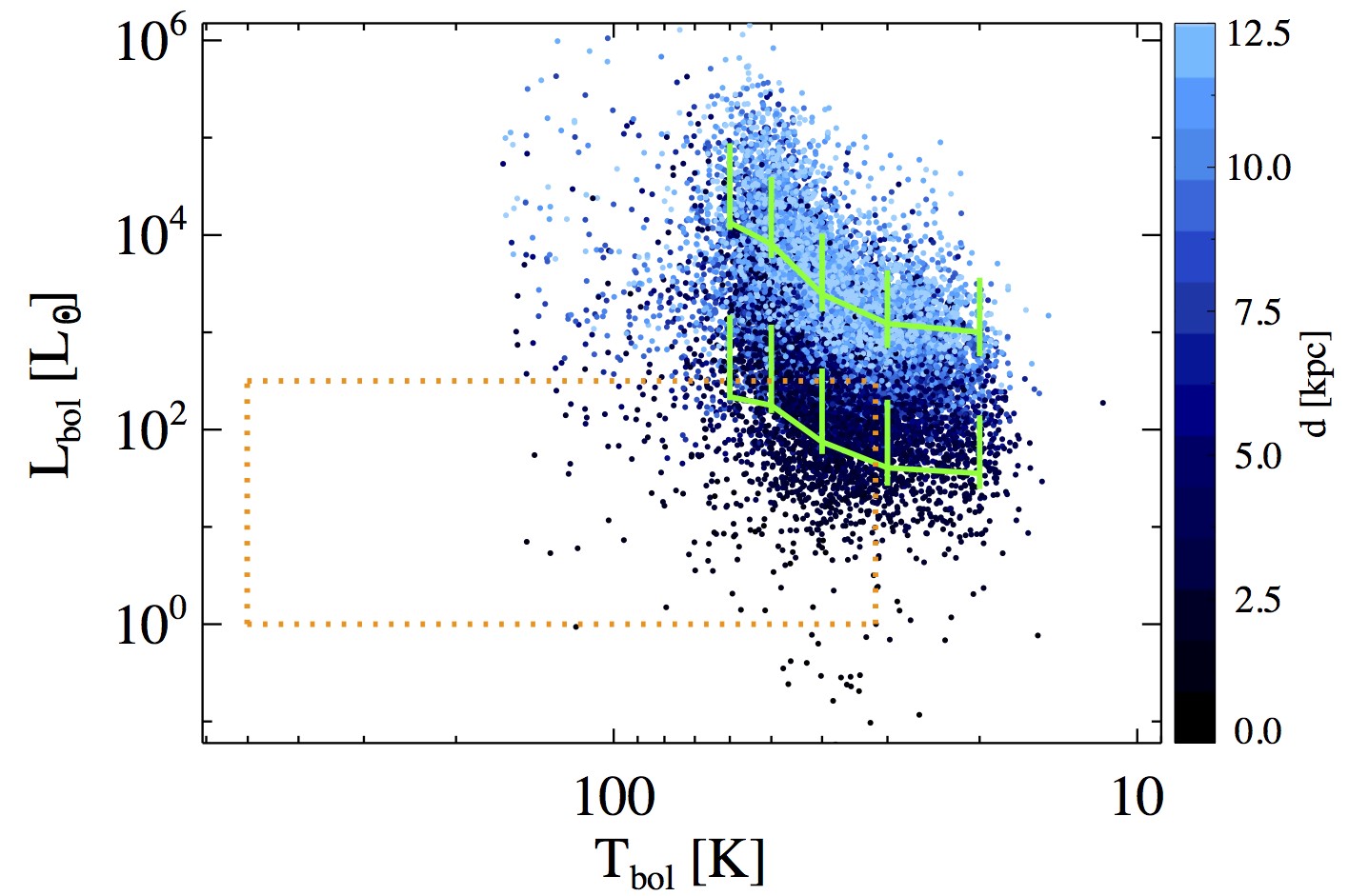}
\caption{Plot of $L_{\mathrm{bol}}$ vs $T_{\mathrm{bol}}$ for proto-stellar sources. 
The dot colour, described by the colour bar on the right, scales with the source heliocentric
distance (the scale saturates at 12.5~kpc). The orange dashed box corresponds to the region
occupied in this diagram by the \textit{Herschel}+\textit{Spitzer} proto-stellar sources
studied in Vela-C by \citet{str15}. As an example, median values of the luminosity
calculated over six bins of $T_\mathrm{bol}$ (starting from 20~K and in steps of 10~K) 
for sources at $d>10$~kpc (upper green broken line) and at $d<3$~kpc (lower green broken 
line), respectively, are shown. The typical distribution of luminosities being strongly
skewed, with a longer tail at high luminosities, it does not make sense to show, as error 
bars, the values of the standard deviation; therefore here we use bars defined by the 15th (bottom
of the bar) and the 85th percentile (top), enclosing 70\% of luminosity values
present in the considered bin of $T_\mathrm{bol}$.}
\label{lumvstbol}
\end{figure}
On the other hand, a direct comparison with the plot of \citet{str15} can be 
performed binning sources with respect to distance, which is rendered in Figure~\ref{lumvstbol} 
through a colour scale. The box enclosing the positions of the proto-stellar cores in the analogous 
plot reported in that paper ($\strafii$~K~$\lesssim T_\mathrm{bol}\lesssim \strafi$~K, 
$\strafiii~L_\odot~\lesssim L_\mathrm{bol}\lesssim \strafiv~L_\odot$) still contains a relevant number 
of sources (\nstraf) from our sample. Nevertheless, a significant fraction of the sources of \citet{str15}
are found to have $T_\mathrm{bol} > 100$~K, while most of our sources lie below that value.
This is essentially due to the wider wavelength range considered by these authors to compute 
$T_\mathrm{bol}$, starting from $3.6~\umu$m, which inevitably leads $T_\mathrm{bol}$ to 
increase according to Equation~\ref{tboleq}.

Grouping sources with comparable distances, a general trend of the luminosity 
to strongly increase at increasing bolometric temperature is seen in the range 
$15~$K~$\lesssim T_{\mathrm{bol}}\lesssim 65$~K in which sample is large enough to reconstruct
such a behaviour. This can be seen through two examples we give in Figure~\ref{lumvstbol}, 
considering sources in the two very different distance ranges, namely $d<3$~kpc and $d<10$~kpc. 
However, in such examples, 
the log-log slope of $L_\mathrm{bol}$ vs $T_\mathrm{bol}$ is everywhere shallower than 6, which is 
the value expected for a perfect grey body with $\beta=2$ \citep[cf. Equations~35 and 44 of][]{eli16}.
This means that, although the departure from a cold grey body behaviour observed in proto-stellar envelopes
at $\lambda < 160~\umu$m introduces an excess both in $L_\mathrm{bol}$ and $T_\mathrm{bol}$,
the latter is more sensitive than the former to the extension of the SED towards shorter
wavelengths. Notice that this qualitative consideration is given neglecting the further 
effect of the spread in distance underlying the two samples used to draw the curves 
in Figure~\ref{lumvstbol}.

\subsection{Surface density}\label{surfdens_par}

Here we search for possible correlations between the main source evolutionary indicators and 
the source surface density introduced in Section~\ref{mass_par}. Notice that this parameter
is not derived from only photometric measurements, but also from the source 
linear size.

\begin{figure*}
\centering
\includegraphics[width=16cm]{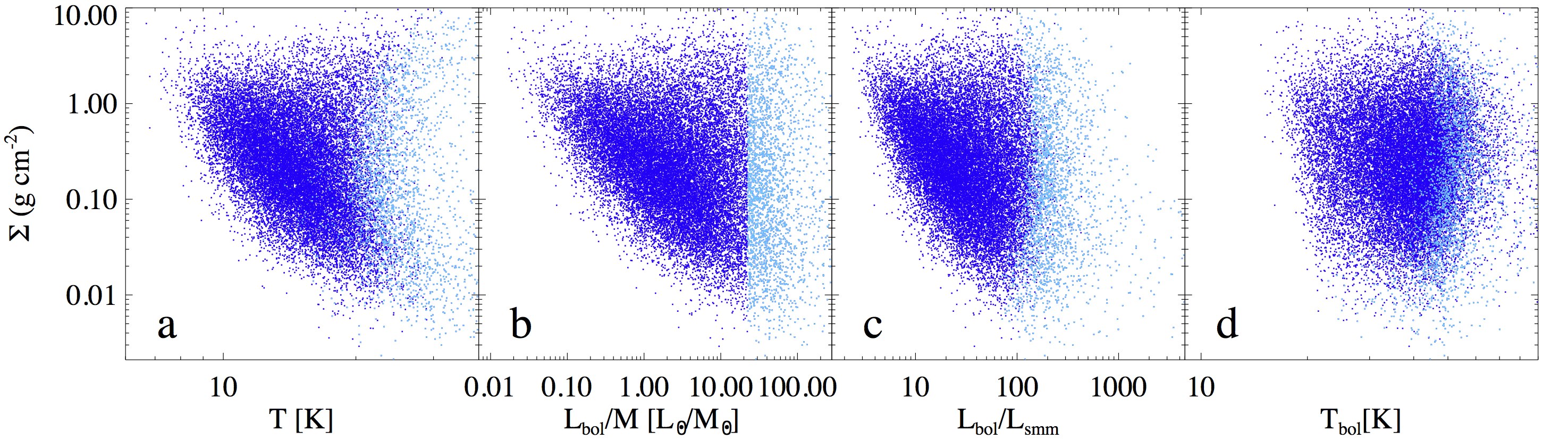}
\includegraphics[width=16cm]{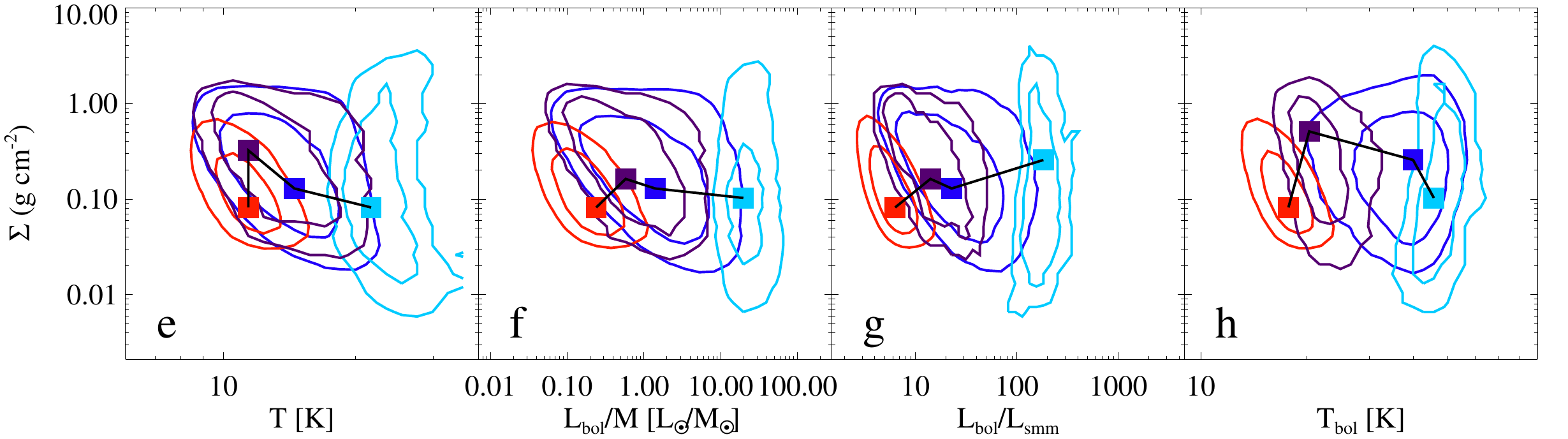}
\caption{Upper panels: relation between clump surface density and, 
from panel \textit{a} to \textit{d}, dust temperature, $L_{\mathrm{bol}}/M$,
$L_{\mathrm{bol}}/L_{\mathrm{smm}}$, and bolometric temperature, respectively. Blue dots 
represent proto-stellar clumps, except for the sub-sample of sources compatible with a 
\ion{H}{ii} region, plotted in cyan. Lower panels (\textit{e}-\textit{h}): the same 
diagrams as in the upper panels, but with source density contours enclosing the 90\% and 
the 20\% of sources, respectively. Contours describing the sample of pre-stellar 
sources and the sub-sample of MIR-dark proto-stellar sources are also plotted in red 
and dark purple, respectively. In absolute terms, hence, the contours
drawn for different populations correspond to different values: for example, 
\numpredens~being the total number of pre-stellar sources considered for building this plot,
the external contour is drawn in order to contain \numpredensninety~sources, and
the internal one to contain \numpredenstwenty~sources; the total numbers of proto-stellar,
candidate \ion{H}{ii} regions, and MIR-dark sources are \numprotodens, \numhiidens, and 
\nummirdarkdens, respectively (notice that the second and third population are
sub-samples of the first one),
from which the values corresponding to the source density contours can be easily derived
as in the example above. Finally, in each panel the peaks of the source 
distributions of the four aforementioned evolutionary classes are reported with filled
circles of same colours of the contours, and connected by a black line.}
\label{sigmevol}
\end{figure*}

In Section~\ref{mass_par} we have shown that proto-stellar sources are found
to be, on average, slightly smaller and denser than starless ones (Figures~\ref{sizefig} and 
\ref{masssize}). A the same time, clump column density is observed to decrease 
with increasing temperature \citep[e.g. IRDCs,][]{per10b}, in such a way that the 
combination of these two opposing effects contribute to the overlap between the surface 
density distributions of the starless and proto-stellar populations seen in Figure~\ref{masssize}. 
In Figure~\ref{sigmevol}, panel $a$, surface density vs dust temperature relation is plotted for 
proto-stellar sources. The bulk of sources show the aforementioned trend, but with increasing 
temperatures the spread in surface density increases, so that a remarkable fraction of 
sources which appear to be, at the same time, dense and warm is found: for instance, considering
proto-stellar sources with $T>25$~K, \ctempsigmlow~have $\Sigma \leq 0.1$~g~cm$^2$ and
\ctempsigmupp~have $\Sigma > 0.1$~g~cm$^2$ (\ctempsigmlowperc\% and \ctempsigmuppperc\% 
of the total, respectively). In particular, candidate \ion{H}{ii} regions are found at 
the highest temperatures, but the position in the plot is not indicative of 
any particular trend, spanning the entire range of surface densities, from~0.01 
to~10~g~cm$^{-2}$.

Not surprisingly, a similar behaviour is found for $\Sigma$ vs the $L_{\mathrm{bol}}/M$ 
ratio (Figure~\ref{sigmevol}, panel $b$), given the tight correlation between $T$ and 
$L_{\mathrm{bol}}/M$ seen in Figure~\ref{evol_temp}. The plot in panel $c$ of $\Sigma$ 
vs $L_{\mathrm{bol}}/L_{\mathrm{smm}}$ is very similar to the previous one, as expected. 
Finally, a larger spread with respect to previous panels is found in the $\Sigma$ vs the 
$T_{\mathrm{bol}}$ plot. To highlight a possible evolutionary sequence emerging from these 
diagrams we report them in the bottom panels of Figure~\ref{sigmevol}, showing the source 
density contours not only for the populations of in panels $a$-$d$, but also for
pre-stellar and MIR-dark proto-stellar sources. The pre-stellar population occupies a 
relatively small region in the left part of the diagrams, while apparently no large 
differences are found between contours of MIR-dark and overall proto-stellar sources. This 
confirms the impression obtained in previous sections, namely that the MIR-dark sources do not 
necessarily constitute the ``earliest-stage tail'' of the proto-stellar population.
Some degree of segregation is found only with respect to the bolometric temperature (panel~$d$),
as already suggested by Figure~\ref{tbol}. Finally, the \ion{H}{ii} region candidates 
occupy the right portion of all plots, showing the largest scatter in surface density.
Despite the large scatter and partial overlap among different classes of objects, if we 
consider - in each plot - only the peaks of the source density contours, an evolutionary 
sequence through various classes might be identified: MIR-dark sources are typically denser 
than pre-stellar but also generic proto-stellar ones, and \ion{H}{ii} region candidates
are normally found at lower densities than the generic proto-stellar sources. On
average proto-stellar sources are denser than the starless ones, and characterised 
by a wide range of densities, often showing a surface density increase in the early stages 
of star formation, followed by a drop with increasing age 
\citep[corroborating the results of][]{gia13,guz15}, although cases of evolved sources 
with large surface densities are also found.

\subsection{A tool for source classification}

Having examined the clump physical parameters as evolutionary 
diagnostics, we show a ``radar-plot'' visualization of these, which allows
a powerful and compact way to show multivariate data. In particular, 
we chose five distance-independent quantities, namely $T$, $L_\mathrm{bol}/M$, 
$L_\mathrm{bol}/L_\mathrm{smm}$, $T_\mathrm{bol}$, and $\Sigma$. 

\begin{figure}
\centering
\includegraphics[width=8.5cm]{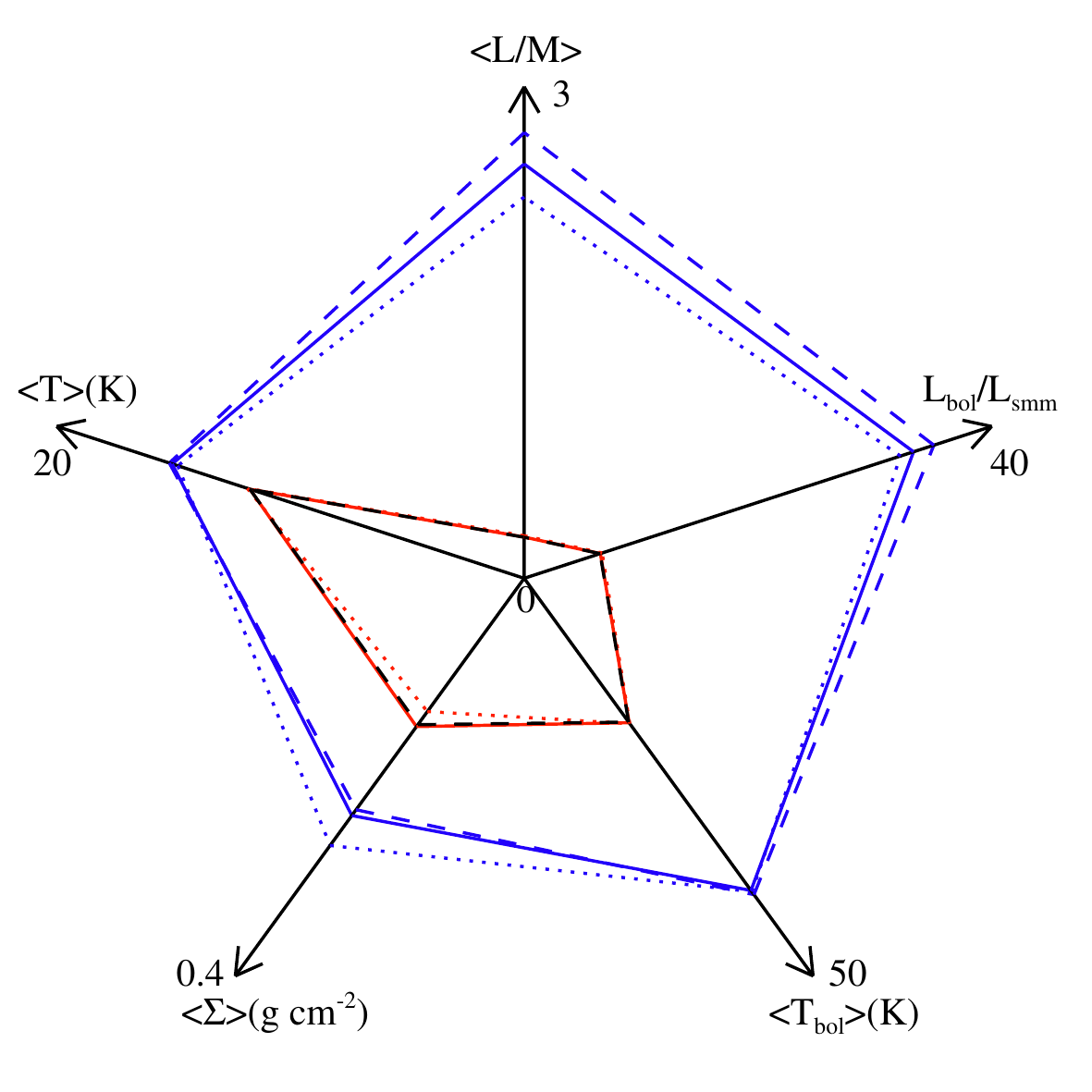}
\caption{Radar plot for the median of five physical parameters (see text) for
the classes of pre- and proto-stellar clumps (red and blue lines, respectively). 
Solid lines represent the statistics of the entire populations, whereas separate 
statistics for the fourth and the first Galactic quadrants (see Section~\ref{IvsIV})
are represented with dashed and dotted lines, respectively. Scales on each axis are
linear, ranging from 0 to the value specified at the end.}
\label{radar}
\end{figure}

In Figure~\ref{radar} the medians of the five indicators for the whole classes of 
pre- and proto-stellar clumps are reported on the five axes of the radar plot, and 
points are connected through a polygonal line to form a pentagon. One can notice that
the pentagon corresponding to the proto-stellar sources includes the pentagon representing 
the pre-stellar ones, since the values of the considered age indicators are typically 
larger for the former class of objects. Such a behaviour is also estimated, independently, 
for the fourth and the first Galactic quadrant, as we will discuss in Section~\ref{IvsIV}.

\begin{figure}
\centering
\includegraphics[width=8.5cm]{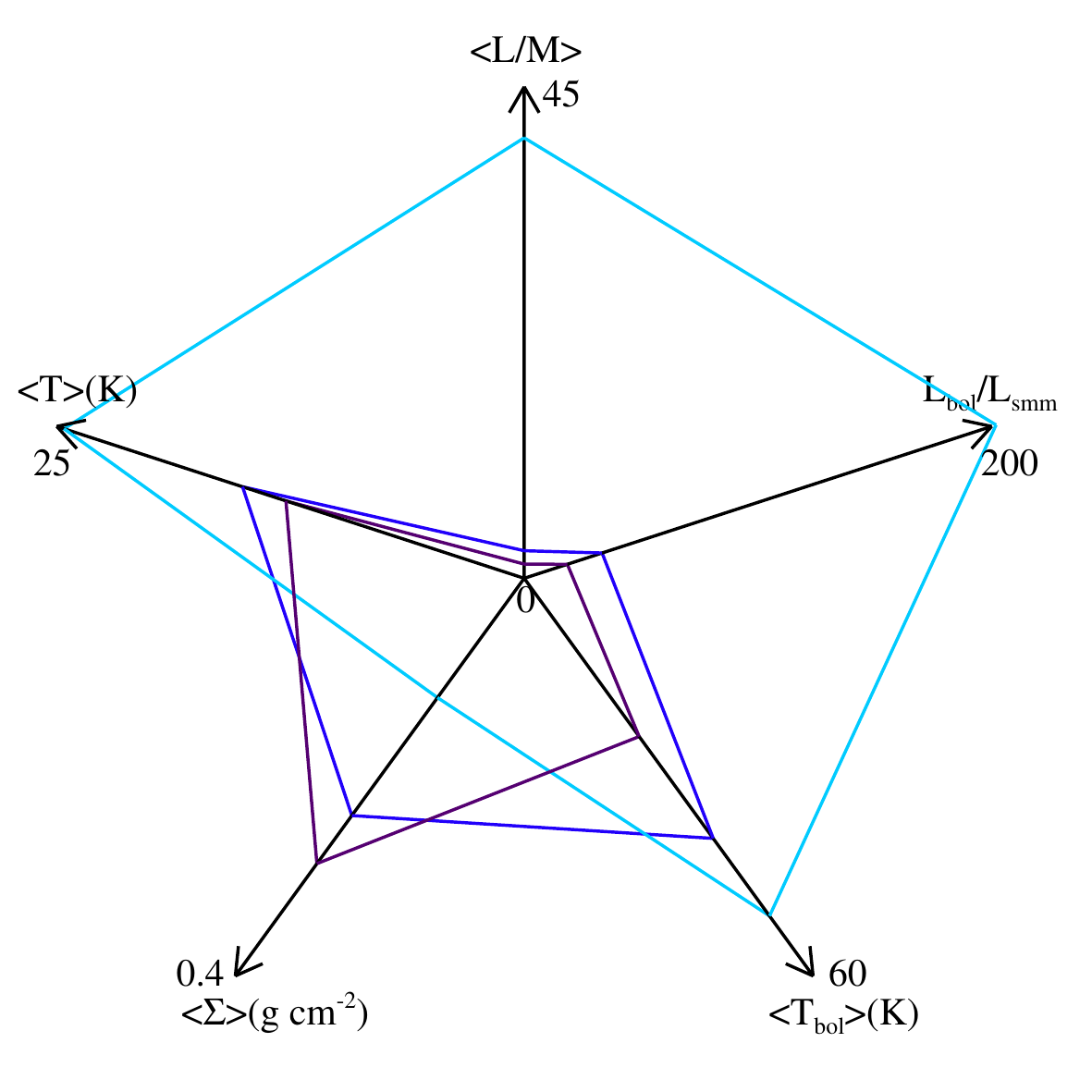}
\caption{As in Figure~\ref{radar}, but for proto-stellar sources (blue line) and for
the two extracted populations of MIR-dark sources (dark purple line) and \ion{H}{ii}-region 
candidates (light blue line), respectively.}
\label{radarhii}
\end{figure}

The radar plot of the proto-stellar sources can be further compared with that of the 
two sub-samples we discussed in previous sections as possible extremes of their age distribution: 
the MIR-dark and the candidate \ion{H}{ii} regions, respectively. In Figure~\ref{radarhii}
all evolutionary indicators appear decreased for the MIR-dark sample, and enhanced 
for the candidate \ion{H}{ii} regions,
with respect to the overall proto-stellar population, except for the surface density which, 
as seen in Section~\ref{surfdens_par}, exhibits the opposite global trend.

The metric described above could be used in future, for instance, to characterise the 
properties of clump populations associated with IRDCs \citep[e.g.,][]{tra15}, lying on 
(vs off) filaments \citep[e.g.,][]{sch14}, located in the outer (vs inner) Galaxy, 
etc. Moreover, this representation of the median (or mean) properties of a sample can 
be applied to a single source, allowing one to to immediately assess if and how this 
approaches or deviates from the global behaviour of a certain class of sources it, in 
principle, belongs to. 

\section{Source spatial distribution along the Galactic Plane}\label{sourcegal}

\subsection{First vs fourth Galactic quadrant}\label{IvsIV}

\begin{figure}
\centering
\includegraphics[width=8.0cm]{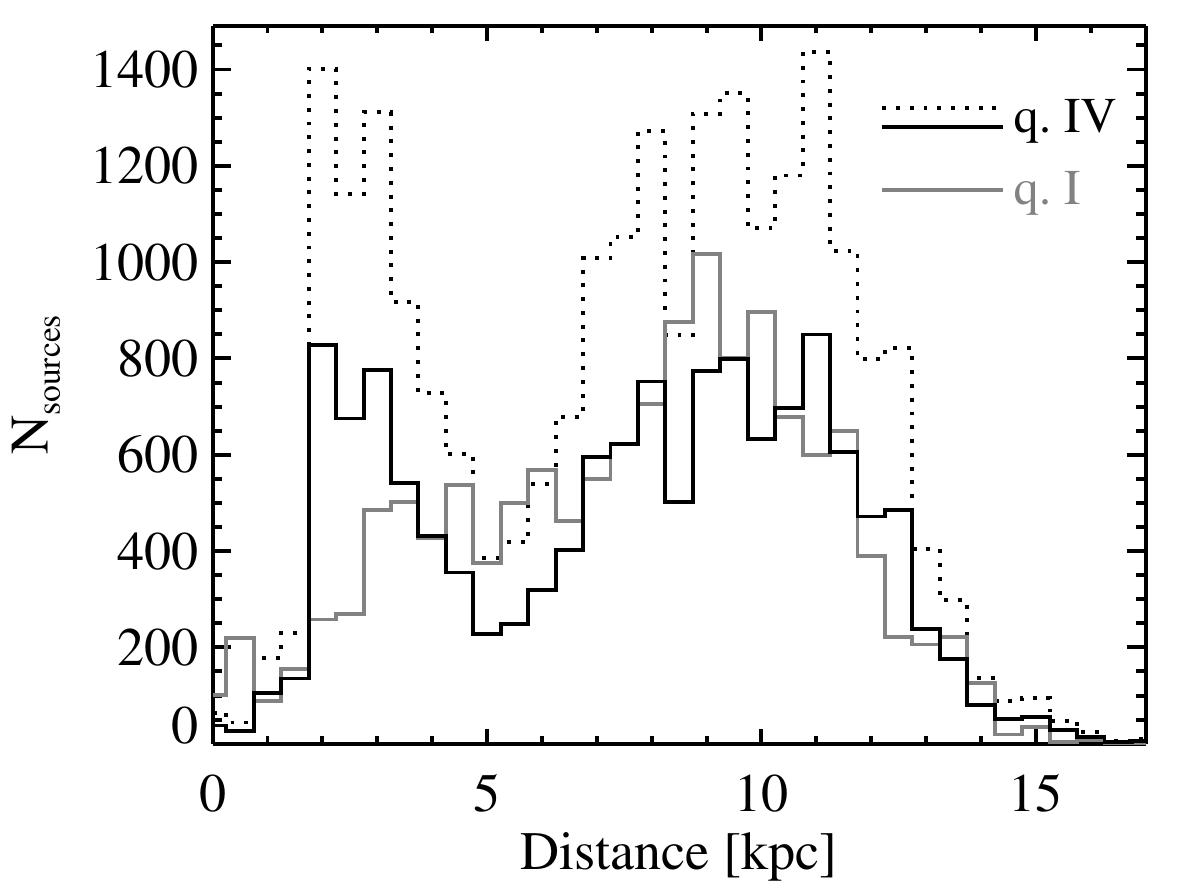}
\caption{Source heliocentric distance distributions (cf. Figure~\ref{sizefig}, left),
obtained separately for the fourth (black) and the first (grey) quadrant, respectively.
To better compare the two distributions, the original fourth quadrant histogram (dotted
line) has been normalized (solid line) by the ratio of the total source counts in the
two quadrants.}
\label{dist1q4q}
\end{figure}

So far we considered the global statistics of the Hi-GAL sources regardless of their
position in the Galactic plane. In this section the statistics of Hi-GAL source properties 
is presented in bins of Galactic longitude, and discussed in view of a comparison between 
the fourth and first quadrant. To introduce this analysis, with particular 
respect to the discussion of distance-dependent observables, first of all we show the 
heliocentric distance distribution in the fourth and the first Galactic quadrants, separately 
(Figure~\ref{dist1q4q}). The main differences, i.e. those extending over more than two bins, are seen 
around 3-4~kpc, where fourth-quadrant sources are predominant in number, and around 4.5-7~kpc, 
where the opposite case is found. However, the overall behaviour of the two distributions is 
similar, thus making a comparison between the two quadrants feasible for 
distance-dependent properties such as mass, linear size and luminosity (see below). 

\begin{figure*}
\includegraphics[width=15.0cm]{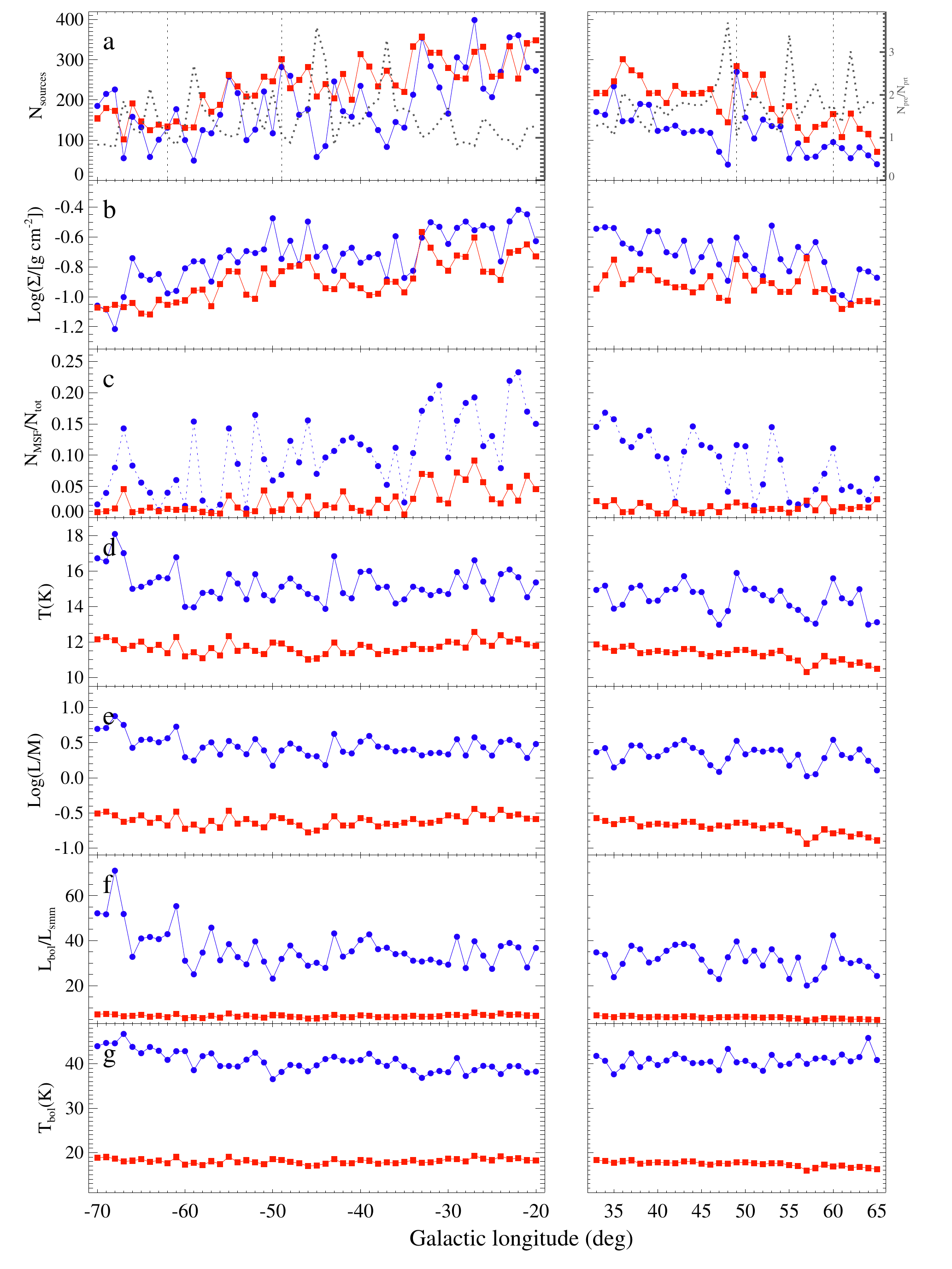}
\caption{Distribution of relevant quantities obtained in this paper for pre- (red 
filled squares) and proto-stellar (blue filled circles) sources as a 
function of the Galactic longitude, averaged over bins of width $1^{\circ}$, 
in the longitude range of the fourth (left panels) and the first (right panels) Galactic quadrants
considered in this work: ($a$) source number (where the where the dotted black line represents 
the pre- over proto-stellar source number ratio reported on the right axis); ($b$) median of decimal 
logarithm of surface density; ($c$) fractional number of dense clumps fulfilling conditions for massive star formation \citep[according to][]{kru08} ; ($d$) median temperature; 
($e$) median of decimal logarithm of $L_{\mathrm{bol}}/M$; ($f$)
median of the $L_{\mathrm{bol}}/L_{\mathrm{smm}}$ ratio; ($g$) median of bolometric temperature.}
\label{longtutti}
\end{figure*}

Figure~\ref{longtutti} presents the starting points for discussing the 
distribution of sources and their properties in the plane. Panel~$a$ 
shows how sources are distributed in longitude. A relation with the trends 
found in single-band catalogues by \citet{mol16a} is expected despite the SED 
filtering carried out in this work. In fact, for each physical quantity we have considered,
$i$) a positive correlation is found moving towards the inner Galaxy, and $ii$) as shown 
by \citet{mol16a}, for both pre- and proto-stellar sources localized excesses are found 
at positions corresponding to spiral arm tangent points or star forming complexes. 

\citet{eno08} used the number ratio between starless and proto-stellar cores to 
roughly estimate, through a proportionality factor, the lifetime of the cores 
they detected in nearby low-mass star forming regions. Because, in our case, we are 
dealing with clumps instead of cores, we cannot apply a similar prescription directly,
although useful considerations can be made concerning clumps. Our clump statistics gives 
a ratio of $\sim 1$ which, as found by \citet{svo16} on their sample of 4683 clumps 
identified at 1.1~mm in the first quadrant, would be suggestive of comparable lifetimes 
for the two classes \citep[see also][]{mot07,rag13}. Here, thanks to large 
statistics, we are able to describe this quantity, $N_{\mathrm{pre}}/N_{\mathrm{prt}}$
as a function of Galactic longitude (Figure~\ref{longtutti}, panel~$a$). Pre-stellar 
sources are in excess with respect to proto-stellar ones almost everywhere, 
except for a few cases where $N_{\mathrm{pre}}/N_{\mathrm{prt}}<1$ in the 
fourth quadrant. In general, this ratio is lower in the fourth 
quadrant, i.e. \numratioiv~vs \numratioi~for the first quadrant.

\begin{figure}
\centering
\includegraphics[width=8.0cm]{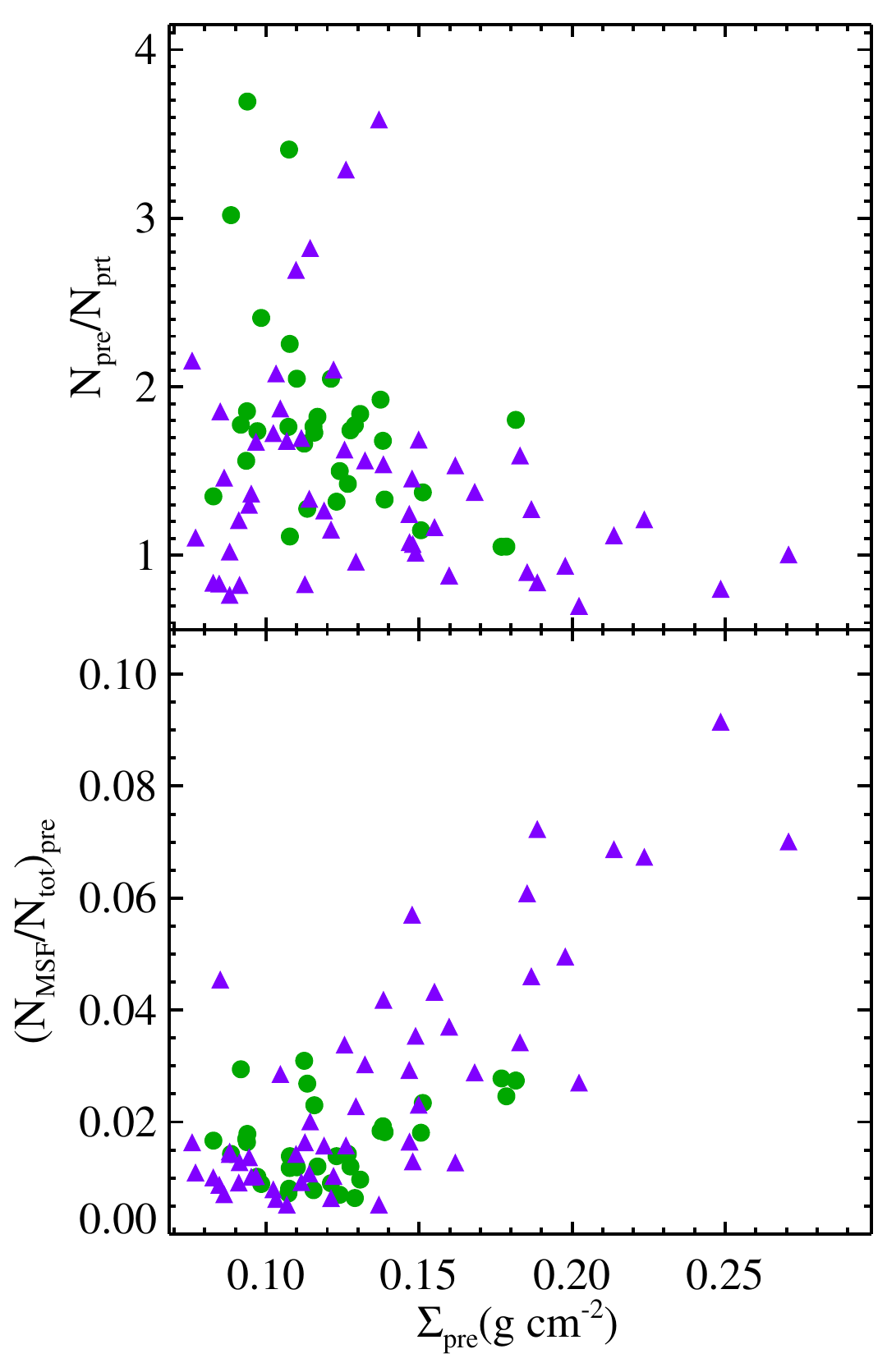}
\caption{Number ratio of pre-stellar on proto-stellar sources (top), and of sources 
compatible with massive star formation on the total (bottom), as a function of median 
surface density of pre-stellar sources, estimated in the same 1-degree bin of Galactic 
longitude. Data are taken from Figure~\ref{longtutti}, panels~$a$, $b$, and $c$. Data 
from the fourth and first Galactic quadrants are represented with purple triangles and 
green circles, respectively.}
\label{eoratiodens}
\end{figure}

To infer relative clump lifetimes from this number ratio is not trivial, since time scales
can also depend on the mass. Indeed the deficit of pre-stellar sources with respect to
proto-stellar ones at both the lowest and the highest masses found at a given heliocentric 
distance was extensively discussed in Section~\ref{cmfpar}, concluding that in particular, 
it should be due to rapid evolution of the most massive, but also densest pre-stellar clumps 
from the quiescent phase to star forming activity.
A synoptic look at panels $a$ and $b$ (the latter showing the median surface density) 
of Figure~\ref{longtutti} seems not to confirm such a scenario, since no evident 
correlation is seen between $N_{\mathrm{pre}}/N_{\mathrm{prt}}$ and density. However, 
plotting the former against the latter (Figure~\ref{eoratiodens}, top panel), values 
of $N_{\mathrm{pre}}/N_{\mathrm{prt}}>2$ are found only for lower median surface densities 
of pre-stellar sources ($\Sigma \lesssim 0.15$~g~cm$^2$). Furthermore, at higher densities 
($\Sigma \gtrsim 0.2$~g~cm$^2$), only cases with $N_{\mathrm{pre}}/N_{\mathrm{prt}}<1.3$
are found. Unfortunately we can not consider this as a strong argument, since the 
inverse of these statements is not found to be true in general. Future studies, focused 
on a more accurate identification and analysis of specific regions characterised by a very 
high or a very low $N_{\mathrm{pre}}/N_{\mathrm{prt}}$, will better assess the relation
with pre-stellar clump density.
 
\begin{figure}
\centering
\includegraphics[width=8.0cm]{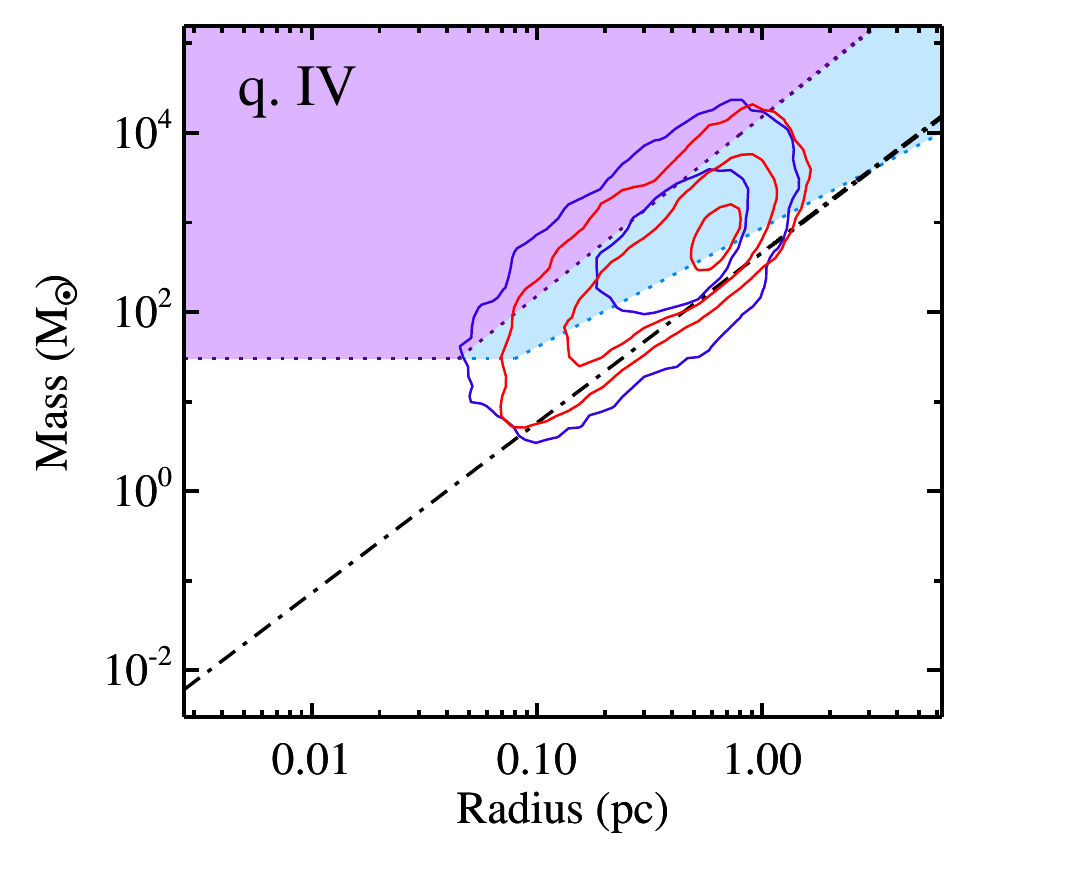}
\includegraphics[width=8.0cm]{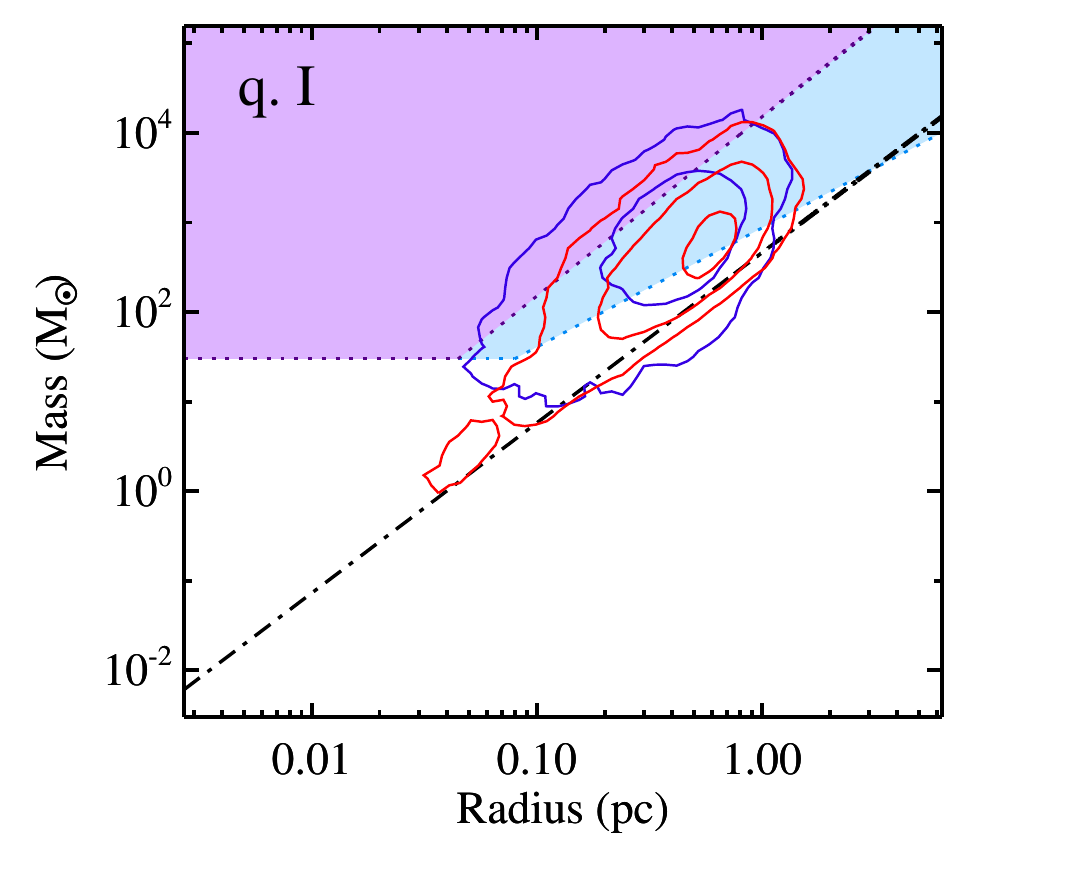}
\caption{The same contour plots as in bottom left panel of Figure~\ref{masssize}, but 
obtained separately for sources in the fourth (top) and in the first (bottom) Galactic 
quadrant, respectively.}
\label{masssize4th}
\end{figure}

The fraction of sources exceeding the threshold necessary for massive star formation 
($N_{\mathrm{MSF}}/N_{\mathrm{tot}}$) according to the prescription of \citet{kru08} 
(see Section~\ref{mass_par}) is shown in panel $c$ of Figure~\ref{longtutti}, again 
separately for pre- and proto-stellar sources. As explained in Section~\ref{mass_par}, 
only starless sources are suitable for this analysis, and this fact is highlighted 
in the plot by connecting the proto-stellar distribution through a dashed line instead 
of a solid one. A few remarks can be made about this plot: 
\begin{itemize}
 \item The fraction of pre-stellar clumps potentially able to form massive stars 
 $(N_{\mathrm{MSF}}/N_{\mathrm{tot}})_\mathrm{pre}$ is generally higher in the fourth 
 rather than in the first quadrant. This can be seen also by plotting the mass vs 
 radius diagram for the two quadrants separately, as in Figure~\ref{masssize4th}: both 
 quadrants denote, with respect to the two adopted density thresholds, a similar 
 ability to form high-mass stars, but the fraction of pre-stellar sources in the 
 fourth quadrant populating the \citet{kru08} area is slightly larger.
 \item In the first quadrant, a relevant amount of pre-stellar sources which are 
 massive and dense enough to give rise to possible high-mass star formation are found.
 In particular while \citet{gin12}, based on BGPS data, claim that no quiescent clumps 
 with $M>10^4~M_{\odot}$ and $r<2.5$~pc are found between $6^{\circ}<\ell<90^{\circ}$
 \citep[confirmed by][after an accurate analysis in the range $10^{\circ}<\ell<65^{\circ}$]{svo16},
 in our catalogue we find \ginsburgfound~sources, located in the range 
 $\ginsburglonmax^{\circ} < \ell < \ginsburglonmin^{\circ}$, that fulfil these 
 criteria. To explain such a discrepancy, one should perform a dedicated analysis, 
 which is beyond the scope of the present paper. Here we limit ourselves to 
 emphasising two aspects. First, for all sources of this group the ambiguity 
 about the heliocentric distance has been solved in favour of the ``far'' solution (the minimum
 distance in the group being $d=\ginsburgmindist$~kpc, and \ginsburgfoundten~sources out of 
 \ginsburgfound~have a distance in the catalogue $d>10$~kpc. A different distance estimate 
 has a linear effect on radius, but a quadratic effect on the mass, and these in turn could 
 potentially affect source classification. Second, the masses of \citet{gin12} are 
 estimated for a constant temperature of $T=20$~K (a value which is far from being 
 representative of a starless clump, as seen in this paper), while in our case 
 they are derived from a grey-body fit simultaneously with temperature: for the 
 \ginsburgfound~sources in question, the derived temperatures are significantly 
 lower than 20~K (with an average of \ginsburgtemp~K), therefore leading to much higher values 
 of mass. Indeed, at 1.1~mm \citep[the wavelength used by][]{gin12}, assuming a grey body 
 temperature of 20~K instead of 13~K underestimates the mass by a factor $\sim 2$,
 independently from the choice of $\beta$. \citet{svo16} used a more
 accurate estimate of temperature, derived from NH$_3$ line observations of each source, 
 which, in global terms, is still higher for pre-stellar sources (a median of 13.96~K against 
 \medtpe~K in our case, see Section~\ref{dtemp}). In addition to this, in the 
 particular case of our \ginsburgfound~sources, grey body temperatures are found to be 
 generally lower, with \ginsburgtempncold~of them having $T<10~K$. Such a temperature
 can be genuine but could be also attributed to a SED affected by problems in the original photometry
 or by relevant deficit of flux at the shortest wavelengths involved in the fit, 
 due to multiplicity (see Section~\ref{filtering}). A dedicated analysis is required for 
 this group of sources to better define their physical conditions and confirm whether 
 they might be genuine progenitors of massive proto-clusters, or simply statistical
 fluctuations in a huge catalogue.
 \item In the fourth quadrant, $(N_{\mathrm{MSF}}/N_{\mathrm{tot}})_\mathrm{pre}$
 is found to increase towards inner longitudes ($\ell \gtrsim -35^{\circ}$).
 \item As expected, in general there is a direct relation between $N_{\mathrm{MSF}}/N_{\mathrm{tot}}$
 for pre-stellar sources and the corresponding median surface density (for pre-stellar
 sources), shown in Figure~\ref{eoratiodens}, bottom panel .
 \end{itemize} 

\begin{figure}
\centering
\includegraphics[width=9.0cm]{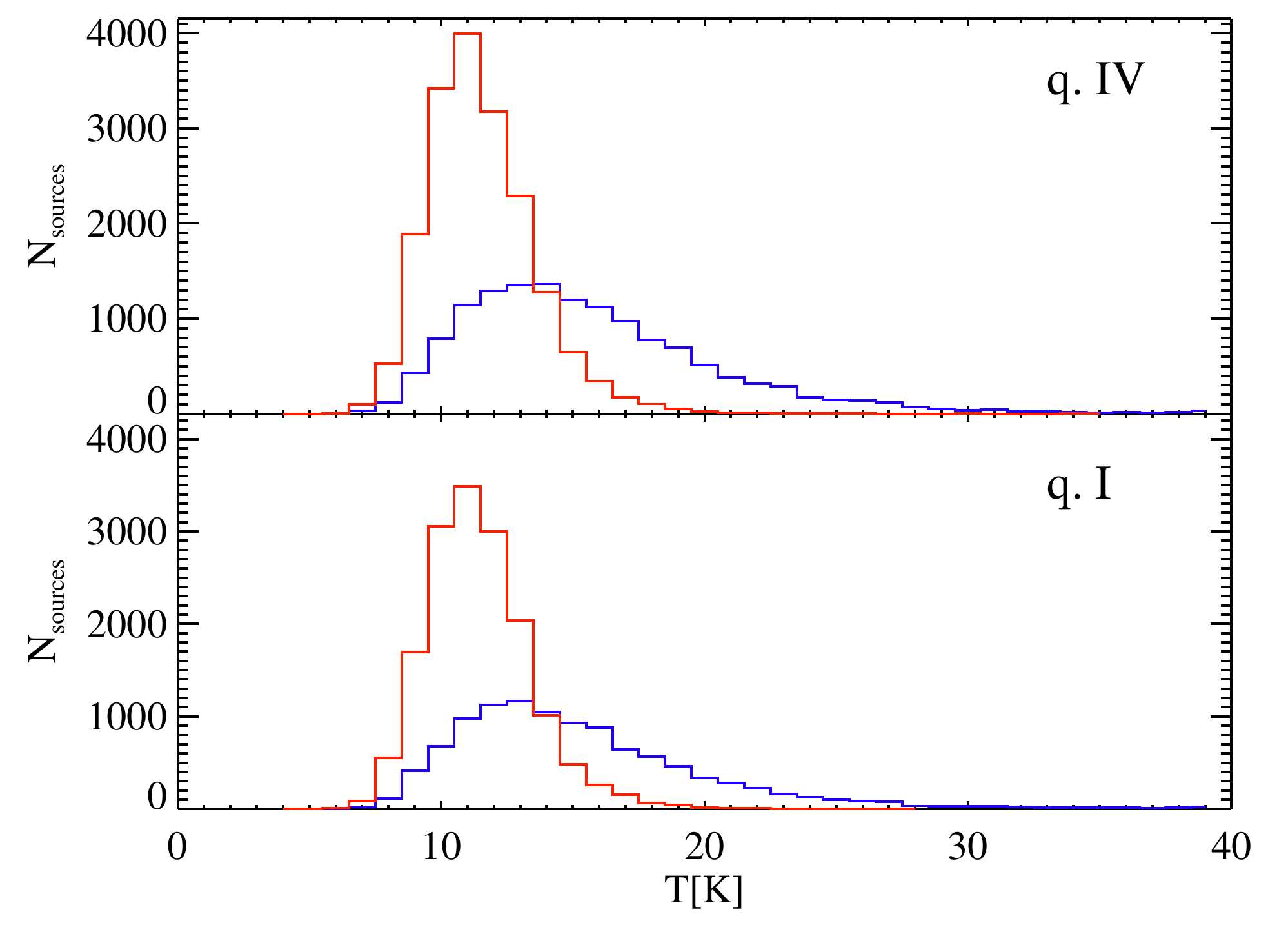}
\caption{The same histogram as in Figure~\ref{hist_temp}, but obtained separately for sources 
in the fourth (top) and in the first (bottom) Galactic quadrant, respectively.}
\label{hist_temp_1q_4q}
\end{figure}

We can now start analysing the behaviour of the evolutionary diagnostics. We begin with dust 
temperature, whose distribution is shown in Figure~\ref{longtutti}, panel $d$. We notice that
the two Galactic quadrants do not present significant differences, as can also be seen in 
Figure~\ref{hist_temp_1q_4q}, similar to Figure~\ref{hist_temp} but with the temperature 
distributions given separately for the fourth and first quadrant. If we compare the top 
and bottom panels of this figure, the distribution of proto-stellar sources in the fourth 
quadrant peaks at slightly higher temperatures with respect to the equivalent population 
in the first quadrant, with the average and median temperatures of
$\bar{T}_{\mathrm{prt,IV}}=$\avetpoIV~K and $\tilde{T}_{\mathrm{prt,IV}}=\medtpoIV$~K in the 
fourth quadrant, and $\bar{T}_{\mathrm{prt,I}}=\avetpoI$~K and 
$\tilde{T}_{\mathrm{prt,I}}=\medtpoI$~K in the first quadrant, respectively. Notice, however,
that the associated standard deviations of ~5~K in all cases make this difference poorly 
significant from the statistical point of view. In the following, however, further slight 
differences between source properties in the two quadrants, globally indicating a more evolved 
stage of sources in the fourth quadrant, will be discussed.

No significant differences are found for the pre-stellar sources in the two quadrants
($\bar{T}_{\mathrm{pre,IV}}=\avetpeIV$~K vs $\bar{T}_{\mathrm{pre,I}}=\avetpeI$~K,
and $\tilde{T}_{\mathrm{pre,IV}}=\medtpeIV$~K vs $\tilde{T}_{\mathrm{pre,I}}=\medtpeI$~K,
respectively).


Figure~\ref{longtutti}, panel $d$ shows the median temperature as a function of the Galactic 
longitude. It can be seen that three strong local peaks ($\tilde{T} \geq 17$~K) at $-68^{\circ}$, 
$-61^{\circ}$, $-43^{\circ}$ are found in the fourth quadrant, while two dips ($\tilde{T} \leq 13$~K) 
at 47$^{\circ}$, 57$^{\circ}$ are found in the first one. However, these do not correspond to large 
source counts (Figure~\ref{longtutti}, panel $a$), indicating that they should not be considered 
as the main cause of the overall discrepancy between the two quadrants which, therefore is 
more indicative of a general trend.

Peaks and dips in the temperature distribution are found in the same longitude bins
for $L_{\mathrm{bol}/M}$ and $L_{\mathrm{bol}}/L_{\mathrm{smm}}$ (Figure~\ref{longtutti},
panels $e$ and $f$, respectively). This is not surprising, given the tight relation
between the temperature and these two indicators as suggested by Figure~\ref{evol_temp}. 
A weaker correlation is seen with bolometric temperature (Figure~\ref{longtutti},
panel~$g$).

\begin{figure}
\centering
\includegraphics[width=8.0cm]{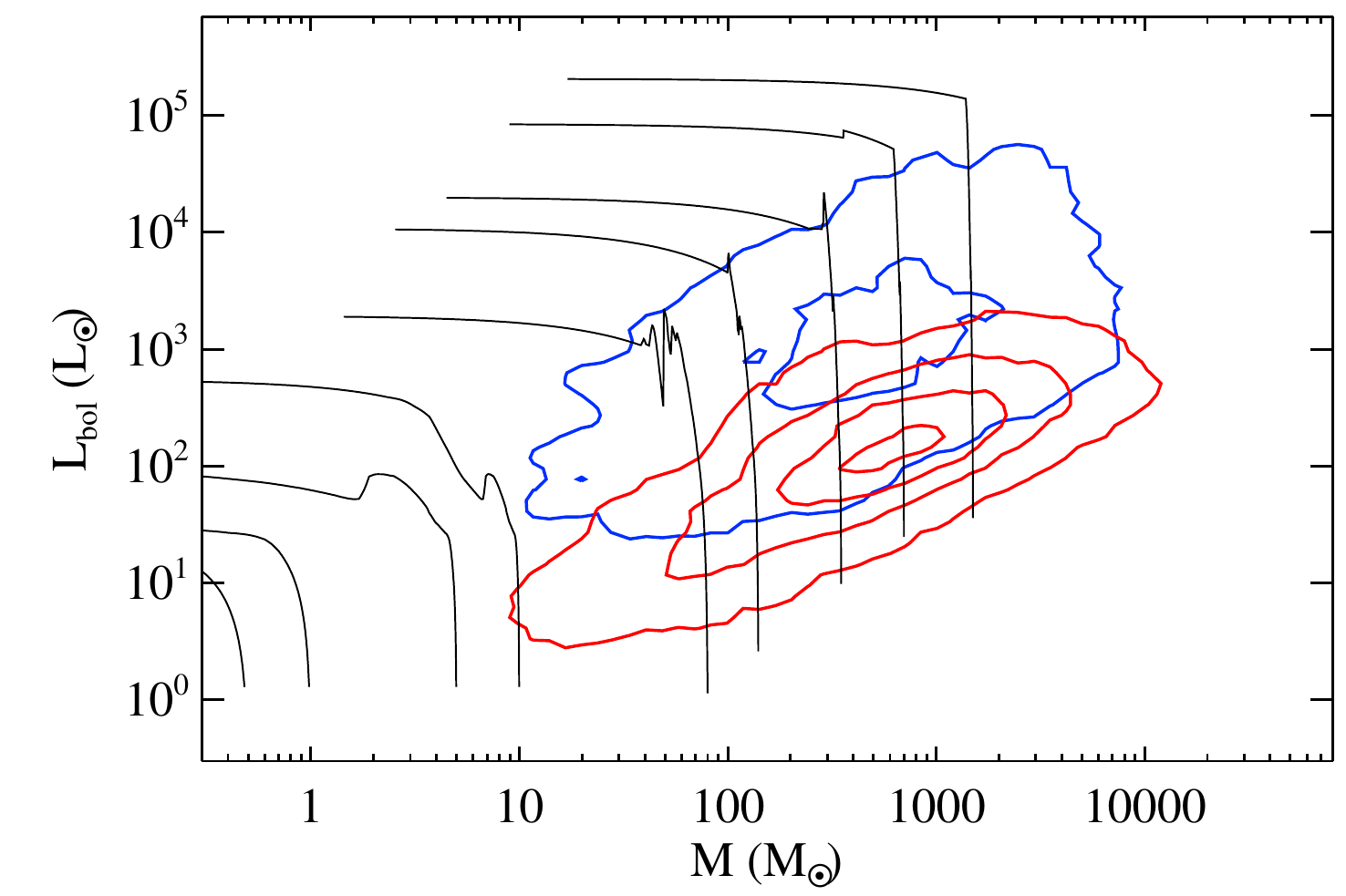}
\includegraphics[width=8.0cm]{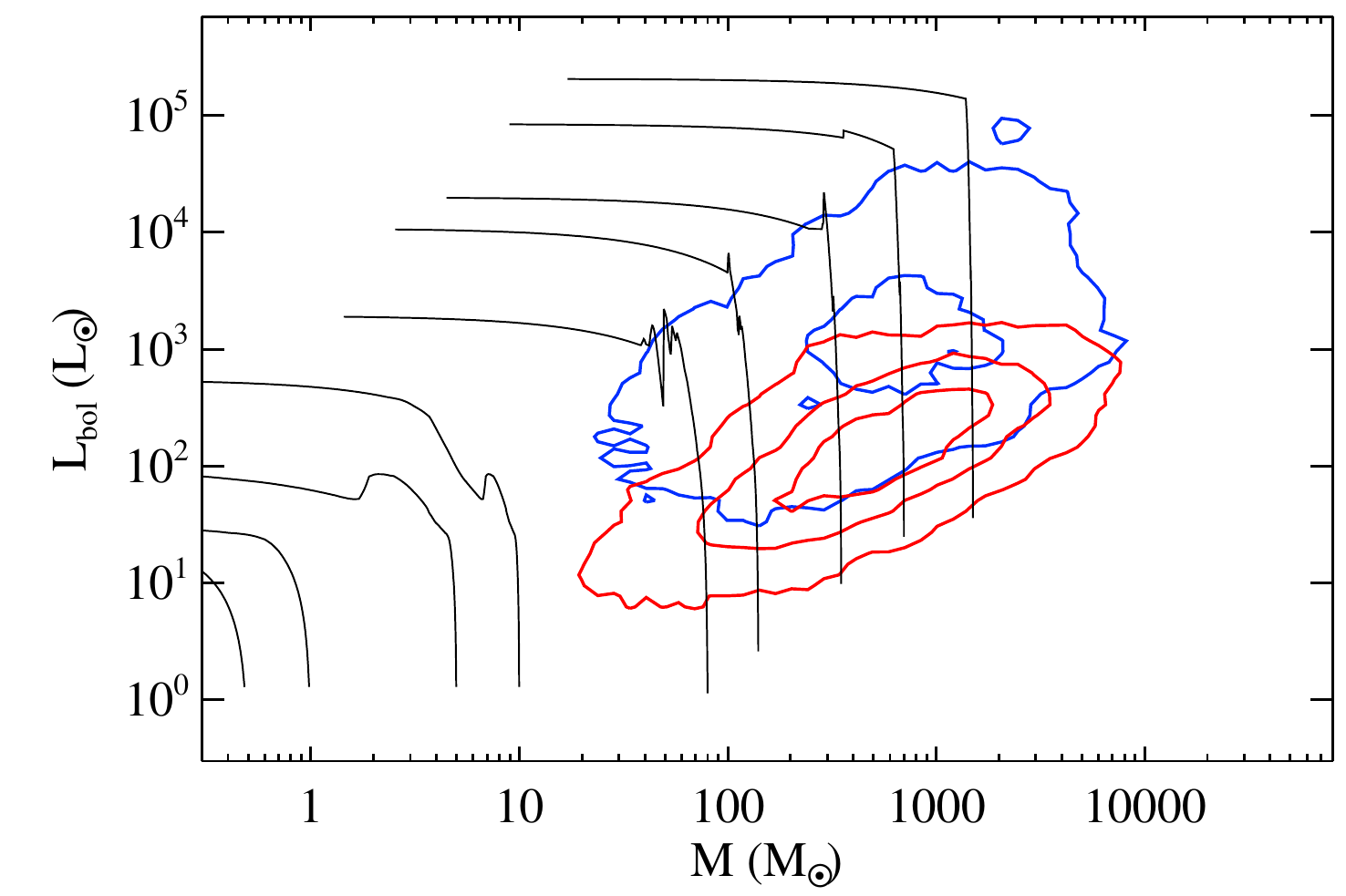}
\caption{The same plot of $L_{\mathrm{bol}}/M$ as in Figure~\ref{lvsm}, but obtained 
separately for sources in the fourth (top) and in the first (bottom) Galactic quadrant, 
respectively.}
\label{lvsm_I_IV}
\end{figure}

Except for the features mentioned above, the global $L_{\mathrm{bol}}/M$ behaviour
in the two quadrants does not appear very different, as corroborated by two observational facts:
\begin{itemize}
\item In Figure~\ref{lvsm_I_IV}, in which the $L_{\mathrm{bol}}$ vs $M$, in form of 
density contours, is plotted separately for the two quadrants: the overall source 
distribution appears to be concentrated in the same region of the diagram, with respect 
to the evolutionary tracks of \citet{mol08}.
\item The mean and median values of this ratio for proto-tellar objects, namely 
$[(L_{\mathrm{bol}}/M)_{\mathrm{prt,IV}}]_{\mathrm{med}}=\lmrIV~L_{\mathrm{\odot}}/M_{\mathrm{\odot}}$
and $[(L_{\mathrm{bol}}/M)_{\mathrm{prt,I}}]_{\mathrm{med}}=\lmrI~L_{\mathrm{\odot}}/M_{\mathrm{\odot}}$, 
respectively, are very similar. They are reported also in Figure~\ref{radar}, together with 
other evolutionary indicators. The charts corresponding to the proto-stellar population of 
the two quadrants show only slight differences although, interestingly, these appear to be 
systematic: the fourth quadrant has larger $T$, $L_\mathrm{bol}/M$, $L_\mathrm{bol}/L_\mathrm{smm}$, 
and $T_\mathrm{bol}$, and lower $\Sigma$, which unequivocally indicate a (slightly) more 
advanced stage of evolution.
\end{itemize}

\subsection{Spiral arms and star formation}\label{spiralarms}
The physical properties discussed in the previous section provided a first 
``1-dimensional'' picture of how the sources of the Hi-GAL catalogue are 
distributed across the Milky Way. A more complete analysis also has to incorporate
source distances to investigate their position in the Galactic plane, 
with respect to spiral arm topology, as already suggested by
Figure~\ref{galpos}. 
In the interest of completeness, we briefly touch here on this topic, while for more 
details and and an extensive discussion we refer the reader to \citet{rag16}.

\begin{figure*}
\centering
\includegraphics[width=16cm]{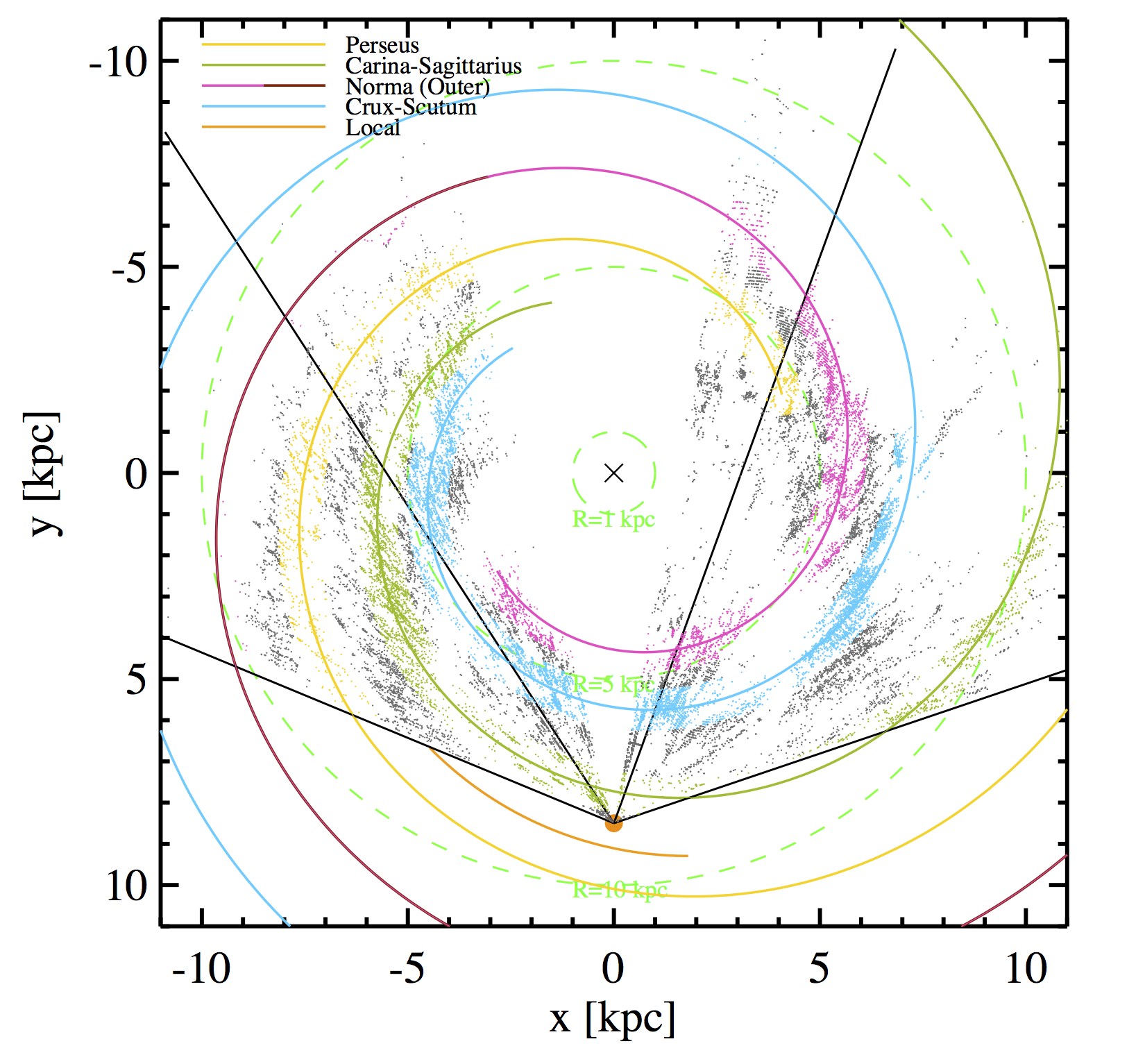}
\caption{The same as Figure~\ref{galpos}, but with different colours for dots representing
sources: positions within a distance 0.5~pc from the theoretical position of an arm
are assigned to it and displayed with the same colour (with no reference to the starless
vs proto-stellar classification). The remaining sources, found at ``inter-arm'' locations,
are displayed in grey.}
\label{armpos}
\end{figure*}

In Figure~\ref{armpos}, as in Figure~\ref{galpos}, we show the overall distribution of
the Hi-GAL clumps along the Galactic plane as well as the location of the spiral arms according
to the model of \citet{hou09}. In this new figure, however, we also make an attempt
to associate clumps with spiral arms. To this end, we use an arm width of 1~kpc, then a 
source-to-arm association distance of 0.5~kpc \citep[cf.][]{ede13}. According to this 
criterion, \narm~sources out of \ndist~considered for this analysis (i.e., those provided 
with a heliocentric distance estimate) are associated with the arms. In addition, 
\niarm~are classified as ``inter-arm'', thus a $\sim 2:3$ ratio, 
significantly different from the 1:7 expected by simulations of \citet{dob07}, but in 
better agreement with the observational evidence of \citet{ede13}, namely 21:38 for 
BGPS sources between $\ell=37^{\circ}.8$ and $42^{\circ}.5$. However, we should 
make the reader aware that this association between sources and arms and the consequent
results are strongly affected by heliocentric distance determination, which is still a
critical point in our analysis. While a new set of Hi-GAL source distances based on an 
improved algorithm is expected (Russeil et al., in preparation), cross-checks can 
be carried out (although on a relatively small number of cases) with distances quoted for
objects in common with surveys such as ATLASGAL \citep{wie15} or BGPS \citep{ell15} (see
Appendix~\ref{champepos}). In this way a distance estimate external to our catalogue 
and considered more reliable can be assumed to easily re-scale the distance-dependent
source properties quoted in our catalogue.

As already evident from Figure~\ref{galpos}, overdensities of source counts do not always 
correspond to arm locations. This certainly also depends on the spiral arm model of choice 
\citep[see, e.g.,][]{rag16} but, even more, is primarily inherent to the set of estimated 
distances. As a matter of fact, different spiral arm models will provide different solutions 
so, for a certain model, a source will result being ``on-arm'' while, for another, it might turn 
out to be ``off-arm''. Moreover, the on-off location of clumps with respect to spiral arms will 
also strongly depend on kinematic distances, which are notoriously limited - by their intrinsic 
uncertainties - in their ability of delineating the morphology of the arms. Therefore, the 
prescription we adopted to associate Hi-GAL clumps to spiral arms produces sharp and unnatural 
separations between on-arm and off-arm populations (see Figure~\ref{armpos}), which affects a 
possible comparison of the two, as well as the remaining analysis. This aspect was already 
evident in \citet{rom09}, who 
examined the same spectroscopic survey, namely GRS, that we use here to determine distances in the 
first Galactic quadrant: they found that only 63\% of $^{13}$CO emission can be associated to spiral 
arms, while several bright and large structures fall in ``inter-arm'' gaps.

\begin{figure*}
\centering
\includegraphics[width=16cm]{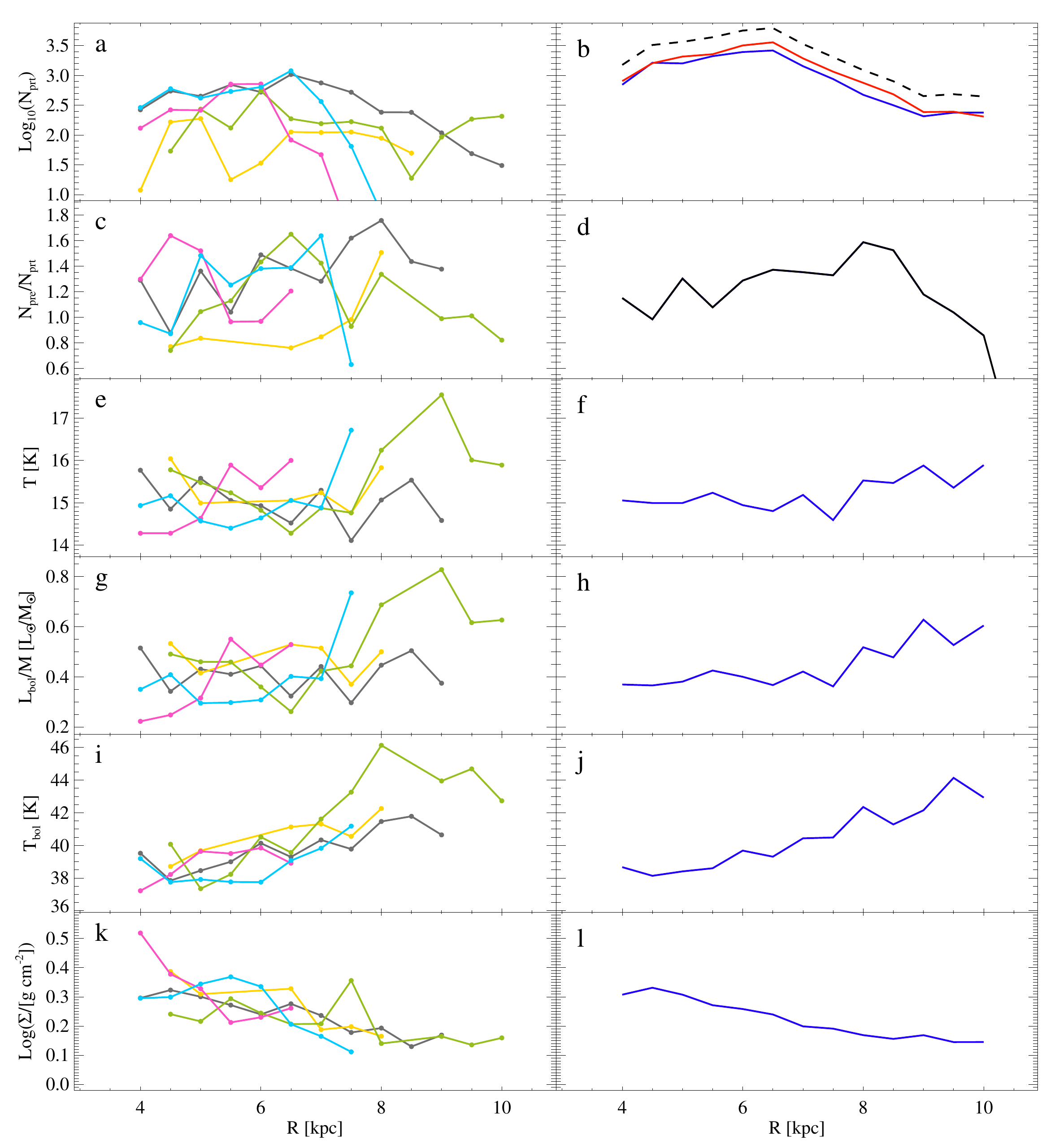}
\caption{Number and physical properties of proto-stellar clumps with a distance estimate, in 
bins of Galactocentric radius (bin width is 0.5~kpc). In left panels, different quantities
are plotted separately by spiral arms (see Figures~\ref{galpos} and~\ref{armpos} 
for colour coding) with the addition of ``inter-arm'' sources (in grey). In right 
panels the overall sample of proto-stellar sources is considered. In detail, panel $a$ shows 
the logarithm of the number of proto-stellar sources, while panel $b$ shows the sum of 
them (blue solid line); the counts of pre-stellar and the sum of pre-+proto-stellar are also 
reported as red solid and black dashed lines, respectively. In panel $c$ the number ratio of 
pre- over proto-stellar sources is reported, while in panel $d$ the same quantity
is computed and reported for the entire source population (black line). Panels $e$, $g$,
$i$, and $k$ are similar to panel $c$, but for dust temperature, logarithm of the luminosity/mass
ratio, bolometric temperature, ad logarithm of surface density respectively. Correspondingly, 
panels $f$, $h$, $j$, and $l$ report the same distributions, but calculated over the entire 
population of proto-stellar sources.}
\label{rgalarms}
\end{figure*}

With these caveats in mind, we present some statistics based on grouping sources
belonging to the same arm, and distributed at a different Galactocentric radius $R$. 
In Figure~\ref{rgalarms}, panel~$a$, first of all we show the number statistics of 
proto-stellar objects, which the other panels are based on. This information 
is useful to understand the reliability of statistics presented in the other panels: 
clumps are counted in bins of 0.5~kpc of Galactocentric radius, and bins poorly 
populated (i.e. $<50$ sources) are ignored in the following. The overall 
number of proto-stellar sources as a function of $R$ is reported in panel~$b$, 
together with the same curve for pre-stellar sources (used to build the plot in 
panel~$d$) and for the sum of the two classes. A direct comparison of the latter 
with the Galactocentric radius distribution of ATLASGAL sources presented by 
\citet{wie15} is not straightforward. This is because, in that case, the plot 
is split between the first and fourth Galactic quadrants, and because counts 
are expressed in surface density, therefore normalised by the area. As in \citet{wie15}, 
a peak is found around 4.5~kpc, corresponding to the peak of the light blue line in panel 
$a$, so that we agree with those authors that this enhancement corresponds 
to the intersection of the Crux-Scutum arm and the Galactic bar in the first quadrant 
\citep[see also][]{ngu11,mot14}. Another peak (the largest we find) is encountered 
around 6.5~kpc, and again it is mostly due to the Crux-Scutum arm, but in particular 
to its portion in the fourth quadrant. In both cases, the curve of ``inter-arm'' 
clumps closely follows that of the Crux-Scutum arm and strongly contributes 
to the aforementioned peaks, corroborating the fact that, in many locations, the 
distinction between ``on-arm'' and ``inter-arm'' locations is quite artificial 
due to the method we used to identify them. Finally, at $R\sim 10$~kpc 
\citep[this distance is not probed by][]{wie15}, sources of the Carina-Sagittarius 
arm produce a peak which is mirrored in the overall source distribution 
as a plateau interrupting a decreasing trend found over a range of 2.5~kpc.

In panel $c$ the ratio between pre-stellar and proto-stellar clumps
$N_{\mathrm{pre}}/N_{\mathrm{prt}}$ (discussed in Section~\ref{IvsIV} as a 
function of Galactic longitude) is shown in bins of $R$ for different arms
and for the ``inter-arm'' sources. Values range from 0.7 to 1.8, but no
specific trends are found either for the behaviour of individual arms or in the
comparison between ``on-arm'' and ``inter-arm'' sources. Therefore, no suggestions
come from this plot about a possible role of spiral arms in triggering or 
regulating star formation, confirming the conclusions of \citet{ede13} and
\citet{rag16}. Analogously, the distributions of three evolutionary indicators 
such as median dust temperature, bolometric luminosity/mass ratio, and 
bolometric temperature of proto-stellar sources (panels~$e$, $g$, and~$i$, 
respectively) show a high degree of scatter and no substantial differences between 
arm and ``inter-arm'' sources. A recognizable trend is present in the 
distribution of median bolometric temperatures, which generally increases at 
increasing $R$ for all arms and for the inter-arm component as well. This is 
more evident in the overall behaviour shown in panel $j$, even ignoring bins at
$R>8$~kpc, which are dominated by the Carina-Sagittarius arm in the fourth quadrant, 
and which are generally characterised by high values for the evolutionary indicators 
as seen also in panels
$e$, $g$, $f$ and $h$, although based on relatively poor statistics (see panel~$a$).
Finally, the distributions of surface densities reported in panel~$k$ show a
slightly decreasing trend with $R$, as recognised in all curves and in the 
overall distribution (panel~$l$). A linear fit to the overall distribution 
between 4.5 and 9~kpc to the latter gives the relation 
$\log_{10}(\Sigma/(\mathrm{g~cm^{-2}}))=\sigmazero - \sigmaslope~(R/\mathrm{kpc})$.
A much stronger decrease of clump surface peak is claimed by \citet{zah16},
who found a drop of three orders of magnitude over the range 0-13~kpc, but this
quantity cannot be directly compared with our average surface density. More
generally, encountering less dense clumps at farther distances from the Galactic 
centre is in agreement with the result of \citet{rag16}: these authors define 
a ``star forming fraction'' as the number ratio (based on the present catalogue)
between proto-stellar and all clumps, and find a general trend decreasing
with increasing $R$ (with a slope of -0.025). In other words, a correlation
between the gradual decrease of star formation activity at large Galactocentric 
radii and the decrease of the clump densities (i.e. of material available for 
gravitational collapse) can be hypothesised.


\section{Summary}
We presented the physical catalogue of Hi-GAL compact sources extracted in 
the inner Galaxy (in the longitude range between $\ell = -71^{\circ}$ and
$\ell=67^{\circ}$). 
First we described how sources were selected from the Hi-GAL five-band 
photometric catalogue and how their physical properties were estimated.
To this aim we illustrated how: 
\begin{enumerate}
\item Five single-band photometric catalogs of \citet{mol16a} were merged, based 
      on simple positional associations. A total of \totfittedseds~SEDs eligible 
      for grey body fit were selected and considered for subsequent analysis,
      \withdistance~of which included a heliocentric distance estimate
      (no distances are available in the central part of the Galaxy
      $-10.2^{\circ}< \ell < 14.0^{\circ}$). 
\item A targeted flux extraction at 160 and 70~$\umu$m has been performed
      for sources missing a flux at these bands. Sources remained with 
      only three Hi-GAL fluxes have been reported in a ``lower-reliability''
      list. Furthermore, the MSX, WISE, MIPSGAL, ATLASGAL, and BGPS 
      catalogues have been searched for possible geometric counterparts, 
      to extend the spectral coverage of the SEDs.   
\item After SED building and filtering, the \nproto~sources containing a flux at 70~$\umu$m 
      have been classified as proto-stellar, while the remaining \nstarless~sources as starless.
      Furthermore, based on subsequent SED fitting and mass computation, starless sources 
      have been classified as gravitationally bound (pre-stellar) or unbound (\npre~and 
      \nunb, respectively). Both these criteria are applied directly and uniquely on photometric
      data, and for this reason are prone to be refined in future using other 
      observational evidence. Furthermore, they can not remove completely a certain degree of contamination
      among different evolutionary classes.
\item A grey body spectrum has been fitted to SEDs, to obtain the average temperature
      and the total mass of a clump (in case of available distance, otherwise at least
      the clump surface density has been derived, taking the source diameter at 250~$\umu$m
      as a reference size). The bolometric luminosity has been derived for sources 
      provided with a distance estimate, while the bolometric temperature has been 
      obtained for all sources.                      
\end{enumerate}

Once the physical properties of the sources were derived, we explored the distributions 
of these observables, their mutual relations with a possible evolutionary scenario, 
and their possible connection with the Galactic large scale structure. Main results are
summarised below:
\begin{enumerate}\addtocounter{enumi}{4}
\item Based on their size, the sources in this catalogue are in most cases classifiable as 
      clumps. For a given distance, proto-stellar clumps are found to be on average 
      more compact than starless ones. Consequently, the former are found to 
      be generally denser than the latter.
\item A significant amount of sources, both pre- and proto-stellar, is found to fulfil
      one or more criteria for compatibility with high-mass star formation. This sample can be 
      extracted for large programs of follow-up observations aimed to clarify the internal
      structure of dense clumps and put observational constraints on models of high-mass
      star formation.
\item The mass function of proto-stellar sources falling in a relatively narrow ($\Delta d=0.5$~kpc) 
      distance range appears generally wider than that of pre-stellar sources, due to completeness 
      effects at low masses, and to relative deficit of pre-stellar clumps at high masses (in turn 
      likely due to evolutionary effects). As a consequence of this, the power-law slope of the 
      pre-stellar function is systematically  steeper than that of the proto-stellar one. Finally, 
      no systematic bias seems to affect the mass function slope at increasing heliocentric distance.      
\item The clump average temperature, estimated over the range $160~\umu$m$\leq \lambda
      \leq 500~\umu$m (so with no reference to the presence of counterparts at shorter
      wavelengths) and representing the physical conditions of the outer part of the 
      clump structure, acts as an evolutionary indicator: the median temperature for
      pre-stellar clumps is \medtpe~K, while for proto-stellar sources is \medtpo~K.
      However a high degree of overlap between the two populations remains and 
      the combined use of further evolutionary indicators is then recommended to
      reduce such degeneracy.
\item We used the bolometric luminosity and its ratios with the clump mass and the sub-millimetre 
      luminosity, together with the bolometric temperature, as further evolutionary 
      indicators. An acceptable degree of separation between pre- and proto-stellar 
      populations is found by analysing the distributions of these observables. Since the 
      proto-stellar sources span a wide range of evolutionary stages, we tried to 
      identify possible candidates to represent both the earliest and the latest 
      stages that we are able to probe with Hi-GAL. On the one hand, we found that sources 
      dark in the MIR may play the former role only in part, while, on the other hand, 
      sources with $L_{\mathrm{bol}}/M> \cesalim L_{\odot}/M_{\odot}$ represent
      the right tail in the distribution of all evolutionary indicators, and are compatible,
      for temperature and colours, with the stage of \ion{H}{ii} regions, although 
      this requires further observational evidences to be gathered.
\item The behaviour of surface density with respect to clump evolution shows, in general, 
      an increase from the pre- to the proto-stellar phase, and a decrease (but with a 
      large spread of values) in the most evolved proto-stellar objects, corresponding
      to the envelope clean-up phase.
\item Regarding the source distributions with Galactic longitude, local 
      excesses of sources are encountered in correspondence with spiral arm tangent 
      points or star forming complexes. A low number ratio between pre- and proto-stellar 
      sources, evaluated in bins of longitude, found in correspondence with high median 
      surface density in the same bins ($\Sigma > 0.2$~g~cm$^{-2}$) may suggest
      a short lifetime of high-density clumps in the pre-stellar stage. Despite of this,
      we find a conspicuous number of pre-stellar cores compatible with high-mass
      star formation, especially in the fourth quadrant. 
\item No large differences are found between medians of clump evolutionary indicators in 
      the fourth and in the first quadrant. However, although differences are quite small, 
      median temperature, luminosity over mass ratio, bolometric over sub-mm luminosity 
      ratio, and bolometric temperature are larger in the fourth than in the first quadrant.
\item Although the set of distances we adopt does not produce, in some regions of the 
      Galactic plane, a source pattern showing well definite spiral arms, we made a 
      tentative assignment of sources to spiral arms modelled by \citet{hou09}. ``On-arm'' 
      and ``inter-arm'' populations show no relevant differences either in the pre- over 
      proto-stellar number ratio or in the median values of the evolutionary indicators. 
      This result could suggest a negligible impact of spiral arms in triggering star 
      formation. However, this result is biased by the uncertainties affecting 
      source distance estimations and consequently association with spiral arms.
\item While temperature and luminosity over mass ratio do not show clear trends as
      a function of the increasing Galactocentric radius, a slightly increasing trend      
      is found for median bolometric temperature, and a slightly decreasing one for the
      median surface density.
\end{enumerate}

In conclusion, the aim of this paper is to give a first look at
the huge amount of information contained in the Hi-GAL physical catalogue of the 
inner Galaxy. Papers based on data taken from this catalogue have been already 
published \citep{rag16,per16}, as well as observational programs submitted 
for (and also partially observed with) ALMA. A number of papers exploiting this
catalogue, or aiming at completing it in the outer Galaxy, are in preparation at
present, and will contribute to deepen and refine the conclusions of this work.

\section*{Acknowledgements}
The authors thank the anonymous referee for her/his careful reading of the 
manuscript and insightful comments. This work is part of the VIALACTEA 
Project, a Collaborative Project under Framework 
Programme~7 of the European Union, funded under Contract~\#607380 that is hereby 
acknowledged. \textit{Herschel} Hi-GAL data processing, maps production and source 
catalogue generation is the result of a multi-year effort that was initially funded 
thanks to Contracts I/038/080/0 and I/029/12/0 from ASI, Agenzia Spaziale Italiana. 
\textit{Herschel} is an ESA space observatory with science instruments provided by 
European-led Principal Investigator consortia and with important participation from 
NASA. PACS has been developed by a consortium of institutes led by MPE (Germany) and 
including UVIE (Austria); KUL, CSL, IMEC (Belgium); CEA, OAMP (France); MPIA (Germany);
IAPS, OAP/OAT, OAA/CAISMI, LENS, SISSA (Italy); IAC (Spain). This development has been 
supported by the funding agencies BMVIT (Austria), ESA-PRODEX (Belgium), CEA/CNES (France), 
DLR (Germany), ASI (Italy), and CICYT/MCYT (Spain). SPIRE has been developed by a 
consortium of institutes led by Cardiff Univ. (UK) and including Univ. Lethbridge 
(Canada); NAOC (China); CEA, LAM (France); IAPS, Univ. Padua (Italy); IAC (Spain);
Stockholm Observatory (Sweden); Imperial College London, RAL, UCL-MSSL,
UKATC, Univ. Sussex (UK); Caltech, JPL, NHSC, Univ. Colorado (USA). This
development has been supported by national funding agencies: CSA (Canada);
NAOC (China); CEA, CNES, CNRS (France); ASI (Italy); MCINN (Spain);
Stockholm Observatory (Sweden); STFC (UK); and NASA (USA).

\bibliographystyle{mnras}

\appendix
\section{Description of physical catalogue}\label{catdescription}
The HI-GAL physical catalogue for the inner Galaxy is hosted in the VIALACTEA knowledge base
\citep[VLKB,][]{mmol16}, and is arranged in two tables (high- and low-reliability SEDs) both 
containing the same columns, defined as follows:
\begin{itemize}
\item Column [1], \textit{ID}: running number (starting from 1 in the high-reliability
table and continuing in the low-reliability one).
\item Column [2], \textit{DESIGNATION}: string composed by ``HIGAL'', ``BM'' (which stays 
for ``band-merged'') and Galactic coordinates of the sources, chosen as the coordinates of
the shortest-wavelength available Hi-GAL counterpart.
\item Columns [3], \textit{GLON}, and [4] \textit{GLAT}: Galactic longitude and latitude, respectively,
assigned to the source, chosen as the coordinates of the shortest-wavelength 
available Hi-GAL counterpart.
\item Columns [5], \textit{RA}, and [6], {DEC}: the same as in columns [3] and [4], respectively, 
but for source Equatorial coordinates.
\item Column [7], \textit{DESIGNATION\_70}: designation of the PACS $70~\umu$m counterpart (if available), 
as defined in the catalogue of \citet{mol16a}. The null string (in case of a missing counterpart at this 
band) is ``-''.
\item Column [8], \textit{F70}: flux density (hereafter flux) of the PACS $70~\umu$m counterpart (if available), 
in Jy, as quoted by \citet{mol16a}. The null value is 0.
\item Column [9], \textit{DF70}: uncertainty associated to the flux in column [8] as quoted by 
\citet{mol16a}. The null value is 0.
\item Column [10], \textit{F70\_TOT}: sum of fluxes of all PACS $70~\umu$m counterparts (if available) lying
inside the half-maximum ellipse of the source detected by CuTEx in the SPIRE $250~\umu$m maps. By definition,
$F_\mathrm{70,tot}\geq F_\mathrm{70}$. The null value is 0. This is the flux at $70~\umu$m actually 
used to estimate the source bolometric luminosity (Section~\ref{sedfit}) and temperature 
(Section~\ref{evolbol}). 
\item Column [11], \textit{DF70\_TOT}: uncertainty associated to the flux in column [10], obtained as the
quadratic sum of uncertainties on single fluxes. The null value is~0.
\item Column [12] \textit{F70\_ADD}: flux of the closest PACS $70~\umu$m counterpart (if available) found 
through targeted source extraction at a detection threshold lower than in \citet{mol16a} where 
$F_\mathrm{70,tot}=0$ (column[11]), as described in Section~\ref{protovsstarl}. The null value is~0.
\item Column [13], \textit{DF70\_ADD}: uncertainty associated to the flux in column [12]. The null 
value is~0.
\item Column [14], \textit{F70\_ADD\_TOT}: sum of fluxes of all PACS $70~\umu$m counterparts (if available) 
found through targeted source extraction at a detection threshold lower than in \citet{mol16a} where 
$F_\mathrm{70,tot}=0$, and lying inside the ellipse at $250~\umu$m (as for column [10]). 
The null value is 0. This is the flux at $70~\umu$m actually used to estimate
the source bolometric luminosity and temperature where $F_\mathrm{70,tot}=0$.
\item Column [15], \textit{DF70\_ADD\_TOT}: uncertainty associated to the flux in column [14], estimated
as for column [11]. The null value is 0.
\item Column [16], \textit{ULIM\_70}: 5-$\sigma$ upper limit in the PACS $70~\umu$m band, estimated where
both $F_\mathrm{70,tot}=0$ (column [10]) and $F_\mathrm{70add,tot}=0$ (column[14]).
\item Columns [17], \textit{DESIGNATION\_160}, [18], \textit{F160}, and [19], \textit{DF160}: 
the same as columns [7], [8], and [9], respectively, but for the PACS $160~\umu$m band.
\item Columns [20], \textit{F160\_ADD}, and [21], \textit{DF160\_ADD}: the same as columns 
[12], and [13], respectively, but for the PACS $160~\umu$m band.
\item Column [21], \textit{ULIM\_160}: 5-$\sigma$ upper limit in the PACS $160~\umu$m band, estimated where
both $F_\mathrm{160}=0$ (column [18]) and $F_\mathrm{160add}=0$ (column [20]).
\item Columns [22], \textit{DESIGNATION\_250}, [23], \textit{F250}, and [24], \textit{DF250}: the 
same as columns [7], [8], and [9], respectively, but for the SPIRE $250~\umu$m band.
\item Columns [25], \textit{DESIGNATION\_350}, [26], \textit{F350}, and [27], \textit{DF350} : the 
same as columns [7], [8], and [9], respectively, but for the SPIRE $350~\umu$m band.
\item Column [28], \textit{FSC350}: SPIRE $350~\umu$m flux ``scaled'' as mentioned in 
Section~\ref{filtering}. Further details on the method are provided, e.g., in \citet{gia12}.
Scaling is not performed when the source size differs by less than a factor $\sqrt{2}$ from
the instrumental beam size at this wavelength.
\item Column [29], \textit{DFSC350}: uncertainty associated to the flux in column [28].
\item Columns [30], \textit{DESIGNATION\_500}, [31], \textit{F500}, [32], \textit{DF500},
[33], \textit{FSC500}, and [34], \textit{DFSC500}: the same as columns~[25], [26], [27], [28],
and [29], respectively, but for the SPIRE $500~\umu$m band.
\item Column [35], \textit{DESIGNATION\_21}: designation of the MSX $21~\umu$m counterpart 
(if available), as defined in the MSX point source catalogue. The null string (in case of a 
missing counterpart at this band) is ``-''
\item Column [36], \textit{F21}: flux of the closest MXS $21~\umu$m counterpart (if available
within the adopted matching radius), in Jy. The null value is 0.
\item Column [37], \textit{DF21}: uncertainty associated to the flux in column [36]. The null value 
is 0.
\item Column [38], \textit{F21\_TOT}: sum of fluxes of all MXS $21~\umu$m counterparts (if available) 
lying inside the ellipse at $250~\umu$m (as done for column [10]). The null value is 0.
\item Column [39], \textit{DF21\_TOT}: uncertainty associated to the flux in column [38], computed
as for column~[11]. The null value is 0.
\item Columns [40], \textit{DESIGNATION\_22}, [41], \textit{F22}, [42], \textit{DF22}, 
[43], \textit{F22\_TOT}, and [44], \textit{DF22\_TOT}: the same as columns [35], [36], [37], 
[38], and [39], but for the WISE $22~\umu$m band.
\item Column [45], \textit{DESIGNATION\_24}: designation of the MSX $24~\umu$m counterpart 
(if available). A string beginning with ``MG'' identifies a source taken from the catalog
of \citep{gut15}, while a string beginning with ``D'' identifies a source specifically 
detected in this work, (cf. Section~\ref{ancillary}). Furthermore, lack of a source
due to saturation is identified with the ``satutared'' string. 
\item Columns [46], \textit{F24}, [47], \textit{DF24}, [48], \textit{F24\_TOT}, and [49], 
\textit{DF24\_TOT}: the same as columns [36], [37], [38], and [39], respectively, but for 
the MIPSGAL $24~\umu$m band. 
In column [46], in case of saturation (see column [45]) a null value -999 is quoted.
\item Column [50], \textit{DESIGNATION\_870}: designation of the ATLASGAL $870~\umu$m counterpart 
(if available). A string beginning with ``G'' identifies a source taken from the catalog
of \citet{cse14} while the string "CuTEx" identifies a source specifically 
detected for this work, as explained in Section~\ref{ancillary}. The null string (in case of a 
missing counterpart at this band) is ``-''.
\item Columns [51], \textit{F870}, and [52], \textit{DF870}: the same as columns [46] and [47],
respectively, but for the ATLASGAL $870~\umu$m band. 
\item Column [52], \textit{DESIGNATION\_1100}: designation of the BGPS $1100~\umu$m counterpart 
(if available), as defined in the BGPS catalogue \citep{gin13}. The null string (in case of a 
missing counterpart at this band) is ``-''.
\item Columns [53], \textit{F1100}, and [54], \textit{DF1100}: the same as columns [46] and [47],
respectively, but for the BOLOCAM $1100~\umu$m band. 
\item Columns [54], \textit{DFWHM250}: circularised and (if the circularised size exceeds the 
instrumental beam size of a factor $\sqrt{2}$) beam-deconvolved size of the sources as estimated 
by CuTEx in the $250~\umu$m band, in arcseconds. 
\item Columns [55], \textit{DIST}: kinematic distance of the source, in pc \citep{rus11}. In case 
of distance ambiguity, it represents the final choice between the ``near'' and the ``far'' estimates,
reported in the two next columns. The null value, in case of unavailable distance estimate, 
is 0.
\item Columns [56], \textit{NEAR\_DIST}, and [57], \textit{FAR\_DIST}: ``near'' and ``far'' 
kinematic distance estimates of the source, in pc. The null value is 0.
\item Column [58], DIST\_FLAG: flag indicating the quality of the distance ``near''/``far''
ambiguity solution. If an external indicator is used to take the decision, the flag is ``G'', otherwise
it is ``B'', and the ``far'' distance is assigned to the source by default (see Section~\ref{dist_sect}).
The null value, in case of unavailable distance estimate, is ``-''.
\item Column [59], \textit{DIAM}: source linear diameter, in pc, obtained combining columns~[54]
and [55].
\item Column [60], \textit{M\_LARS}: Larson's mass, in Solar masses, evaluated as described in 
Section~\ref{protovsstarl}. The null value, in case of unavailable distance, is 0.
\item Column [61], \textit{FIT\_TYPE}: flag indicating if the expression of the grey body fitted 
to the source SED is given by Equation~\ref{gbthick} (``thick'' case, ``Tk'' flag) or 
Equation~\ref{gbthin} (``thin'' case, ``Tn'' flag).
\item Column [62], \textit{EVOL\_FLAG}: flag indicating the evolutionary classification of the
source (0: starless unbound; 1: pre-stellar; 2: proto-stellar).
\item Column [63], \textit{MASS}: clump total mass, in units of Solar masses, derived fitting a grey body 
to the source SED. In case of unavailable distance, the fit is performed anyway assuming a virtual 
distance of 1~kpc, and the corresponding mass is quoted as a negative value.
\item Column [64], \textit{DMASS}: uncertainty associated to the mass in column [63].
\item Column [65], \textit{TEMP}: dust temperature of the clump, in~K, derived from the grey-body 
fit.
\item Column [66], \textit{DTEMP}: uncertainty associated to the temperature in column [65].
\item Column [67], \textit{LAM\_0\_TK}: value of $\lambda_0$ (see Equation~\ref{gbthick}), 
in $\umu$m, derived from the grey-body fit. The null value, corresponding to the value ``Tn'' 
of the flag FIT\_TYPE, is 0.
\item Column [68], \textit{L\_BOL}: bolometric luminosity, in units of solar luminosity, estimated as 
described in Section~\ref{sedfit}. As in the case of the mass, for sources devoid of distance estimate
a luminosity corresponding to the virtual distance of 1~kpc is calculated and quoted as a negative 
value. 
\item Column [69], \textit{LRATIO}: ratio between bolometric luminosity in Column~[67] and
its fraction computed over the range $\lambda \geq 350 \umu$m.
\item Column [70], \textit{T\_BOL}: bolometric temperature, in~K, calculated based on 
Equation~\ref{tboleq}.
\item Column [71], \textit{SURF\_DENS}: surface density, in g~cm$^{-2}$, calculated dividing 
the mass in Column~[63] by the area of the circle with the diameter in Column~[59]. Where the 
distance is unavailable, this quantity can be evaluated anyway, assuming a whatever virtual 
distance for intermediate calculations, and starting from Column~[54] instead of [59].
\end{itemize}

\section{Possible Mid-infrared associations of starless Hi-GAL sources}\label{mirfakes}
In Section~\ref{protovsstarl} we defined the classification of Hi-GAL sources
in starless and proto-stellar, based on the absence or the presence of a detection at 
70~$\umu$m, respectively. For the former ones, possible associations with MIR 
point sources are considered spurious, and are not taken into account for deriving 
the bolometric luminosity and temperature. To ascertain how reliable is this 
assumption, we performed two different tests.

\begin{figure}
\centering
\includegraphics[width=8.0cm]{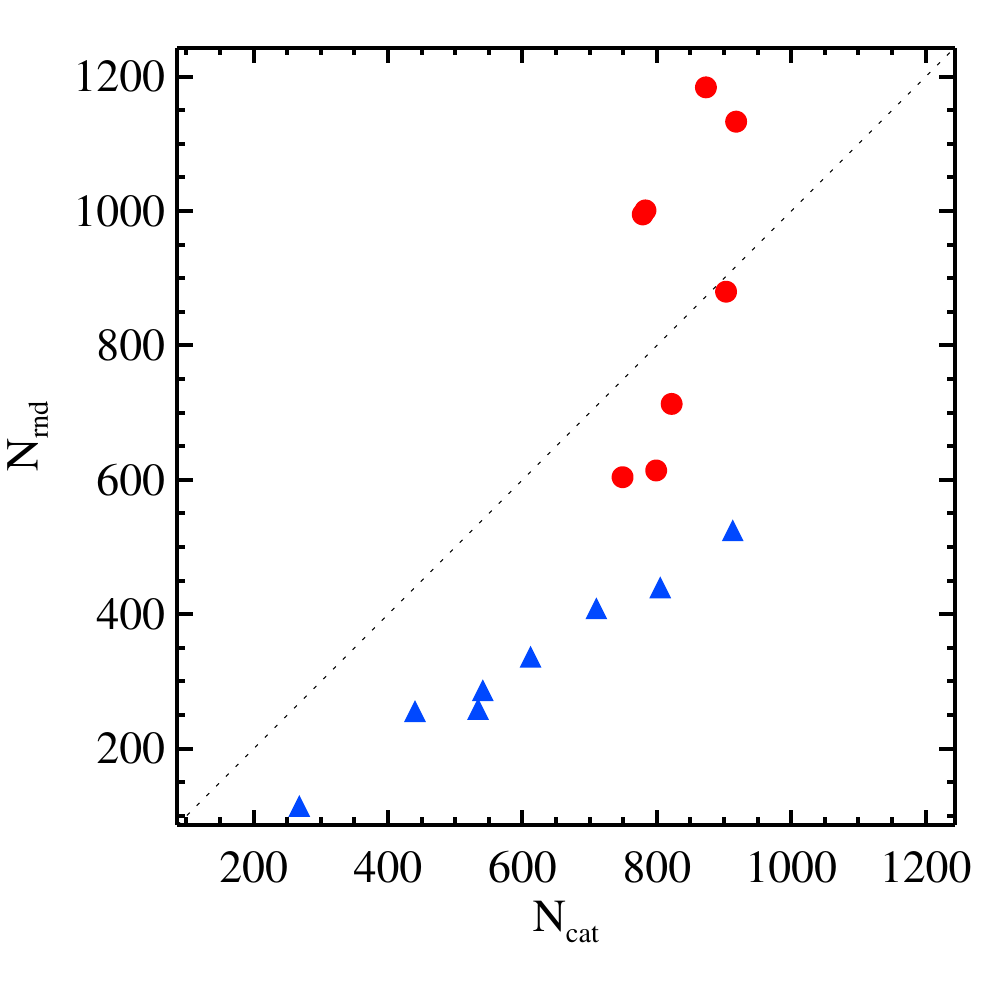} 
\caption{Comparison between the number $N_{\mathrm{cat}}$ of Hi-GAL sources (in each 
of eight different test fields, see text) found to be associated to MIR counterparts 
(being the total number of MIR sources $N_{\mathrm{MIR}}>N_{\mathrm{cat}}$) and the 
number $N_{\mathrm{rnd}}$ of associations established between the same Hi-GAL sources
and $N_{\mathrm{MIR}}$ positions randomly dispersed in the same field. Red circles
represent the statistics of starless sources detected at 160~$\umu$m, while blue
triangles represent proto-stellar sources.}\label{mir_associations}
\end{figure}

The first test consisted in checking the impact of possible foreground MIR 
point sources on the incidence of chance associations with Hi-GAL sources. We 
chose eight Hi-GAL $4^{\circ} \times 1.4^{\circ}$ fields, centered on longitudes 
$\ell=-60^{\circ},-40^{\circ},-20^{\circ},-10^{\circ},10^{\circ}, 20^{\circ},
40^{\circ},60^{\circ}$, respectively, and latitude $b=0^{\circ}$, so well 
inside the PACS-SPIRE common science area. First, we identified all sources 
having a detection at 160~$\umu$m and at 21 and/or 22 and/or 24~$\umu$m, but with
no detection at 70~$\umu$m, i.e. the sub-sample of the entire starless population 
which might be misclassified in case of a too faint flux at 70~$\umu$m. We further 
restricted the investigated box area to the limits given by the maximum and 
minimum of longitude and latitude of the selected sources, and counted the total 
amount $N_{\mathrm{MIR}}$ of MSX, WISE, and MIPSGAL sources falling within such 
area. Then we randomly dispersed $N_{\mathrm{MIR}}$ points across the area, and 
performed the association with the selected starless Hi-GAL sources as described 
in Section~\ref{ancillary}. Finally, for each field we compared the number of 
associations $N_{\mathrm{rnd}}$ found with that of the actual associations 
between the selected Hi-GAL starless sources and MIR catalogue sources 
($N_{\mathrm{cat}}$); in Figure~\ref{mir_associations}, red circles represent the 
obtained statistics, with $N_{\mathrm{rnd}}/N_{\mathrm{cat}}$ being $>75\%$ 
in all cases. This suggests that most of the associations found between 
starless Hi-GAL sources and MIR catalogues can be explained as a chance alignment
along the line of sight. 

On the other hand, MIR counterparts found for proto-stellar sources should be 
considered as a more genuine effect. In fact, repeating the procedure described
above on proto-stellar sources found in the test fields, we find an association
rate smaller than $60\%$ in all cases, suggesting that, unlike the starless case,
the real spatial disposition of MIR sources follows more closely the one of 
Hi-GAL proto-stellar objects, so cases of chance associations are expected to 
be less frequent.

The second test directly involves the photometry of the sources. The question 
arises whether the failed detection at 70~$\umu$m is caused by the fact that the 
SED is in the Wien regime of the grey body at this wavelength, emitting
a too small flux to be detected with PACS, or it is due to lack of sensitivity, 
such that the emission at 70~$\umu$m is in excess of that expected from a grey body 
at given temperature, albeit remaining below the flux threshold of the instrument.

\begin{figure}
\centering
\includegraphics[width=8.5cm]{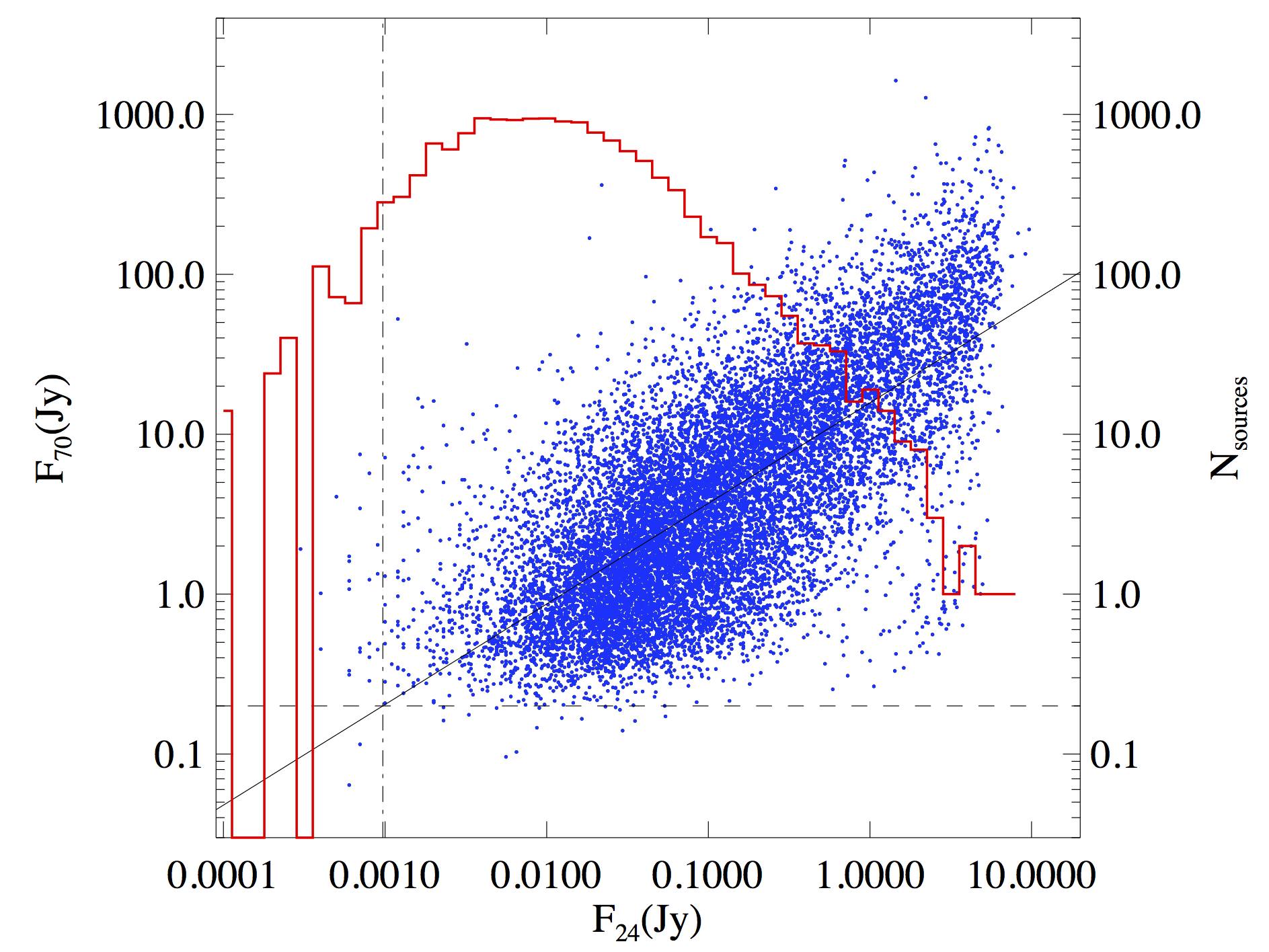} 
\caption{Scatter plot of flux at $70~\umu$m vs flux at $24~\umu$m for proto-stellar
sources of the Hi-GAL catalogue provided with both these fluxes (blue dots). The solid 
line represents the power-law fit to the displayed points (see text). The horizontal
black line represents the limit flux of 0.2 Jy at $70~\umu$m, while the vertical
dotted-dashed line corresponds to the abscissa of the intersection of the previous
line with the power-law fit. The red line is the histogram of flux at $24~\umu$m
of sources with detection at $160~\umu$m and no detection at $70~\umu$m, plotted 
with respect to the $y$ axis reported on the right side of the box.}\label{proto70vs24}
\end{figure}

To investigate this issue, first we plot $F_{70}$ vs $F_{24}$ for proto-stellar
sources (Figure~\ref{proto70vs24}). An overall correlation between the two fluxes
is seen \citep[cf with Spitzer literature, e.g.][]{you05,str10}, so that for this
class of objects low PACS fluxes are expected in correspondence of low MIPS fluxes. 
Fitting a power law $F_{70}= a\,(F_{24})^b$ to the plotted data gives $a=\afitpar$ 
and $b=\bfitpar$. Most of our sources have $F_{70}>0.2$~Jy, so that the corresponding
$F_{24}$ suggested by the fit would be \fmipspacs~Jy. This implies that, in principle, 
a source with $F_{24}>\fmipspacs$~Jy should be observable with PACS at 70~$\umu$m as well.
Nevertheless, only the \mipsfrac\% of sources detected at 24~$\umu$m and 160~$\umu$m and
not detected at 70~$\umu$m (histogram in Figure~\ref{proto70vs24}) has 
$F_{24}<\fmipspacs$~Jy, while for the rest a counterpart at 70~$\umu$m, according
with the general trend of the other sources, should be detectable. Even increasing, 
to be more conservative, the threshold on $F_{70}$ to the 90\% completeness limit
for this band quoted by \citet{mol16a}, namely $\sim 0.5$~Jy, only \mipsfraccompl\% of 
MIPS sources would be fainter than the corresponding limit flux $F_{24}=\fmipspacscompl$~Jy, 
leaving the remaining ones to be likely chance associations.

We conclude that both tests discourage to consider a positional match between a Hi-GAL 
source undetected at 70~$\umu$m and a MIR source as a genuine physical association in 
most cases. It is noteworthy that in literature, before the release of \textit{Herschel} catalogues, 
the pre- vs proto-stellar nature of sub-mm sources was ascertained just through simple
spatial association with MIR counterparts, hence suffering of possible chance alignment 
contamination \citep[e.g.,][]{cse14}, all the more considering the large gap in wavelength 
between the two bands associated ($\lambda \sim 20~\umu$m to $\lambda \geq 870~\umu$m).

\section{Distance bias on source property estimation and interpretation}\label{appdist}
Whereas studying a single star forming region or complex it is possible to assume
a single global distance estimate for its compact source population, a Galactic
plane survey as Hi-GAL inevitably contains an extraordinary variety of distances
superposed along the same line of sight. In Section \ref{physsize}
we already illustrated how such a variety produces a large spread of source physical
sizes corresponding to the angular extent of the compact sources, in turn implying
substantial consequences on their structural classification. Other source physical
properties, however, may turn out to be strongly biased by distance effects. A
systematic study of these effects will be published in \citet{bal17}. 
Here we limit ourselves to a few considerations related to this issue.

\subsection{Source confusion and classification}\label{confusion}
The first issue originates from the simple concept of perspective confusion of two
(or more) sources with a given physical separation, as their heliocentric distance
increase, so that their angular separation decreases accordingly. Keeping in mind
the classification of proto-stellar vs. starless sources introduced in
Section~\ref{protovsstarl}, one can imagine the basic case in which two sources,
quite close each other in the sky and belonging to these two different classes,
are virtually placed at an increasing heliocentric distance and ``re-observed'',
until they get confused at the \textit{Herschel} resolution. The 70~$\umu$m flux
determining the proto-stellar classification
of the former source would be, in this case, assigned to the new unresolved source
including also the original starless companion, making it globally flagged as
proto-stellar. On the one hand, this classification would remain true in principle,
as such a structure, in fact, would have a proto-stellar content. On the other
hand, the mass actually involved in the star formation activity would be
(even significantly) smaller than the value quoted for the entire clump mass.
This would result, at large distances, in an overestimate of the mass assigned 
to proto-stellar structures \citep[see also][]{hat08}.

To try to quantify this effect, we performed a simple test,
considering all the sources of our Hi-GAL catalogue located at $d<4$~kpc,
and ideally moved all of them at a larger distance $d\textsf{'}$, estimating the
corresponding decrease of their mutual angular separations. Given a pair of sources
$i$ and $j$, located at their original distances $d_i$ and $d_j$ and separated
in the sky by an angle $\varphi_0$, if they are virtually moved away to
a distance $d\textsf{'} > \max(d_i,d_j)$, the new simulated angular separation would become
\begin{equation}
\varphi\textsf{'}=\varphi_0\frac{\left(d_1+d_2\right)}{2 d\textsf{'}}\;.
\end{equation}
Probing virtual distances in steps of 1~kpc, starting from $d\textsf{'}=5$~kpc, and assuming that two
(or more) sources get confused when their mutual angular separation becomes smaller than the
SPIRE beam at $250~\umu$m (i.e. for $\varphi\textsf{'} < 18\arcsec$), a trend for the ratio
(both in number and in total mass) of sources classified as proto-stellar over the whole
source sample has been estimated.

\begin{figure}
\centering
\includegraphics[width=7.0cm]{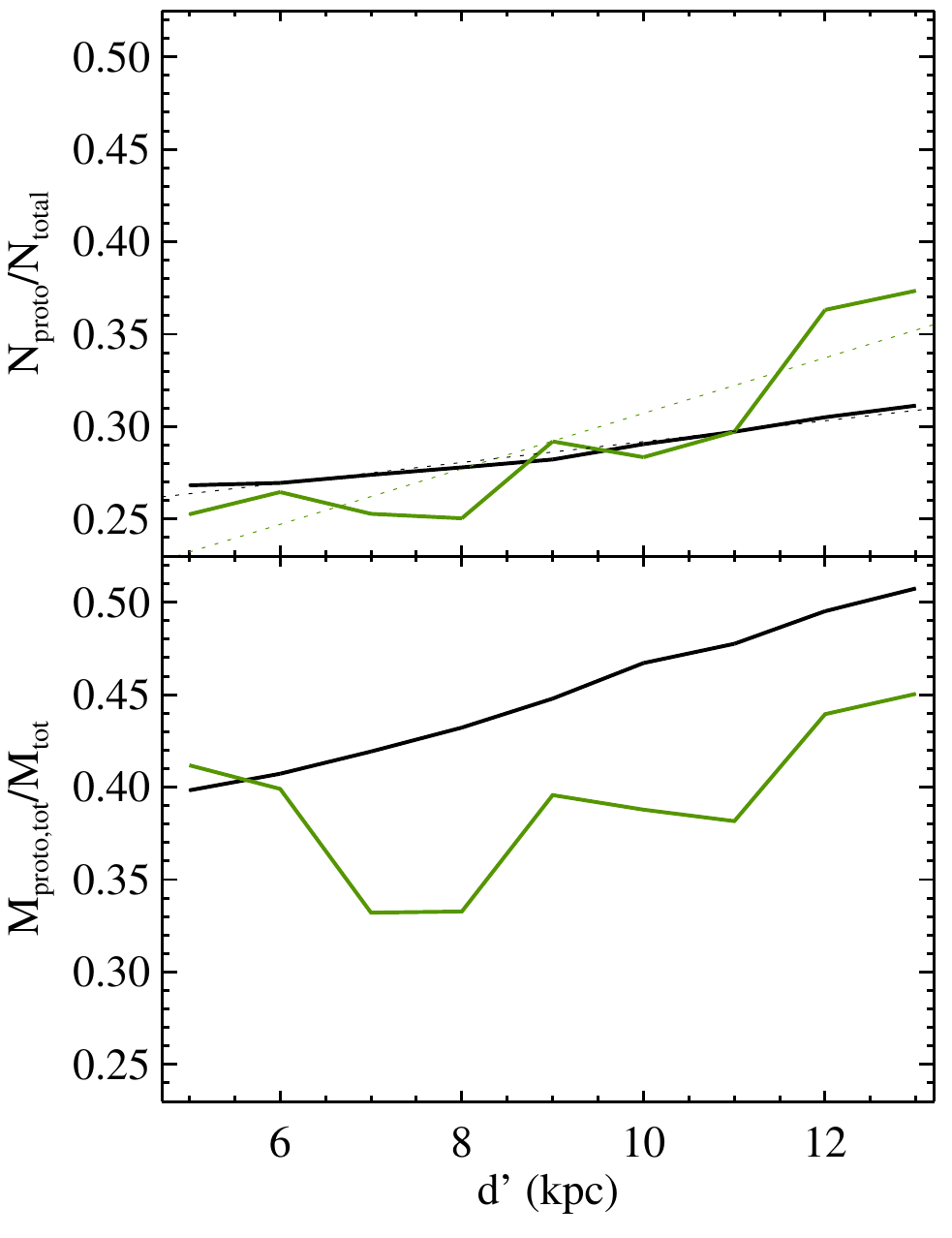}
\caption{Distance bias affecting the estimate of the global fraction of proto-stellar sources,
both in number (top) and mass
(bottom). All the Hi-GAL sources located within $d=4$~kpc have been virtually moved to larger
distances, starting from $d\textsf{'}=5$~kpc, and their mutual separation reevaluated accordingly.
Sources getting confused after this operation define a new unresolved source which assumes a
proto-stellar character if at least one of its original components was proto-stellar; the new
mass is simply calculated as the sum of the masses of the original sources. Top: black solid line,
fraction of proto-stellar sources over the total amount of sources, as a function of the new
simulated distance. Black dotted line: linear best fit of the previous one. Green solid line:
the same quantity calculated for the real Hi-GAL sources encountered at the considered distances
(in bins 1~kpc-wide, centered at multiples of 1~kpc). Green dotted line: linear best fit to the
previous one. Bottom: the solid lines are the same as in the top panel, but for the proto-stellar
fraction in mass.}\label{pretoproto}
\end{figure}

An increment of such fractions is clearly visible in both panels of Figure~\ref{pretoproto} 
(top for the number ratio, and bottom for the mass ratio, respectively).
A linear best-fit suggests a slope of $0.005$~kpc$^{-1}$ for the
proto-stellar fraction in number. This trend can be used to correct global properties
(as for instance the SFR) which sensitively depend on the estimate of the
proto-stellar population. Indeed, in Figure~\ref{pretoproto}, we displayed the same quantity
also for the real population of sources, i.e. the proto-stellar fraction for sources of our
catalogue encountered at the probed distances (i.e. located within bins of width 1~kpc, centered
on the various values of $d\textsf{'}$). The observed behaviour in this case is not as smooth as
in the simulated case, since it depends on peculiar environments encountered throughout
the plane in the considered distance bin. However, a generally increasing trend is found,
as expected, for the number ratio.

The observed trend of the mass fraction with the distance, instead, is not clearly increasing
(Figure~\ref{pretoproto}, bottom panel), suggesting that it can not be simply treated adding
up the masses of the single sources that are going to be merged, since the SED shape
(determined by the peak position) of the single sources and the one of the resulting merged source
can differ remarkably. Indeed, the resulting mass of the merged source would be the sum of the
original masses only if all the original SEDs corresponded to the same temperature.

\begin{figure}
\centering
\includegraphics[width=7.0cm]{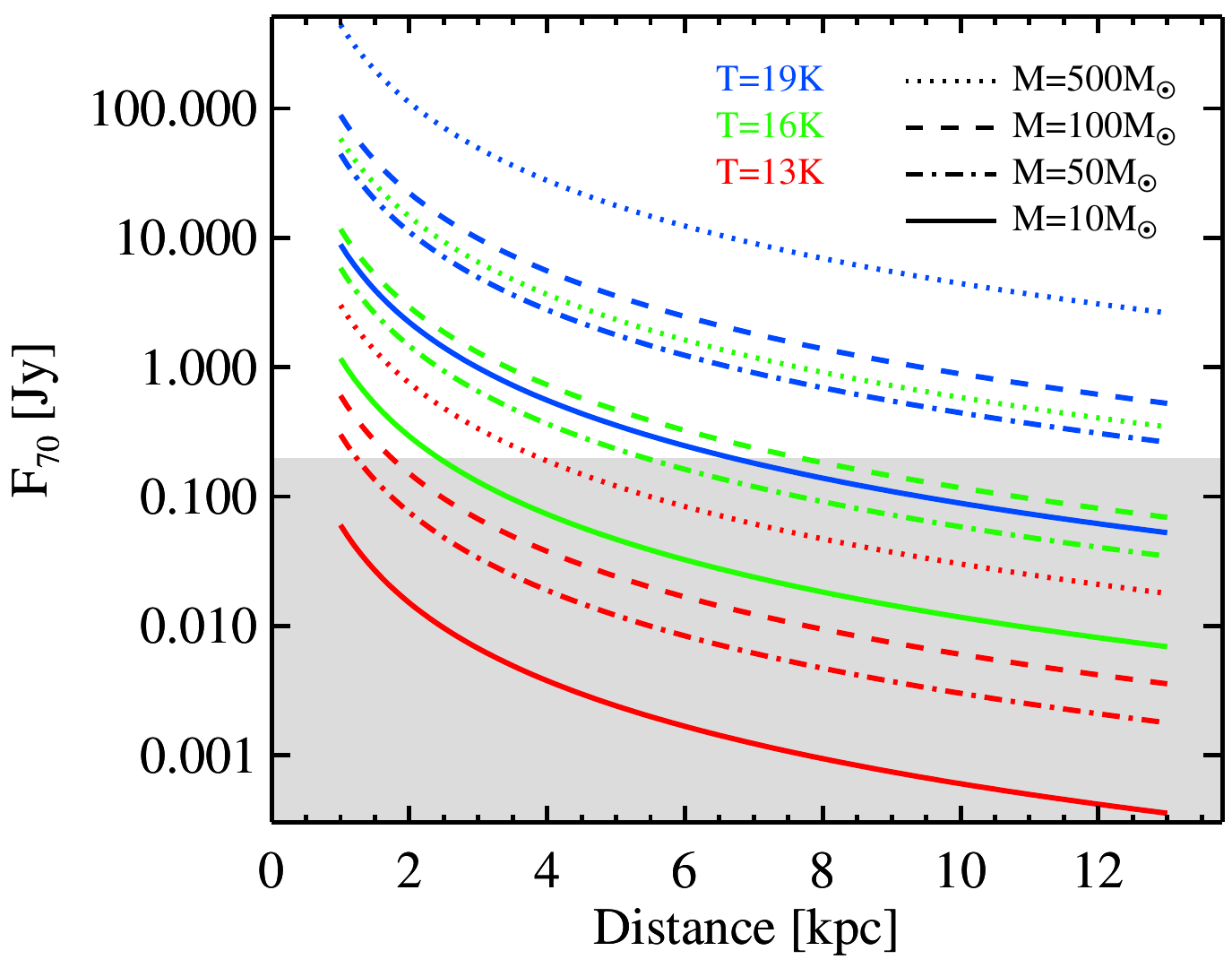}
\caption{Decrease of the grey-body flux at 70~$\umu$m (Equation~\ref{gbthin})
for a set of masses and temperatures (dust opacity parameters are those used in this paper),
as a function of heliocentric distance. The correspondence between the three probed temperatures and
line colours, and the one between the four probed masses and the line style are shown
in the legends in the upper part of the plot. The grey shaded area, bordered on the top
on the assumed PACS sensitivity limit of 0.2~Jy at 70~$\umu$m, corresponds to the
condition in which the grey body source is not expected to be detected with \textit{Herschel}
in Hi-GAL. Considering, for example, a real proto-stellar source emitting
a flux 10 times larger than the one expected from the grey body which best fits the SED at 
$\lambda \geq 160~\umu$m, the curves plotted for 100 and $500~M_\odot$ would assume
in this case the role of those plotted for 10 and $50~M_\odot$, respectively, to 
assess the detectabilityof such source.}\label{sens70}
\end{figure}

\begin{figure}
\centering
\includegraphics[width=7.0cm]{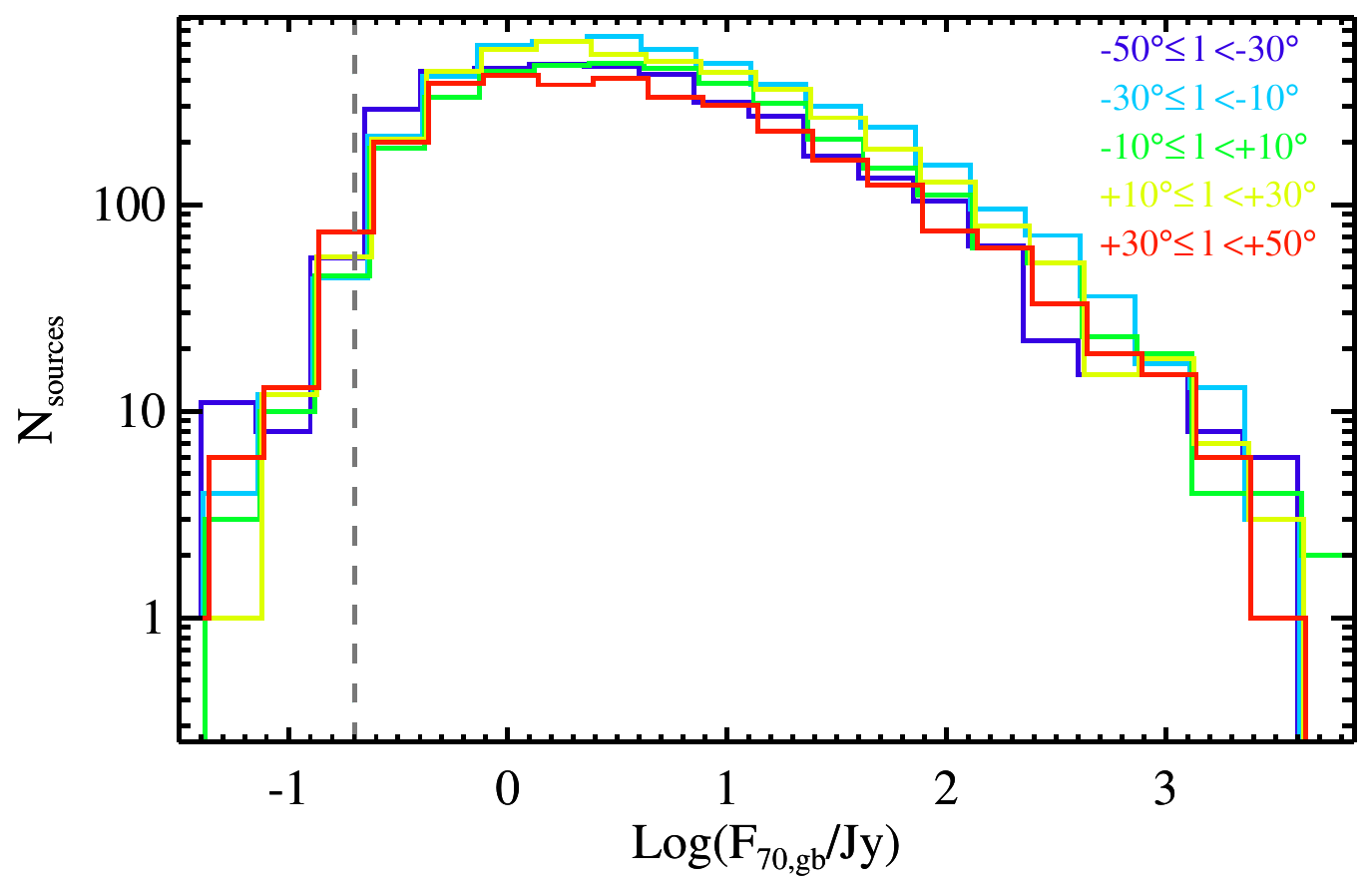}
\caption{Distributions of 70~$\umu$m fluxes used in this paper, obtained separately
over Galactic longitude bins of width $20^\circ$ (the bins are reported at the top right of
the plot, together with the bin-colour correspondence.). The assumed PACS sensitivity limit 
of 0.2~Jy at 70~$\umu$m is reported as a vertical grey dashed line.}\label{histo70}
\end{figure}

As already mentioned in Section~\ref{protovsstarl}, the confusion effect discussed above
is in competition with the possible inability of detecting emission at 70~$\umu$m from distant
and relatively small proto-stellar clumps, leading to their misclassification as starless sources. In 
Figure~\ref{sens70} we propose a simple exercise to show how the grey-body flux at 70~$\umu$m, obtained 
with Equation~\ref{gbthin} exploring a grid of values of mass and temperature and using the same 
dust parameters reported in Section~\ref{sedfit}, decreases as a function of source heliocentric 
distance. We compare these trends with the typical PACS sensitivity limit at this band found 
in our catalogue, which 
mildly depends on the line of sight and we assume to be 0.2~Jy, as indicated by Figure~\ref{proto70vs24}. 
To better investigate a possible dependence of this limit on the Galactic longitude 
\citep[although][have already shown that the PACS~70~$\umu$m band is the least affected by this effect 
among the \textit{Herschel} bands]{mol16a} 
in Figure~\ref{histo70} we plot the distributions of the 70~$\umu$m fluxes used in this 
paper, built for five $20^\circ$-wide different chunks of longitude. For all the five histograms, the
bin containing the 0.2~Jy flux is the smallest containing a statistically significant ($N>10$) number 
of sources. With respect to this value, in Figure~\ref{sens70} one sees that a grey body with 
$M\leq 50~M_\odot$ and $T \leq 16~$K could not be detected at $d \gtrsim 5$~kpc, while only more 
favourable parameter combinations (such as, for example, $M\geq 500~M_\odot$ and $T \geq16~K$, or
$M\geq 50~M_\odot$ and $T>19$~K) can remain detectable up to $d=13$~kpc, a value which is representative
of very far objects in our catalogue. Clearly, in the approach followed in this paper, consisting of
modeling the portion of the SED at $\lambda \geq 160~\umu$m with a grey body and the flux observed
at 70~$\umu$m ($F_\mathrm{70,obs}$) as a simple upper limit to better constrain the fit, this flux 
is expected to in excess with respect to the grey body ($F_\mathrm{70,gb}$) best-fitting the SED at 
the longer wavelengths. Obviously, this adds a further degree of freedom, so that, 
for example, one can review the curves plotted in Figure~\ref{sens70} for the case $M=500~M_\odot$
as those of a source with $M=50~M_\odot$ and $F_\mathrm{70,obs}/F_\mathrm{70,gb}=10$, and
so on (we estimate that, for sources present in our catalogue, $F_\mathrm{70,obs}/F_\mathrm{70,gb}<10$
is found in around half of the cases in which $F_\mathrm{70,obs}$ is available). In conclusion,
due to the difficulty in describing $F_\mathrm{70,obs}$ through a simple model, it is not possible
to predict accurately its value starting from the rest of the SED, nevertheless we expect that a 
minor but significant fraction of relatively low-mass (and/or low-temperature) clumps observed 
at large distances and flagged as starless sources are actually misclassified due a selection effect
un the flux at 70~$\umu$m.

\subsection{Mass completeness limits}\label{appmasslim}
\begin{figure}
\centering
\includegraphics[width=8.0cm]{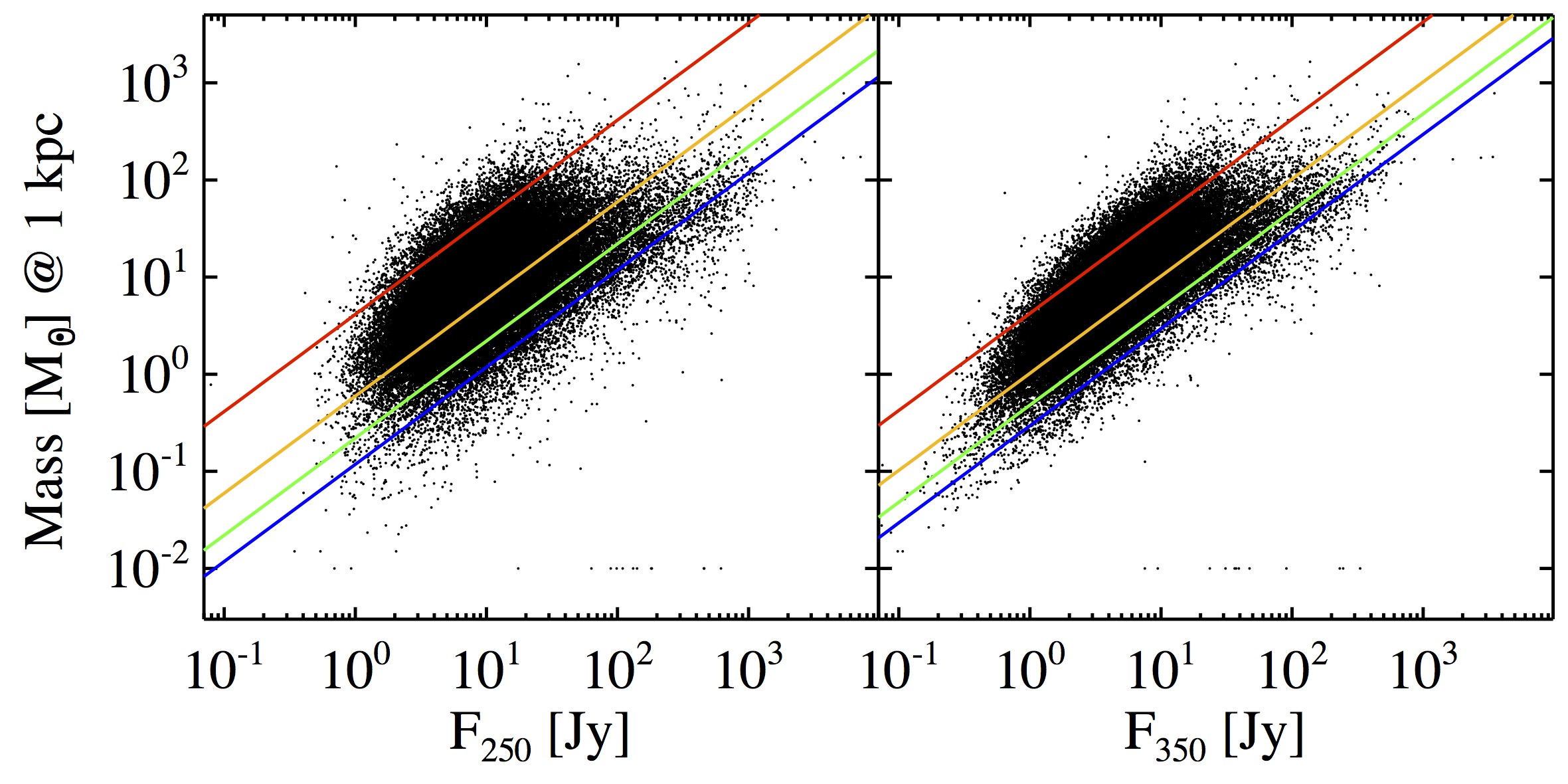}
\caption{\textit{Left}: relation grey body mass vs flux at $250~\umu$m for all the sources included 
in the Hi-GAL physical catalogue. The same relation for a grey body is represented with solid lines 
for four different temperatures: 10 (red), 15 (orange), 20 (green), and 25~K (blue), respectively. 
\textit{Right}: the same as in the left panel, but for the flux at $350~\umu$m.}
\label{mass350}
\end{figure}

A discussion of the mass completeness limits for the Hi-GAL sources requires as a first step
to identify the most suitable band for estimating masses. Intuitively,
wavelengths at which the emission is optically thin so that the integrated flux can be considered
proportional to the amount of matter in the clump, should be taken into account. The 500~$\umu$m
band might be the preferred one, but according to the constraints stated in Section~\ref{filtering},
the flux at this band may be not present in all the considered SEDs, while fluxes at 250 and
350~$\umu$m are always present by construction. In Figure~\ref{mass350} we show the relation between
the masses of the catalogue sources and their fluxes at these two bands; to remove the dependence on
the distance, the masses have been scaled to the same ``virtual'' distance $d_v=1$~kpc, through a
factor $(d_v/d)^2$ (or, for sources having no kinematic distance estimate, imposing their distance
to be just $d_v$). Of course, from the analytic point of view such relation is expected to depend
on the temperature, so that also lines corresponding to grey bodies at different temperatures and
$\beta=2$ are overplotted. As expected, the level of spread is smaller (than a tighter correlation 
is found) in the mass vs $F_{350}$ plot, which we adopt hereafter for the following analysis of the 
completeness limit.

\citet{mol16a} provided 90\% completeness limits for their photometric catalogues at all the five 
Hi-GAL bands, subdivided by tile. Thus, given a completeness limit $F{_\mathrm{compl,350}}$ at 
350~$\umu$m (as a function of the considered tile, so, roughly, of the Galactic longitude), the 
90\% mass completeness limit will depend on the source temperature (according to the grey body 
law) and on its distance (being $M_{\mathrm{compl}}$ proportional to $d^2$). Collapsing all the 
involved constants and unit conversion factors, such dependency can be condensed as
\begin{equation}\label{mcomp350}
\frac{M_{\mathrm{compl}}(T,d)}{\mathrm{M_{\odot}}}=0.0705\: \left(\frac{F{_\mathrm{compl,350}}}{\mathrm{Jy}}\right)\: \left(\frac{d}{\mathrm{kpc}}\right)^2\: \left(\exp{\left(\frac{41.1094}{\left(T/\mathrm{K}\right)}\right)-1}\right) 	\; .
\end{equation}

\begin{figure}
\centering
\includegraphics[width=8.5cm]{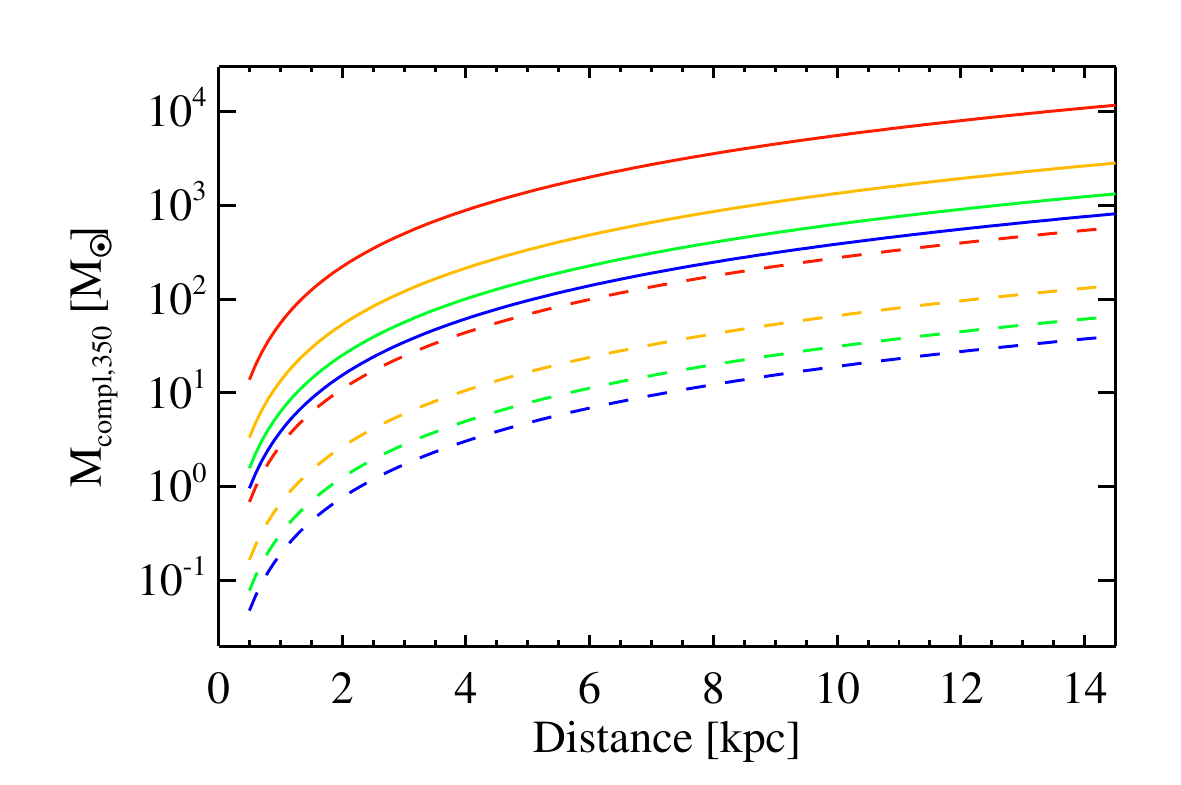}
\caption{Curves of the 90\% mass completeness limit based on the photometry at 350~$\umu$m, 
as a function of the source distance, according to Equation~\ref{mcomp350}. Two values of 
$F{_\mathrm{compl,350}}$ are probed (see text): solid and dashed curves correspond, 
to 13.08 and 0.65~Jy, respectively. Finally, four values of the temperature parameters 
are probed, namely the same as in Figure~\ref{mass350}, represented using the same colour 
encoding.}
\label{mcomp}
\end{figure}

In Figure~\ref{mcomp}, the behaviour of the mass completeness limit based on Equation~\ref{mcomp350}
is reported, showing how it varies as a function of the distance for two values of
$F{_\mathrm{compl,350}}$, namely the maximum and the minimum ones found by \citet{mol16a} 
in their Hi-GAL completeness analysis (13.08 and 0.65~Jy, respectively), and for four 
temperatures between 10 and 25~K. Using the Hi-GAL physical catalogue, therefore, will require
to take into account such mass completeness limits.

\section{Clump property consistency with respect to previous surveys}\label{champepos}
A blind comparison between our clump property catalogue with the results of previous 
surveys of the Galactic ISM can turn out to be misleading if no attention is paid to the assumptions
such catalogues are built on. The main one is the definition itself of the source typology, 
in turn depending on the characteristics of the exploited observations (wavelength domain, tracer, 
resolution, sensitivity, etc.).

We reaffirm here that this paper deals with the properties of \textit{Herschel} compact sources (i.e.
with an angular extent no larger than a few tens of arcsecs) whose SEDs are eligible for modified
black body fit at $\lambda \geq 160~\umu$m. Depending on the distance, these sources can correspond
to pre- or proto-stellar cores, or (in the most likely case) to larger clumps, or even to entire 
clouds (Section~\ref{physsize}). The wavelength range we consider for the grey-body fit allows us 
to obtain global/average properties of the cold envelope component of the sources. Instead, for 
obtaining the bolometric luminosity we consider also fluxes at shorter wavelengths, to take
into account the emission from the possible proto-stellar content of the Hi-GAL source. We finally
recall that, through appropriate assumptions and operations (flux scaling), we estimate the source
properties as referred to a volume corresponding to the deconvolved source size observed 
at~$250~\umu$m.

An interesting comparison can be carried out with the data coming from the Census of High- and
Medium-mass Protostars (CHaMP) survey \citep{bar11,ma13}. The 303 CHaMP sources were identified
in HCO$^+$(1-0) Mopra observations in the coordinate range $280^{\circ} < \ell < 300^{\circ}$,
$-4^{\circ} < b < 2^{\circ}$ \citep{bar11}. Masses were estimated from HCO$^+$(1-0), while
collecting \textit{MSX}, \textit{IRAS}, and SIMBA-SEST photometry SEDs were built to estimate
the bolometric luminosity and, at $\lambda \geq 60~\umu$m, the temperature of the cold envelope
\citep{ma13}.

\begin{figure*}
\centering
\includegraphics[width=15.0cm]{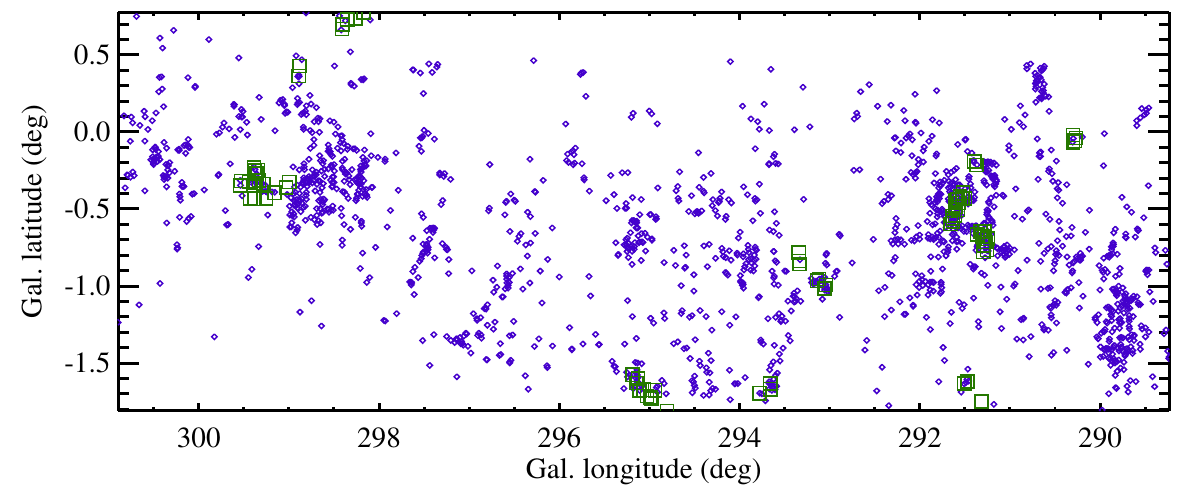}
\caption{Spatial dispposition of Hi-GAL proto-stellar sources (blue diamonds) and CHaMP ones
(green squares), respectively, in the common surveyed area.}
\label{champ_pos}
\end{figure*}

The most striking difference between the CHaMP and Hi-GAL source global behaviour lies in the
statistics of $L_{\mathrm{bol}}/M$ \citep[][their Figure~2, $d$, and our Figure~\ref{lmratio},
respectively]{ma13}. In their case, $L_{\mathrm{bol}}/M$ distribution peaks at few tens
$L_{\odot}/M_{\odot}$, while in our case, for the proto-stellar population, the peak position 
is found to be smaller of about one order of magnitude.
To check this discrepancy, we searched for the Hi-GAL proto-stellar sources having a positional 
match with the ones of the CHaMP catalogue. In the common surveyed area, we found 6513 and 77 
sources for Hi-GAL and CHaMP, respectively, with 69 matches within a searching radius of 
$2\arcmin$. Cases of multiple
associations were resolved keeping only the Hi-GAL closest counterpart of each CHaMP source. In
this way, distance-independent quantities such as $T$ and $L_{\mathrm{bol}}/M$ can be directly
compared. Instead $M$ and $L_{\mathrm{bol}}$ depend on the distance adopted in the two different
surveys, so that the best way to compare them is to rescale these quantities to report them to 
the same virtual distance $d_v$, similarly to the procedure adopted in the previous section.

\begin{figure*}
\centering
\includegraphics[width=15.0cm]{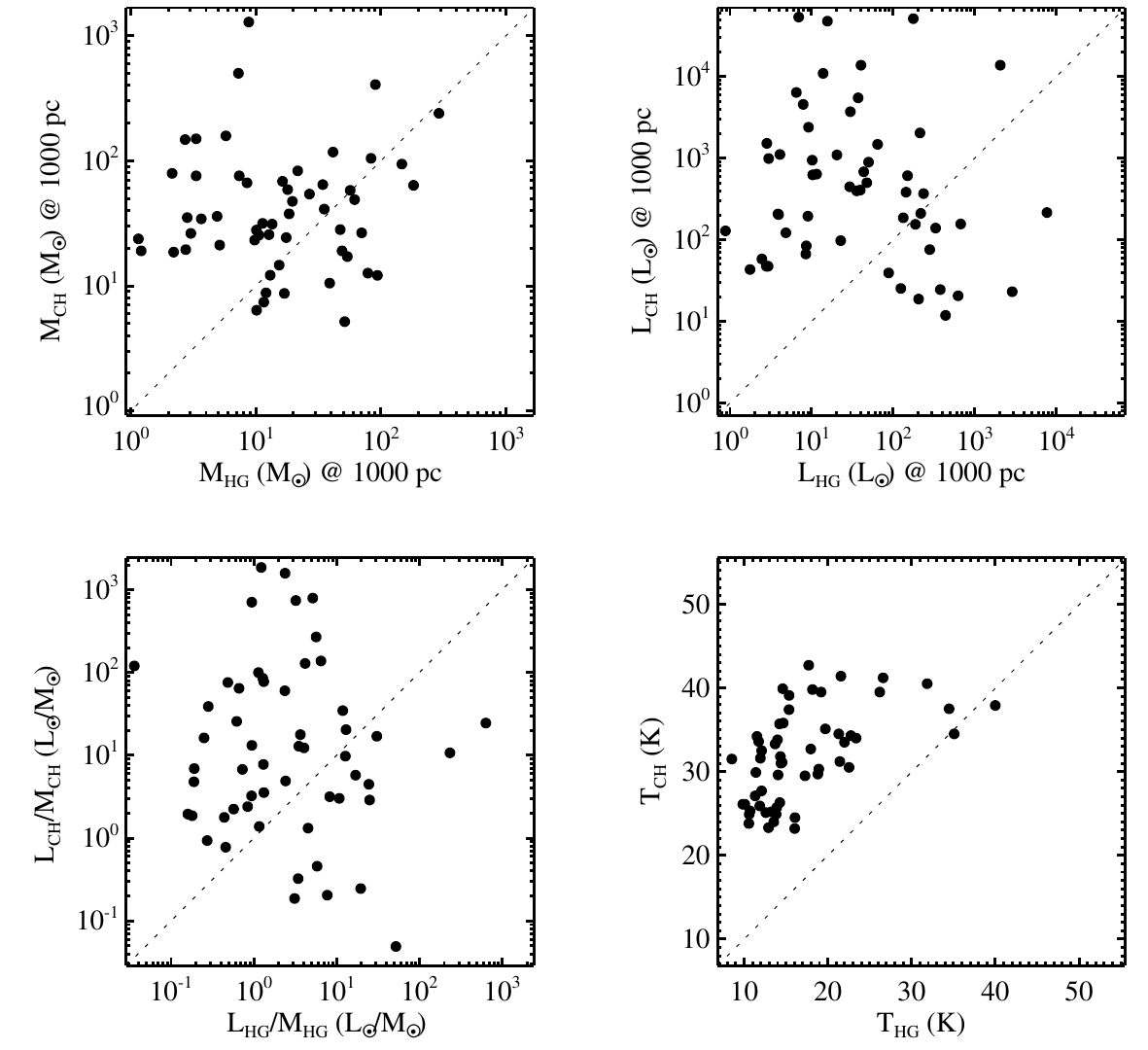}
\caption{Comparison of physical properties derived for Hi-GAL and CHaMP compact sources.
\textit{Top, left}: comparison of masses for sources found in both surveys; masses are scaled
to a common distance of 1~kpc to allow unbiased comparison. \textit{Top, right}: the same as
in the previous panel, but for bolometric luminosities. \textit{Bottom, left}: comparison of
$L_{\mathrm{bol}}/M$ ratios (distance-independent) for the sources of the two previous panels.
\textit{Bottom, right}: the same as in the previous panel, but for temperatures. In all panels 
the bisector corresponding to the 1:1 relation is represented as a dotted line.}
\label{champ_props}
\end{figure*}

Figures~\ref{champ_pos} and~\ref{champ_props} illustrate a comparison of the two surveys, in
the overlap area. In the former, the angular disposition of the sources is shown, while in
the latter masses and luminosities, scaled at $d_v=1$~kpc, are compared (top left and right panels,
respectively). Both quantities show a wide spread around the 1:1 relation. While masses were
determined using different methods, the differences in luminosity can be deduced from the different
way the SEDs were built: on the one hand CHaMP uses \textit{IRAS} fluxes, which are expected to come
from an area in the sky remarkably larger than that of the typical Hi-GAL sources (so the resulting
luminosity tends to be significantly larger), on the other hand such CHaMP SEDs do not cover the
crucial range $100-500~\umu$m in which emission from cold dust peaks, thus potentially neglecting
part of the clump FIR luminosity. The first effect seems to be prevailing in the majority of cases,
being the median of the $L_{\mathrm{bol,CH}}/L_{\mathrm{bol,HG}} \sim 16$. Furthermore, since the
FIR portion of the CHaMP SEDs generally peaks at shorter wavelengths than Hi-GAL SEDs, also the
envelope temperatures are systematically found to be higher in the former case, as clearly shown
in the right lower panel of Figure~\ref{champ_props}. All these contributions lead to shift the
CHaMP distribution of $L_{\mathrm{bol}}/M$ compared with Hi-GAL (left lower panel), so that the
median value for the ratio between $L_{\mathrm{bol}}/M$ of CHaMP and Hi-GAL is $\sim 4$.

\begin{figure*}
\centering
\includegraphics[width=16.5cm]{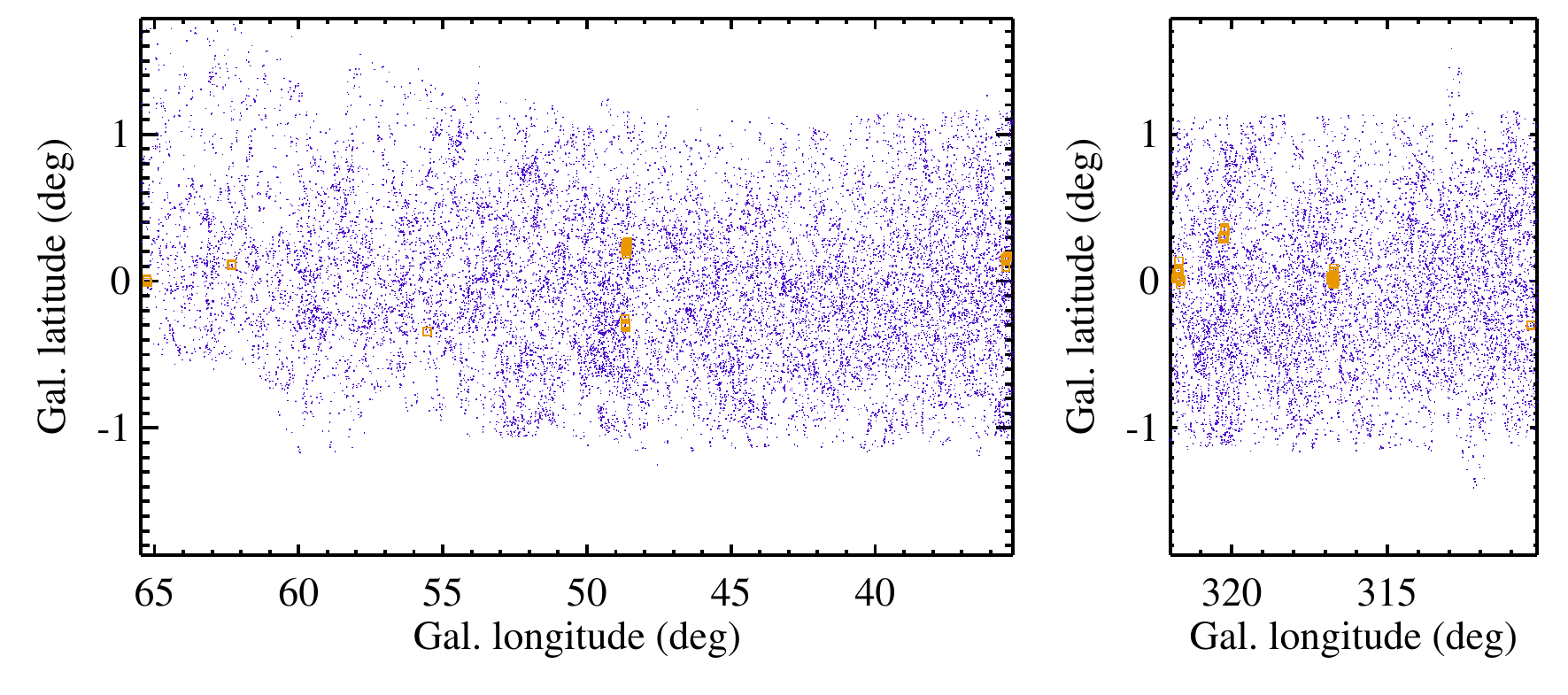}
\caption{
Spatial disposition of Hi-GAL (blue dots) and EPoS (gold squares) sources, respectively, in the
common surveyed areas in the first (\textit{left}) and the fourth (\textit{right}) Galactic
quadrant, respectively.}
\label{epos_pos}
\end{figure*}

A more direct comparison can be carried out with first results of the \textit{Herschel} key-program
The Earliest Phases of Star formation \citep[EPoS,][]{rag12}. This program consisted in a PACS and
SPIRE photometric mapping survey of objects known to be in the cold early phases of star
formation. 60 targets were observed, 45 of which corresponding to high-mass star forming regions,
in which \citet{rag12} found 496 compact sources. Out of these, 90 lie in the portion of
the Galactic plane considered in this paper, (43 in the first quadrant and 47 in the fourth one,
Figure~\ref{epos_pos} $a$ and $b$, respectively). The sources were detected only in PACS images at
70, 100, and 160~$\umu$m, respectively. SPIRE maps were not used either for detecting counterparts
or, even more so, for extracting photometry at these wavelengths. This prevents an exact match
between the 90 EPoS objects and the entries of our Hi-GAL physical catalogue: first, in several
cases clusters of PACS EPoS sources might correspond to a single SPIRE counterpart due to
different resolution and, second, possible Hi-GAL equivalents of EPoS sources might not
survive the selection process described in \ref{filtering} and based on SED regularity between
160 and 500~$\umu$m. In the end, we found 50 matches between the list of \citet{rag12} and our Hi-GAL
physical catalogue, within a searching radius of 1$\arcmin$.

\begin{figure}
\centering
\includegraphics[width=9.0cm]{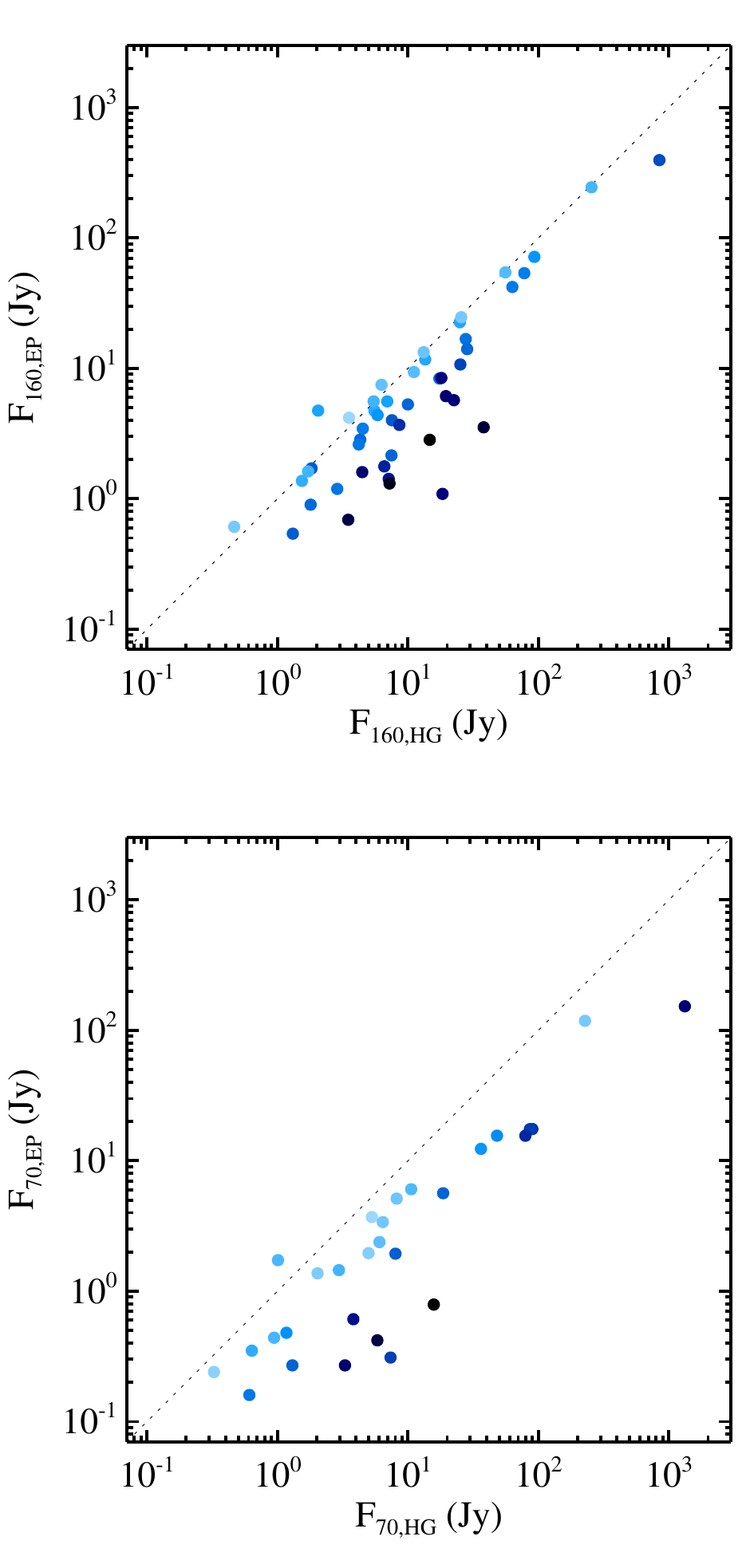}
\caption{Comparison of EPoS vs Hi-GAL integrated fluxes at 70 (\textit{left}) and 160~$\umu$m
(\textit{right}), respectively. Different shades of blue are used to represent the size of the Hi-GAL
source (estimated as the average FWHM of 2-D Gaussian resulting from the fit performed by CuTEX), to
highlight that larger flux discrepancies correspond to larger estimates of the Hi-GAL source sizes:
the lightest blue level corresponds to the smallest sizes found at 70 and 160~$\umu$m (9$\arcsec$
and 12.4$\arcsec$, respectively), while black correspond to the largest sizes (23.7$\arcsec$ and
40.8$\arcsec$, respectively).}
\label{epos_fluxes}
\end{figure}

For these sources, the fluxes at 70 and 160~$\umu$m can be directly compared,
 (Figure~\ref{epos_fluxes}, \textit{left} and \textit{right}, respectively)\footnote{The number 
 of comparable sources is actually smaller than 50, due to possible lack of a flux at 
 70 or at 160~$\umu$m in the Hi-GAL SED.}. 
 In both cases, the Hi-GAL ones appear generally
overestimated with respect to the EPoS ones. The most general reason of this discrepancy has to be
searched in the different way the fluxes were extracted in the two cases. \citet{rag12} carried out
PSF photometry, implicitly assuming a point-like appearance of the sources, while in this work
\textit{compact sources} were extracted with CuTEx, thus considering that source sizes can extend up
to a few PSFs and consequently measuring larger total fluxes over larger areas. This aspect is
emphasised in Figure~\ref{epos_fluxes}, using different colours for denoting the different source sizes
of the Hi-GAL sources; as a general trend the photometric data of the two surveys depart from 
equality as the source size estimates in Hi-GAL depart
from the PSF extent\footnote{Due to PACS on-board coadding, in \textit{Herschel} parallel mode 
at $60\arcsec$~s$^{-1}$ the PSFs turn out to be elongated along the scan direction 
\citep[see, e.g.,][]{mol16a}, being $5.9\arcsec \times 12.1\arcsec$ at 70~$\umu$m and $11.6\arcsec\times
15.7\arcsec$ at 160~$\umu$m (PACS Observer's Manual, v.2.5.1), which correspond to a circularised FWHM of
8.4$\arcsec$ and 13.5$\arcsec$, respectively.}. 


\begin{figure*}
\centering
\includegraphics[width=15cm]{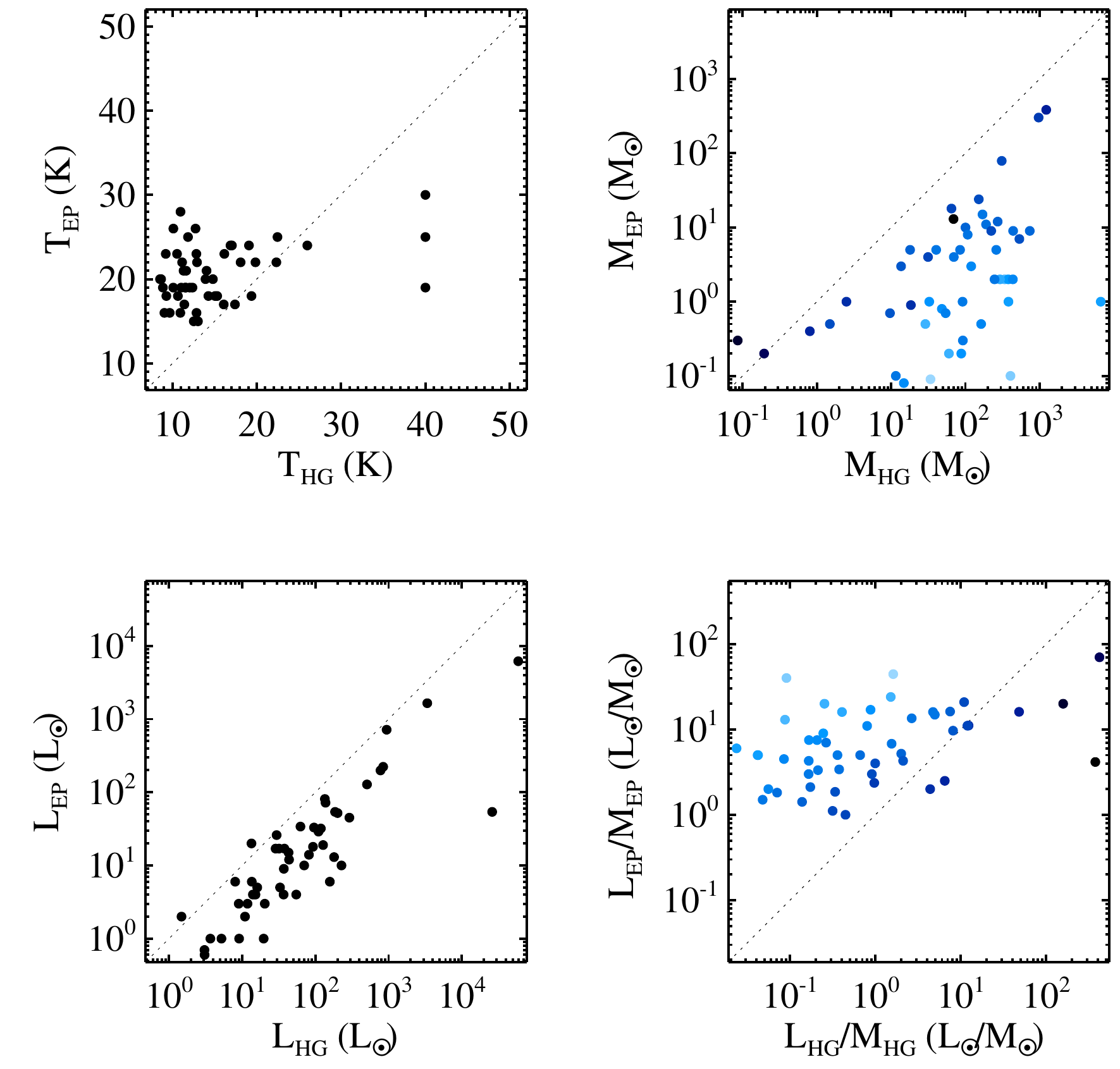}
\caption{Comparison of physical properties derived for Hi-GAL and EPoS \citep{rag12} compact
sources. \textit{Top, left}: comparison of temperatures for sources found in both surveys.
\textit{Top, right}: the same as in the previous panel, but for masses scaled to a common virtual
distance of 1~kpc to allow unbiased comparison. Different shades of blue are used for the symbols to 
represent the $\Delta T=T_\mathrm{HG}-T_\mathrm{EP}$ temperature discrepancy between Hi-GAL and EPoS for 
each source, going from 21~K (black) to -17~K (light blue). \textit{Bottom, left}: The same as in top 
left panel, but for bolometric luminosities scaled to a common virtual distance of 1~kpc. \textit{Bottom,
right}: comparison of $L_{\mathrm{bol}}/M$ ratios for the sources of the previous panels. Different
shades of blue are used as in the top right panel. Finally, in all panels the bisector corresponding
to the 1:1 relation is represented as a dotted line.}
\label{epos_props}
\end{figure*}

Photometric discrepancies are among possible explanations of further differences found comparing 
physical properties of the two source lists, analyzed in Figure~\ref{epos_props}. EPoS source 
temperatures are found to be overestimated compared with those in Hi-GAL (top left panel). This is 
expected due to the different wavelength range explored and considered for the grey body fit, being 
the latter more suited to trace the peak of the cold dust \citep[see a similar discussion in][]{fon05}. This is
also the main reason of generally finding lower mass estimates in EPoS than in Hi-GAL (after
rescaling both at a virtual distance $d_v=1$~kpc), as shown in Figure~\ref{epos_props}, top right panel: 
as a general trend, the larger is the temperature discrepancy, the larger is consequently the mass
discrepancy. EPoS bolometric luminosities are generally smaller than Hi-GAL ones (bottom left
panel), being the median ratio of the two equal to $\sim 0.26$. This is due to the wider
spectral coverage of the SEDs in this paper, potentially going from 21~$\umu$m to 1.1~mm, and also
to the lower fluxes measured in EPoS at 70 and 160~$\umu$m, as described above. Finally, combining
the information contained in the two previous panels, in the bottom right panel we show the
comparison of the $L_{\mathrm{bol}}/M$ ratio for the two surveys: this ratio is generally larger
for EPoS despite the larger luminosities found in Hi-GAL since masses at the denominator
overcompensate for that. Furthermore, largest discrepancies correspond to largest temperature
discrepancies.

\begin{figure}
\centering
\includegraphics[width=7.0cm]{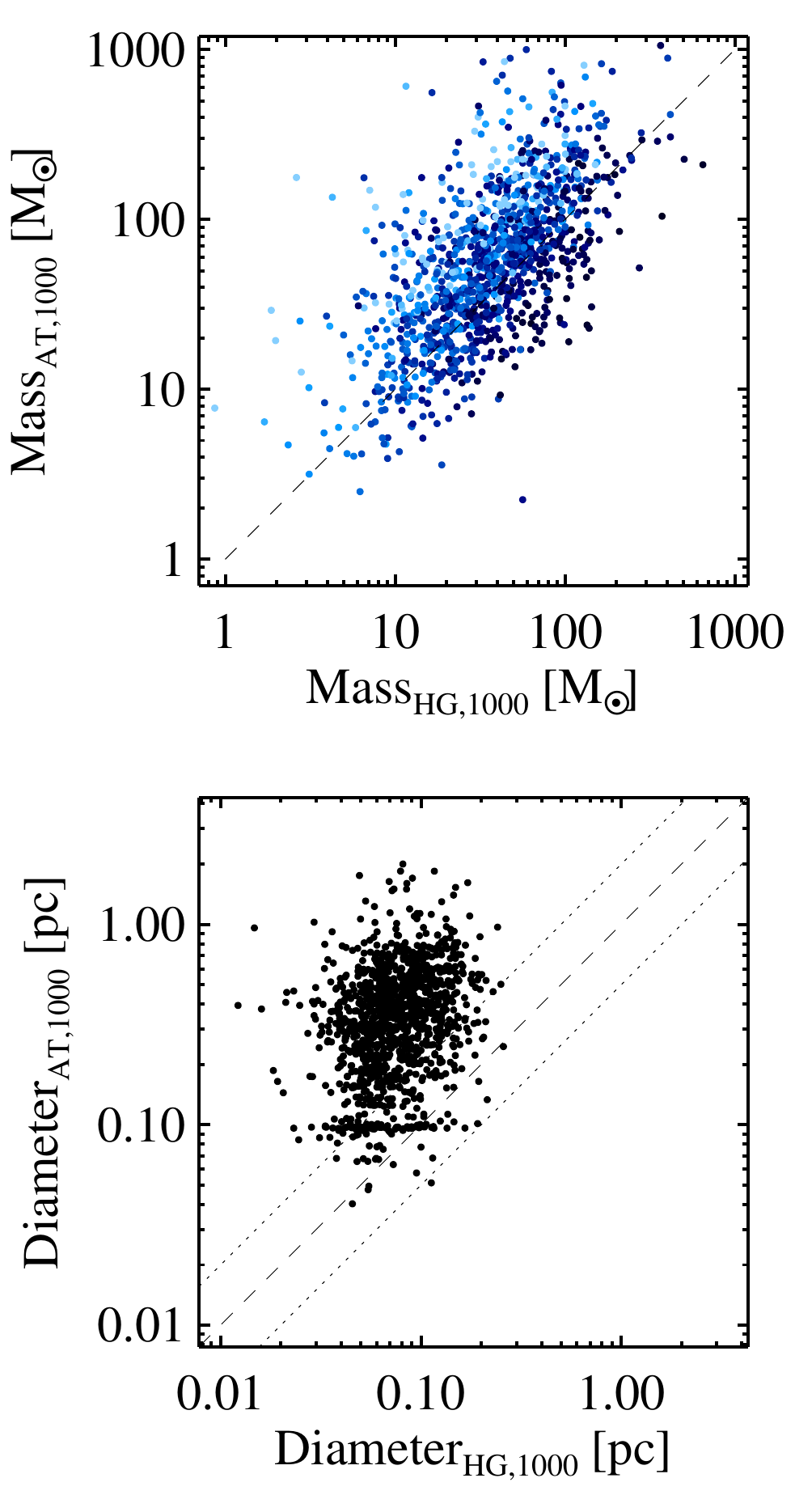}
\caption{Comparison of physical properties derived for Hi-GAL and ATLASGAL \citep{wie15} compact
sources, re-scaled to a common virtual distance $d_v=1$~kpc. \textit{Top}: comparison of masses for 
sources found in both surveys. Different shades of blue are used for the symbols to 
represent the Hi-GAL temperature going from 5~K (black) to 30~K (light blue). The grey dashed line
represents the 1:1 relation. \textit{Bottom}: comparison of diameters, re-scaled to $d_v$. 
The grey dashed line represents the 1:1 relation, while the dotted lines correspond to the 
ratios 2 and 0.5, respectively.}
\label{atlascompare}
\end{figure}

The above comparison shows that both the considered spectral range and adopted photometry 
strategy can lead to quite
different interpretation of the physical and evolutionary status of the same source. This must be
kept in mind in discussing the results presented in this paper as well as in comparing the present 
Hi-GAL catalogue with other catalogues of similar sources. Particular care, for instance, 
has to be taken comparing Hi-GAL clump properties with those derived for two single-band 
sub-millimetre surveys, namely ATLASGAL and BGPS, whose data have been also used in this work.
Starting with the former, we compare Hi-GAL clump masses derived in this work with those derived using only 
the ATLASGAL flux at 870~$\umu$m \citep{wie15} Notice that in this last work the masses are based on the fluxes 
of \citet{con13} and \citet{urq14b} typically much larger than those in the catalogue of \citet{cse14}, 
due to different estimate of the source size. As a consequence of this, we could not establish an 
immediate association between Hi-GAL and ATLASGAL sources of \citet{wie15} simply exploiting the ATLASGAL counterparts already quoted in our catalogue \citep[which instead are taken from][]{cse14}, but we had to 
search for positional matching within a searching radius of 36\arcsec, i.e. the centroid position 
accuracy provided by \citet{wie15}. In Figure~\ref{atlascompare}, top panel, the comparison of masses 
is shown; to make it meaningful, the masses of both lists have been re-scaled to a virtual distance
of 1~kpc, as also done in the previous comparisons with other surveys. Two main aspects are evident in the plot: 
$i$) a certain degree of spread around a linear relation
is present, mostly due to the spread in Hi-GAL temperature, as opposed to the fact that masses
of \citet{wie15} are obtained for a fixed temperature (namely 23.1 and 20.8~K for the fourth and
the first Galactic quadrant, respectively), as highlighted by the colour scale adopted to represent 
the Hi-GAL temperature; $ii$) a remarkable departure of this trend from a 1:1 behaviour, in which the 
masses of \citet{wie15} are larger than the Hi-GAL ones. This is essentially related to the aforementioned 
discrepancy between source sizes. In Figure~\ref{atlascompare}, bottom panel, it can be seen how re-scaled
diameters (or, equivalently, angular sizes) of sources determined by \citet{wie15} are systematically 
and significantly larger than those observed for Hi-GAL sources, so that the fluxes (and, correspondingly 
the masses) are evaluated over larger areas of the sky.

\begin{figure}
\centering
\includegraphics[width=7.0cm]{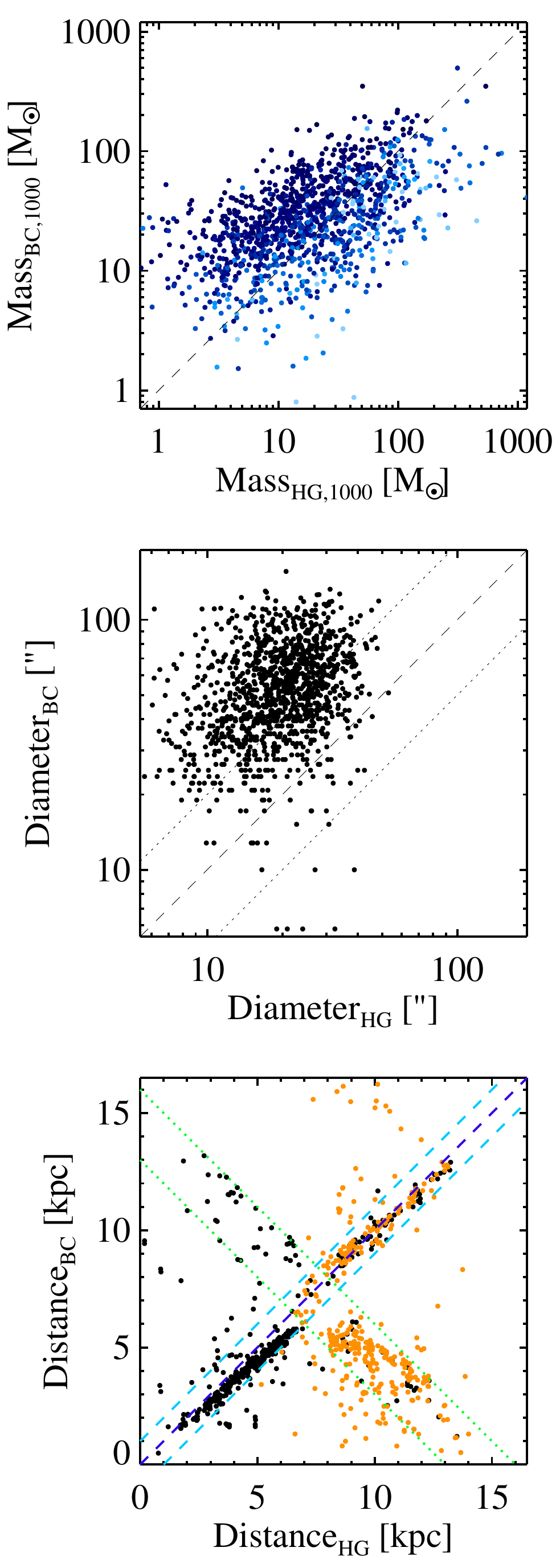}
\caption{\textit{Top, centre}: the same as top and bottom panels of Figure~\ref{atlascompare}, 
respectively, but for BGPS sources \citep{svo16}. 
In the top panel angular beam-deconvolved diameters are compared, instead of linear sizes,
this information being directly available in the catalogue of \citet{svo16}. \textit{Bottom}:
Comparison of heliocentric BGPS and Hi-GAL distances: 
black dots are sources for which the far/near Hi-GAL distance ambiguity
has been fully solved in this work, orange dots for the less reliable
cases in which the far distance has been assumed as final decision due to lack of other indicators. 
The bisector corresponding to the 1:1 relation is represented as a dark blue dashed line;
two further light blue dashed lines are plotted 1~kpc above and beolow it, respectively. 
Two green dotted lines represent the region populated by sources for which the near/far 
distance ambiguity has been solved in opposite ways in the two catalogues.}
\label{bgpscompare}
\end{figure}

In a similar way it is possible to make a comparison of masses derived in this paper and those
of of a sample of massive BGPS sources studied in \citet{svo16}. In this case, the BGPS source designation 
used is the same as in the BGPS catalogue \citep[i.e.][]{gin13}. 
Similarly to the ATLASGAL case, 
the majority of BGPS masses appear in excess compared to Hi-GAL (Figure~\ref{bgpscompare}, 
top), typically as a consequence of a larger angular size assigned to the source (Figure~\ref{bgpscompare}, 
center). Again, the spread in mass differences appears mostly due to the spread of Hi-GAL source grey body 
temperatures.

This comparative analysis with BGPS parameters offers us the opportunity of testing the 
heliocentric distances used in our catalogue ($d_\mathrm{HG}$). The comparison with BGPS 
distances ($d_\mathrm{BC}$), derived by \citet{ell13,ell15}, is shown in the bottom panel of 
Figure~\ref{bgpscompare}. A significant number of points is located close to the bisector
(\nbisector~within $\lvert d_{\mathrm{BC}} - d_{\mathrm{HG}} \rvert < 1$~kpc, out of 
\totplotted~plotted points), 
demonstrating a good agreement between the two methods in these cases. Using a different colour 
for sources of our catalogue for which the near/far distance ambiguity is not solved (so that
the far distance is chosen by default and a bad quality flag is assigned, see 
Section~\ref{dist_sect}), it can be seen that in several cases the validity of such 
choice is supported by a good agreement with a BGPS distance. Departures from the 1:1 behaviour,
following peculiar trends (e.g. the one at 5~kpc$\lesssim d_\mathrm{HG} \lesssim$~7~kpc) are ascribable
to different Galactic rotation models adopted. Another trend is clearly visible along the ``orthogonal'' 
direction, easily explainable with cases in which the near distance estimate has been assigned in one
of the two catalogues, and the far distance in the other. This trend is not as tight as the one 
around the bisector: delimiting by eye the region populated by sources following this trend (see
figure), and neglecting the area corresponding to the intersection with the 
$\pm 1$~kpc-belt around the bisector, we find \northog~sources characterised by this discrepancy. 
Noticeably, in the lower part of this area, corresponding to sources with an assigned far
distance in our catalogue, such assignment is flagged in most cases as ``bad quality''. Finally, 
a total of \remainingdists~sources (i.e. the \remainingperc\% of the sources considered for this test) 
remain out of these two main trends, and correspond to distances that can not be reconciled simply 
changing the near/far decision.


\bsp	

\clearpage

\noindent
Author afilliations\\
\iaps INAF-IAPS, via del Fosso del Cavaliere 100, 00133 Roma, Italy\\
\stsi Space Telescope Science Institute, 3700 San Martin Dr., Baltimore, MD, 21218, USA\\
\liver Astrophysics Research Institute, Liverpool John Moores University, Liverpool Science Park Ic2, 146 Brownlow Hill, Liverpool, L3 5RF, UK\\
\lamm Aix Marseille Univ., CNRS, LAM, Laboratoire d'Astrophysique de Marseille, Marseille, France\\
\mpia Max-Planck Institute for Astronomy, K\"{o}nigstuhl 17, D-69117 Heidelberg, Germany\\
\calte Infrared Processing Analysis Center, California Institute of Technology, 770 South Wilson Ave., Pasadena, CA 91125, USA\\
\unile Dipartimento di Matematica e Fisica, Universit\`a del Salento, 73100, Lecce, Italy\\
\cnrs CNRS, IRAP, 9 Av. colonel Roche, BP 44346, F-31028 Toulouse cedex 4, France\\
\unitou Universit\'e de Toulouse, UPS-OMP, IRAP, F-31028 Toulouse cedex 4, France\\
\nagoya Department of Physics, Nagoya University, Chikusa-ku, Nagoya, Aichi 464-8601, Japan\\
\calg Department of Physics \& Astronomy, University of Calgary, AB, T2N 1N4, Canada\\
\colo Center for Astrophysics and Space Astronomy, University of Colorado, Boulder, CO, 80309, USA\\
\toron Canadian Institute for Theoretical Astrophysics, University of Toronto, McLennan Physical Laboratories, 60 St. George Street, Toronto, Ontario, Canada\\
\leeds School of Physics and Astronomy, University of Leeds, Leeds LS2 9JT, UK\\
\cardiff School of Physics and Astronomy, Cardiff University, Cardiff CF24 3AA, Wales, UK\\
\greno IPAG, University Grenoble Alpes, 38000, Grenoble, France\\
\cea AIM Paris-Saclay, CEA/IRFU - CNRS/INSU - Univ. Paris Diderot, Service d'Astrophysique, CEA-Saclay, 91191 Gif-sur-Yvette Cedex, France\\
\oafi INAF, Osservatorio Astrofisico di Arcetri, Largo E. Fermi 5, 50125, Firenze, Italy\\
\koln I. Physik. Institut, University of Cologne, Z\"ulpicher Strasse, 50937, K\"oln, Germany\\
\eso European Southern Observatory, Karl Schwarzschild str. 2, 85748, Garching, Germany\\
\mpir Max-Planck-Institut f\"ur Radioastronomie, Auf dem H\"ugel 69, 53121 Bonn, Germany\\
\asdc ASI Science Data centre, 00044 Frascati, Roma, Italy\\
\esac ESA/ESAC, PO Box 78, Villanueva de la Ca\~{n}ada, 28691, Madrid, Spain\\
\unife Dipartimento di Fisica e Scienze della Terra, Universit\`{a} degli Studi di Ferrara e Sezione INFN di Ferrara, via Saragat 1, 44100, Ferrara, Italy\\
\porto Instituto de Astrof\'isica e Ci{\^e}ncias do Espa\c{c}o, Universidade do Porto, CAUP, Rua das Estrelas, PT4150-762 Porto, Portugal\\
\unirm Dipartimento di Fisica, Universit\`{a} di Roma ``La Sapienza'', P.le Aldo Moro 2, 00138, Roma, Italy\\
\diet DIET, Universit\`{a} di Roma ``La Sapienza'', 00185, Roma, Italy\\
\estec ESA, Directorate of Science, Science Support Office, ESTEC/SCI-S, Keplerlaan 1, NL-2201 AZ Noordwijk, The Netherlands\\
\chile Departamento de Fisica, Universidad de Atacama, Copayapu 485, Copiap\'o, Chile\\
\strbg Observatoire Astronomique de Strasbourg, Universit\'e de Strasbourg, CNRS, UMR 7550, 11 rue de l'Universit\'e, 67000, Strasbourg, France\\
\geneve Observatoire de l'Universit\'e de Gen\`{e}ve, 51 chemin des Maillettes, 1290, Sauverny, Switzerland\\
\cas CAS Key Laboratory of Space Astronomy and Technology, National Astronomical Observatories, Chinese Academy of Sciences, Beijing 100012, China\\
\laval D\'epartement de physique, de g\'enie physique et d'optique and Centre de recherche en Astrophysique du Qu\'ebec, Universit\'e Laval, 1045 avenue de la m\'edecine, Québec, G1V 0A6, Canada\\
\ira Italian ALMA Regional Centre, INAF-IRA, Via P. Gobetti 101, 40129 Bologna, Italy\\
\hertf Centre for Astrophysics Research, School of Physics Astronomy \& Mathematics, University of Hertfordshire, College Lane, Hatfield, AL10 9AB, UK\\
\oact INAF - Astrophysical Observatory of Catania, Via Santa Sofia 78, 95123, Catania, Italy\\
\lanca Jeremiah Horrocks Institute, University of Central Lancashire, Preston PR1 2HE, UK\\
\cern CERN, 385 Route de Meyrin, 1217 Meyrin, Switzerland\\
\oana INAF - Astronomical Observatory of Capodimonte, via Moiariello 16, I-80131 Napoli, Italy\\
\oats INAF - Osservatorio Astronomico di Trieste, via G.B. Tiepolo 11, Trieste, Italy\\
\unina Department of Physics ``E.Pancini'', University Federico II, via Cinthia 6, I-80126 Napoli, Italy\\
\sztaki Institute for Computer Science and Control (MTA SZTAKI), Laboratory of Parallel and Distributed Systems, Victor Hugo u. 18-22., Budapest 1132, Hungary\\
\ucl Department of Physics and Astronomy, University College London, London, WC1E 6BT, UK\\
\ral STFC, Rutherford Appleton Labs, Didcot, OX11QX, UK\\
\nobeyama Nobeyama Radio Observatory, 462-2 Nobeyama Minamimaki-mura, Minamisaku-gun, Nagano 384-1305, Japan\\

\label{lastpage}
\end{document}